\documentclass[usenatbib]{mnras}

\date{Accepted 11 February 2020}

\usepackage{amsmath}
\usepackage{amssymb}
\usepackage{graphicx}
\usepackage{xspace}
\usepackage{tabularx}
\usepackage{listings}
\usepackage{color}

\newcommand{\be}{\begin{equation}}
\newcommand{\ee}{\end{equation}}
\newcommand{\bea}{\begin{subequations}\begin{eqnarray}}
\newcommand{\eea}{\end{eqnarray}\end{subequations}}
\newcommand{\eq}[1]{Equation~(\ref{eq:#1})}

\newcommand{\fig}[1]{Figure~\ref{fig:#1}}
\newcommand{\tab}[1]{Table~\ref{tab:#1}}
\newcommand{\s}[1]{Section~\ref{s:#1}}
\renewcommand{\ss}[1]{Section~\ref{ss:#1}}
\newcommand{\sss}[1]{Section~\ref{sss:#1}}
\newcommand{\ap}[1]{Appendix~\ref{a:#1}}
\renewcommand{\aa}[1]{Appendix~\ref{aa:#1}}

\newcommand{\lcdm}{$\Lambda$CDM\xspace}

\newcommand{\msun}{{\rm M}_{\odot}}
\newcommand{\msunh}{h^{-1}{\rm M}_{\odot}}
\newcommand{\mres}{M_{\rm min}}
\newcommand{\mcut}{M_{\rm cut}}
\newcommand{\mvir}{M}
\newcommand{\mpc}{{\rm Mpc}}

\newcommand{\eg}{e.g.,\xspace}
\newcommand{\ie}{i.e.,\xspace}

\newcommand{\sn}{s_{\rm new}}

\newcommand{\sinit}{s_{\rm init}}

\newcommand{\sa}{s_{\rm a}}
\renewcommand{\d}{{\rm d}}
\newcommand{\dx}{\delta x}
\newcommand{\dm}{\delta m}
\newcommand{\ds}{\delta s}
\newcommand{\eqsdef}{Equations~(4)\xspace}
\newcommand{\zf}{z_{\rm form}}

\newcommand{\surfs}{SURFS\xspace}
\newcommand{\shark}{SHARK\xspace}
\newcommand{\colwidth}{0.992\columnwidth}
\newcommand{\mcrit}{M_*}
\newcommand{\cnfw}{c}
\newcommand{\pc}{~percent\xspace}
\renewcommand{\O}{\mathcal{O}}

\title[Structure of Merger Trees]{Characterising the Structure of Halo Merger Trees Using a Single Parameter: The Tree Entropy}

\author[D. Obreschkow et al.]{Danail Obreschkow$^{1,2}$, Pascal J. Elahi$^{1,2}$, Claudia del P. Lagos$^{1,2}$,
\newauthor Rhys J. J. Poulton$^{1,2}$ and Aaron D. Ludlow$^{1,2}$\vspace{1mm}\\
\!\!$^{1}$\,International Centre for Radio Astronomy Research, M468, University of Western Australia, 35 Stirling Hwy, Perth, WA 6009, Australia\\
\!\!$^{2}$\,ARC Centre of Excellence for All Sky Astrophysics in 3 Dimensions (ASTRO 3D)}

\begin{document}

\maketitle

\label{firstpage}

\begin{abstract}
Linking the properties of galaxies to the assembly history of their dark matter haloes is a central aim of galaxy evolution theory. This paper introduces a dimensionless parameter $s\in[0,1]$, the `tree entropy', to parametrise the geometry of a halo's entire mass assembly hierarchy, building on a generalisation of Shannon's information entropy. By construction, the minimum entropy ($s=0$) corresponds to smoothly assembled haloes without any mergers. In contrast, the highest entropy ($s=1$) represents haloes grown purely by equal-mass binary mergers. Using simulated merger trees extracted from the cosmological $N$-body simulation \surfs, we compute the natural distribution of $s$, a skewed bell curve peaking near $s=0.4$. This distribution exhibits weak dependences on halo mass $M$ and redshift $z$, which can be reduced to a single dependence on the relative peak height $\delta_{\rm c}/\sigma(M,z)$ in the matter perturbation field. By exploring the correlations between $s$ and global galaxy properties generated by the \shark semi-analytic model, we find that $s$ contains a significant amount of information on the morphology of galaxies -- in fact more information than the spin, concentration and assembly time of the halo. Therefore, the tree entropy provides an information-rich link between galaxies and their dark matter haloes.
\end{abstract}

\begin{keywords}
cosmology: large-scale structure of Universe -- galaxies: evolution.
\end{keywords}

\section{Introduction}\label{s:introduction}

Progress in observational and theoretical astrophysics over the last few decades has solidified the concept that galaxies form in the gravitational potential wells of dark matter haloes \citep{White1978}. In the standard Cold Dark Matter (CDM) model, these haloes form bottom-up \citep{Peebles1965} through a hierarchical coalescence of \textit{progenitors} into more massive \textit{descendants}, as well as through the accretion of diffuse matter. These processes, known as \textit{mergers} and \textit{smooth accretion}, respectively, are responsible for roughly 2/3 and 1/3, respectively, of the mass growth at any halo mass \citep{Genel2010,Wang2011}.

In the CDM framework of structure formation, the stellar content of galaxies grows via two processes: \textit{internally} by gas cooling and star formation in the host halo; and \textit{externally} by accretion of other galaxies \citep{Guo2008,Zolotov2009,Oser2010,Pillepich2015}. The balance between these two processes affects a number of key observables, such as the morphology \citep{Kormendy2009}, the amount of kinematic support \citep{Lagos2018a} and the structure of the stellar halo \citep{Helmi2018}. Since the coalescence of galaxies is normally triggered by an earlier merger of their haloes, the hierarchical assembly of haloes is pivotal to understanding the evolution of galaxies \citep{Toomre1972,White1991,Conselice2014}.

The rich assembly history of haloes not only dictates the hierarchical growth of galaxies, but it also largely determines the internal structure of the haloes themselves. It is widely accepted that the rich quasi-fractal substructure seen in simulated CDM haloes \citep{Springel2008b} is a biased sub-sample of the progenitor hierarchy \citep[\eg][]{Moore1999,Klypin1999b}. Further, there is good evidence that the approximately universal density profile of haloes, the `NFW profile' \citep{Navarro1996}, is itself a consequence of their universal mass accretion histories (\citealp{Ludlow2013}; but see \citealp{Pontzen2013} for a different view). At very least, there is growing evidence that the internal structure of haloes is largely determined by their assembly history \citep[\eg][]{Navarro1997,Bullock2001,Wechsler2002,Zhao2009,Correa2015a} and retains a memory of the primordial fragments \citep{Gao2008b,Ludlow2016}.

In studying the bottom-up mass assembly of haloes, it useful to represent this history by abstract trees, known as \textit{merger trees}, where the branches denote haloes that join hierarchically \citep{Roukema1993}. In mathematics, trees are defined as graphs in which any two vertices are connected by exactly one path \citep{Mesbahi2010}. Halo merger trees are a subclass of such trees, called \textit{rooted directional in-trees}, where all edges (branches) are directed towards a single \textit{root} vertex. The direction represents the direction of time and the root represents the final halo. Following this definition, halo histories approximated by merger trees cannot contain disjoint branches or haloes that split up as time progresses (but generalisations exist).

It is useful to ascribe a \textit{mass} to each branch in a merger tree, reflecting the physical mass of the halo this branch represents. Graphically, the mass can be visualised by the thickness of the branches \citep[\eg Fig. 6 in][]{Lacey1993}. Halo merger trees then look like real trees, in that they follow Leonardo da Vinci's rule \citep{Richter1970} whereby the mass (or cross-sectional area of real trees) remains conserved at the vertices. Smooth accretion (or merger events that lie below the resolution limit of a simulation) can be accommodated in this representation by allowing the branches to continually increase their mass between mergers. Similarly, it is also possible to account for the potential stripping/evaporation of mass (negative accretion) by decreasing the mass along individual branches.

Despite their importance for galactic evolution, only a few attempts have been made to quantify merger histories (\eg see discussions of the `tree contours', \citealp{ForeroRomero2009}; `bushiness', \citealp{Wang2016b}; and `effective number of streams', \citealp{Wang2017}). It is common to resort to, \eg the progenitor mass function (PMF; \citealp{Bond1991}) and the main branch mass assembly history (MAH; \eg \citealp{Li2007,Correa2015b}). However, these measures discard the chronological ordering of mergers (in the case of the PMF) and all except the main branch in a tree (in the case of the MAH). Moreoever, these measures are themselves complicated mathematical functions, whose statistical comparison is not straightforward. 

The current struggle to quantify merger trees contrasts with the widespread use of classification schemes for single merger events. It is common to label mergers by the number $n$ of coalescing haloes, distinguishing between \textit{binary mergers} ($n=2$) and \textit{multi-mergers} ($n>2$). For binary mergers, the mass ratio $r\equiv m/M\in(0,1]$ is often used to measure the significance of the merger \citep{Fakhouri2010} and divide between \textit{minor} (normally $r<1/3$) and \textit{major} ($r\geq1/3$) mergers; sometimes the dividing threshold is set to 1/10 \citep{Genel2010}. In the case of galaxy-galaxy mergers, it is further common to specify the types \citep[\eg][]{Dubinski1996,Springel2005b}, structure and gas fraction (\eg \textit{wet} versus \textit{dry} mergers) of the objects involved.

\begin{figure}
  \includegraphics[width=\colwidth]{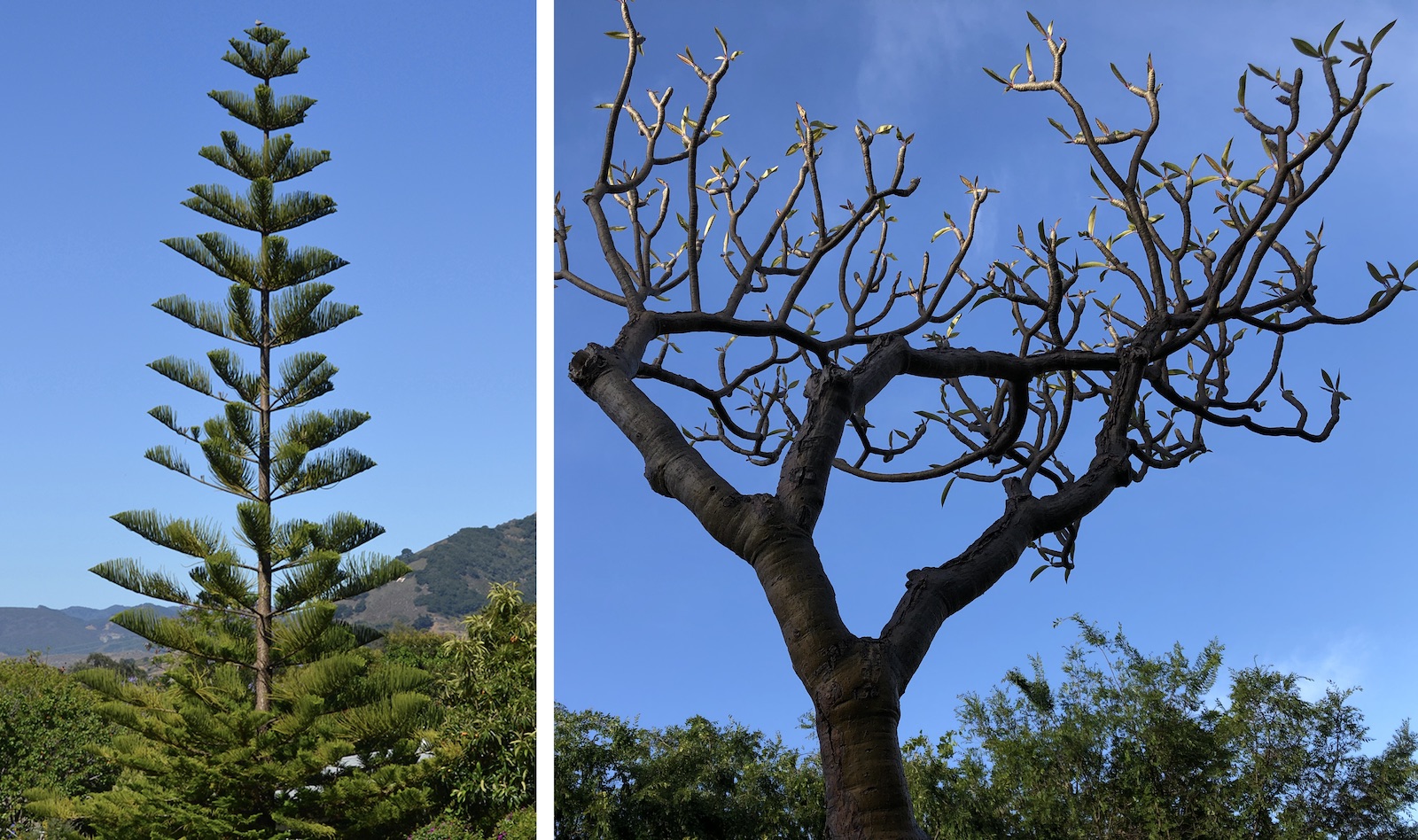}
  \caption{Natural trees that are structurally similar to two extreme types of merger trees. The Norfolk pine (Araucaria heterophylla, left) resembles `minimal' merger trees with no significant mergers in their history; the White Frangipani (Plumeria obtusa, right) is similar to `maximal' merger trees made exclusively of equal-mass binary mergers. By definition, these trees have minimal ($s=0$) and maximal ($s=1$) tree entropy, respectively.}
  \label{fig:australian_trees}
\end{figure}

The goal of this paper is to propose and analyse a parameter $s\in[0,1]$, the \textit{tree entropy}, that extends the mass ratio $r$ of single binary mergers to entire trees (and subtrees thereof), also allowing for multi-mergers and smooth mass changes between merger events. The $r$-value of single binary mergers has two extrema: $r\rightarrow0$ is the limit of a minor merger with a vanishing minor mass ($m\rightarrow0$), whereas $r=1$ denotes a major merger of identical masses ($m=M$). By analogy, we will define the $s$-value such that $s=0$ corresponds to trees without mergers, grown exclusively by smooth accretion, whereas $s=1$ denotes trees grown exclusively by an infinite hierarchy of equal-mass binary mergers. These two types of trees will thus be called \emph{minimal} and \emph{maximal} merger trees, respectively. Structurally, these two types of trees are analogous to the real trees depicted in \fig{australian_trees}. Another biological analogy of minimal versus maximal merger trees is the inheritance of mitochondrial versus nuclear DNA. The former is passed on exclusively by the mothers, whereas the latter is subject to Mendelian inheritance (50\pc maternal, 50\pc paternal)\footnote{Anecdote: An impetus for publishing this work came from the question, whether the histories of galaxies resemble rather mitochondrial or nuclear DNA, asked by Paul Schechter at the conference on `The Physics of Galaxy Scaling Relations and the Nature of Dark Matter' (Queen's University, July 2018).}.

As we shall see, the convention of $s$ for minimal and maximal trees, combined with a list of physical requirements, allows us to propose a heuristic definition for $s$, inspired by Shannon's information entropy. We will show that this definition of $s$ carries new information about the haloes, not yet contained in other standard parameters. Furthermore, $s$ also holds a significant amount of information on the physical properties of the galaxies in the haloes, at least within the physics hardwired in semi-analytic models.

\s{derivation} motivates and formally introduces the tree entropy $s$. The most important properties of $s$ are discussed and visualised using mock trees in \s{properties}. In \s{cdmsim}, we determine the tree entropies in a pure CDM universe, using the \surfs $N$-body simulation suite. Using these simulated data, we analyse the statistics of $s$, its cosmic evolution and its relation to other halo properties. \s{galaxies} focuses on the question of how the global properties of galaxies depend on the tree entropy of their host haloes. This discussion includes an information analysis that compares the information content of $s$ and other halo properties relative to the Hubble morphology. \s{conclusion} summarises the paper with an outlook to future applications.


\section{The `entropy' of merger trees}\label{s:derivation}

The aim of this section is to introduce a parameter $s$ quantifying the complexity of rooted directional trees, such as halo merger trees. We first clarify the definition of such merger trees (\ss{mergertrees}) and then explain the guiding principles for constructing the parameter $s$ (\ss{requirements}), before formally defining this parameter (\ss{definition}). The result-focused reader may skip Sections \ref{ss:mergertrees} and \ref{ss:requirements}.

\subsection{Types of tree representations}\label{ss:mergertrees}

\begin{figure}
  \includegraphics[width=\colwidth]{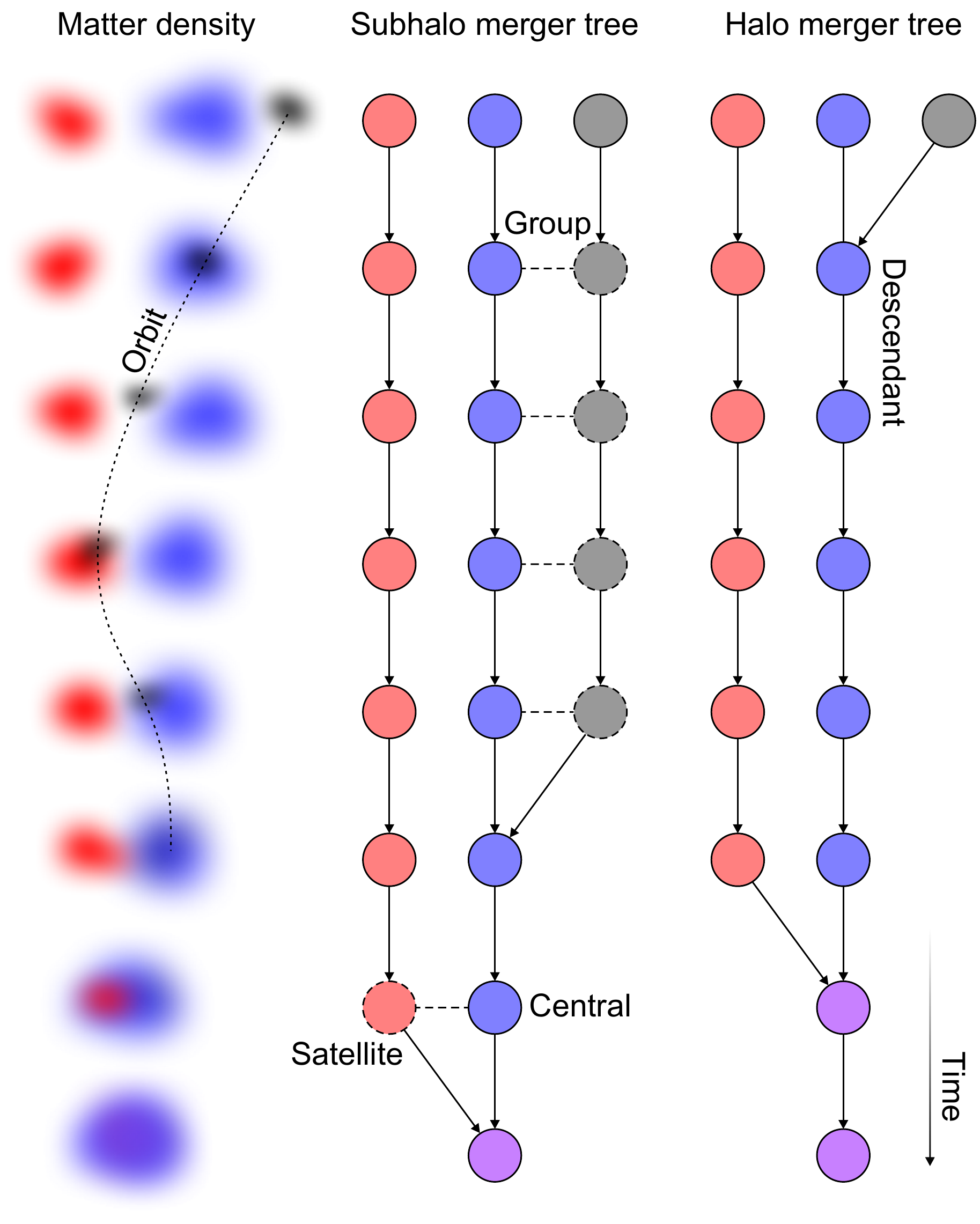}\vspace{-2mm}
  \caption{Different representations of the cosmic evolution of three haloes over eight time-snapshots (top to bottom). As shown on the left, the smallest halo (black) passes through the largest one (blue), then briefly escapes into the nearby intermediate halo (red), before fully merging with the largest one. By construction, the halo tree representation (right) merges structures at their first encounter and does not allow the small halo to escape again. The subhalo representation keeps track of substructure (dashed circles), but most subhalo tree algorithms force substructure to remain linked to a fixed central (horizontal dashed lines).}
  \label{fig:tree_approximation}
\end{figure}

It is important to distinguish between `haloes' and `subhaloes' and their associated `halo merger trees' and `subhalo merger trees', illustrated in \fig{tree_approximation}.

\textit{Halo} merger trees represent the assembly history of first generation haloes, i.e. gravitationally bound structures, which are \emph{not} themselves substructures of larger bound systems. All the self-bound substructure that exists within such haloes is considered a part of the haloes. When two haloes become gravitationally bound to each other they are merged into a single new halo, irrespective of whether one of them temporarily becomes a self-bound substructure within the other. In numerical simulations, haloes are normally identified using a basic friends-of-friends (FOF) algorithm \citep{Davis1985} or derivatives thereof.

By contrast, \textit{subhaloes} refer to self-bound lumps of matter, irrespective of whether they are a sub-system of a larger bound system. Self-bound substructures within a subhalo are considered different subhaloes. There can be multiple generations of subhaloes within subhaloes. In this terminology, the `haloes' (in the above sense) \emph{without} their substructure are called `central' subhaloes (also known as `main', `first generation' and `background' subhaloes), whereas their self-bound substructures are called `satellite' subhaloes (see \citealp{Han2017} for a more in-depth discussion). A plethora of algorithms exist to identify subhaloes in numerical simulations (see overview by \citealp{Onions2012}). They usually combine three-dimensional position-based FOF \citep[e.g.][]{Springel2008} or six-dimensional phase space-based FOF \citep[e.g.][]{Elahi2019} and unbinding techniques with a hierarchical algorithm to identify substructure.

Dark matter simulations can be used to construct merger trees by linking the haloes and subhaloes identified at different output times (`snapshots'). Tree-building algorithms \citep[see overview in][]{Srisawat2013} normally enforce a strict hierarchy with exactly one descendant for each halo/subhalo. In subhalo representations, satellites are normally (but not always, \eg \citealp{Jiang2014}) forced to remain within a fixed central. Enforcing a strict tree and subtree structure is inherently an approximation, since real haloes/subhaloes occasionally split up and/or transition to another tree/branch (\eg the black satellite depicted in \fig{tree_approximation}). However, in reality, only a small fraction ($\sim5$\pc for the SUFRS simulation discussed in \s{cdmsim}) of the haloes and subhaloes show such anomalous behaviour. Most tree-builders also demand that each halo/subhalo has a descendant in the immediate next snapshot, which sometimes requires interpolating across a few snapshots, \eg when a satellite is hard to distinguish from its central subhalo (\eg black satellite in the second snapshot of \fig{tree_approximation}).

Halo merger trees are frequently used in studying the assembly hierarchy of haloes and the evolution of the halo mass function. Examples include the seminal extended Press-Schechter (EPS) formalism \citep{Bond1991}, as well as halo growth studies using statistically generated `Monte Carlo' merger trees and $N$-body simulation-based merger trees (see \citealp{Jiang2014b} for an overview).

Subhalo trees are less adequate to study the hierarchical assembly of haloes, because two coalescing lumps normally first become a central and a satellite. By the time these two subhaloes actually merge into a single subhalo, the satellite may have mostly evaporated into the central and therefore the mass ratio of the merger is strongly dependent on how long satellites are tracked before they merge. In turn, subhalo trees are better suited than halo trees for galaxy evolution studies, in particular for semi-analytic models \citep[\eg][]{Benson2012,Lacey2016,Croton2016,Lagos2018b}, which benefit from tracking the haloes of galaxies that fall into a larger halo (\eg a group halo).

Since this work focuses on mass assembly, we will adopt halo trees rather than subhalo trees. In the halo tree representation, satellite subhaloes (dashed circles in \fig{tree_approximation}) are lost. In order to study the mass assembly history of satellites themselves, it is possible to construct `halo trees' at the generation of the satellite. For example, the `halo tree' of a second generation subhalo (\ie a satellite, whose immediate host is a central), is a progenitor tree made only of second generation subhaloes, where all higher generation substructure is considered part of the second generation subhaloes. This definition is analogous to the standard definition of `halo trees' for first generation haloes shown in \fig{tree_approximation}.

\subsection{Search for a new parameter}\label{ss:requirements}

Let us now turn to the task of introducing a parameter $s(t)$ quantifying the complexity of the mass assembly tree of a halo at a given time $t$. By choice, this parameter should only depend on the geometry of the tree, \ie the descendant-links and progenitor masses, akin to structural parameters in graph theory \citep{Balakrishnan2012}. We deliberately discard any other information, such as time scales, impact factors, angular momentum, tidal dynamics, etc. This will allow us to compare the information contained in the tree geometry to that of more common quantities, such as the mass, spin and concentration (\s{cdmsim}).

In the broadest sense, we would like $s$ to quantify the complexity of the mass assembly hierarchy in a physically `meaningful' sense. By this we understand that $s$ should carry statistical information on the halo's history and on the galaxy/galaxies that reside inside the halo. For instance, we can think of global morpho-kinematic galaxy properties (\eg the bulge-to-disk stellar mass ratio and the related random-to-ordered stellar velocity ratio), which are already known to depend significantly on merger history (see refs. in \s{introduction}). Expressing this idea in explicit terms is, of course, a challenging affair, which requires a mixture of heuristics and prior knowledge of the galaxy-halo connection.

In view of the approximate scale-invariance of cosmic structure \citep{Blumenthal1984,Davis1985}, it makes sense to define $s$ independently of the overall mass scale. In other words, we require $s$ to be invariant under a uniform rescaling of all the progenitor masses of a halo. Since $s$ then only depends on the dimensionless descendant-links and relative progenitor masses, $s$ must itself be a \emph{dimensionless} parameter. Without loss of generality, we can restrict $s$ to the interval $[0,1]$ and request that the two extrema should correspond to haloes that have been the least ($s=0$) and most ($s=1$) strongly reshaped by mergers, respectively. In this way, $s$ will become a natural extension of the mass ratio $r=m/M\in[0,1]$ of a single binary halo-halo merger.

\subsubsection{Self-similar trees}\label{sss:selfsimilar}

\begin{figure}
  \includegraphics[width=\colwidth]{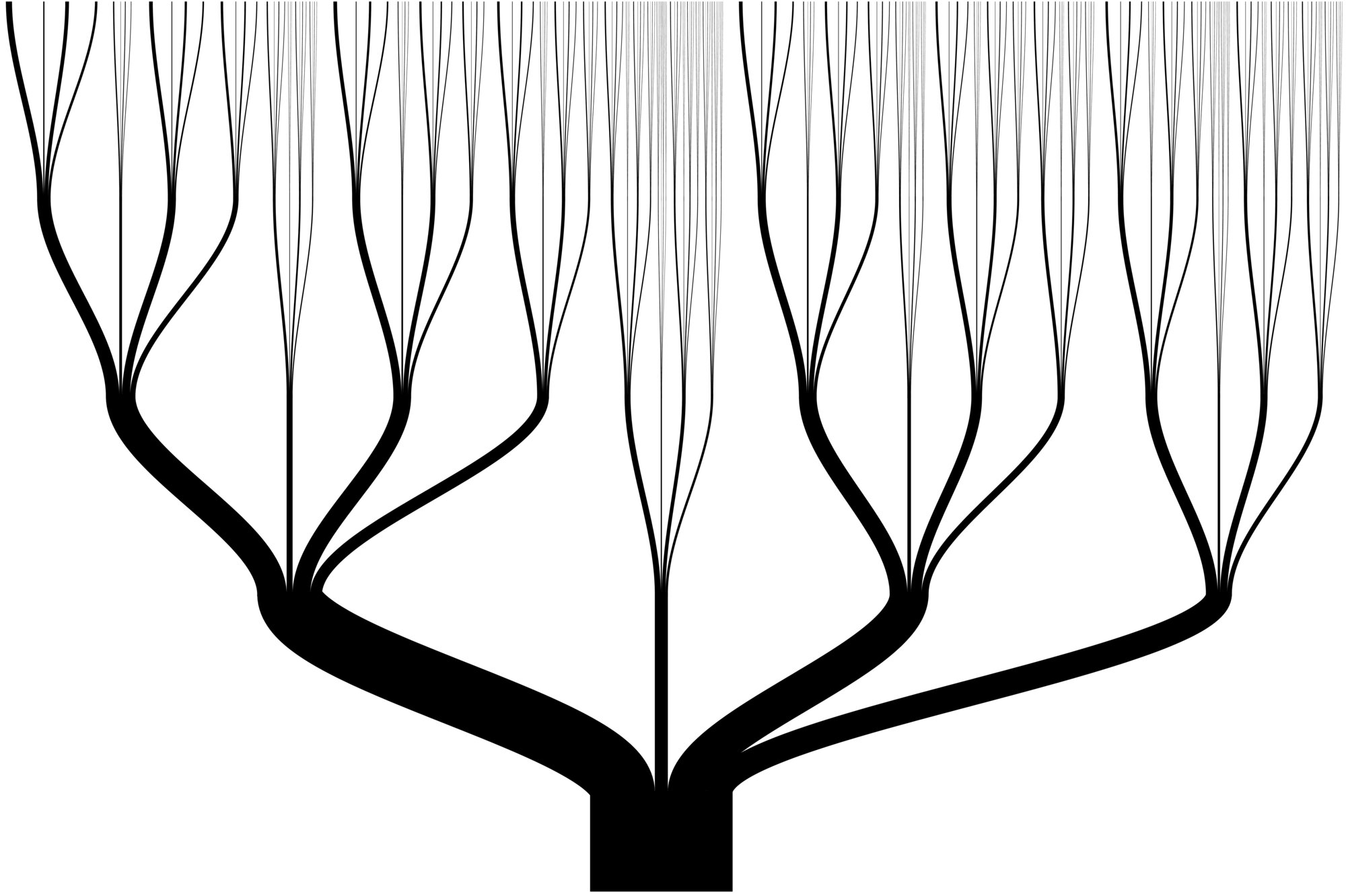}
  \caption{Illustration of a self-similar tree made of 4th order multi-mergers of fixed mass ratio 5:1:3:2. By scale-invariance, the tree parameter $s$ must take the same value anywhere in this tree.}
  \label{fig:self_similar_tree}
\end{figure}

In formalising a definition of the parameter $s(t)$, self-similar merger trees without smooth accretion turn out to be a fruitful starting point. Self-similar trees are hypothetical trees, in which every halo has the same assembly history up to an overall mass scaling (\fig{self_similar_tree}). Our requirement that $s$ should not depend on the overall mass scale then implies that each halo must have the same value of $s$. Thus, self-similar trees reduce the problem of defining a time-dependent tree parameter $s(t)$ to the simpler task of defining a single \emph{time-independent} parameter $s$ for the entire tree. This parameter can only depend on the mass fractions $x_i\equiv(m_i/\sum_{i=1}^n m_i)$ of the $n$ progenitors joining at each node; and a physically meaningful definition should be independent of the mass ordering.

The problem of quantifying the geometry of a self-similar tree thus reduces to assigning a number $s$ to the unordered set $\{x_i\}$, where $\sum x_i=1$. To address this problem, we use the prior knowledge that major mergers (i.e. with comparable values of $x_i$) have much stronger ramifications than minor ones in terms of restructuring the dark matter \citep{Ludlow2012,Klypin2016} and transforming the dynamics and kinematics  of galaxies \cite[e.g.][]{Conselice2014,Lagos2018a}. Translated to our problem, this means that more similar values $\{x_i\}$ should lead to higher values of $s$ than more diverse sets. In this qualitative sense, the problem of defining $s$ is reminiscent of quantifying the \textit{information} of a probability set $\{x_i\}$. This information (\ie the number of bits required to encode events that occur with probabilities $\{x_i\}$), is given by Shannon's information entropy $\mathcal{H}=-\sum x_i\ln x_i$ (with $0\ln0=0$; \citealp{Shannon1948}).

The information entropy $\mathcal{H}$ is maximal if all values $x_i$ are identical and minimal ($\mathcal{H}=0$), if all values of $x_i$ except one vanish (which corresponds to a halo `merging' with zero-mass haloes, \ie no merger at all). While these are desirable properties for our complexity measure, $\mathcal{H}$ has the unfavourable property of diverging as $n\rightarrow\infty$. This not only violates our requirement for $s$ to remain bound to $[0,1]$, but also works against the physical insight that a halo grown from an instantaneous merger of $n\rightarrow\infty$ progenitors is simply a smoothly collapsed halo, which should thus have as low of a complexity as a minimal tree ($n=1$). This requirement could be met by using the normalised information entropy $\mathcal{H}/n$ \citep{Katok2007}, which vanishes both for $n=1$ and $n\rightarrow\infty$. However, in the present context, such a normalisation is not useful, since adding branches with vanishing mass ($x_i\approx0$) to a merger would increase $n$, but would have no physical effect on the halo. Thus $s$ must not explicitly depend on $n$ in this case. A clever way of bypassing the use of $n$ is to raise the normalised masses $x_i$ to the power of a constant $\alpha>1$ in the definition of $\mathcal{H}$. This gives rise to a \textit{generalised information entropy},
\be\label{eq:generalizedshannon}
	H = -f\sum_{i=1}^n x_i^\alpha\ln x_i,
\ee
where the normalisation factor $f=(\alpha-1)e$ (with Euler's constant $e$) ensures that $H$ spans the interval $[0,1]$. For any finite $\alpha>1$, $H$ vanishes for mergers of $n=1$ and $n\rightarrow\infty$ haloes (with non-zero masses); and adding empty branches ($x_i=0$) has no effect on $H$. Similar generalisations of the Shannon entropy \citep[\eg][]{Mathai2007} have been proposed and successfully applied in other fields, for instance in defining income equality metrics in econometry \citep{Cowell1980,Shorrocks1980}. 

Since $H$ has the properties that we would expect from an astrophysically motivated complexity measure of self-similar merger trees, we heuristically define that the parameter $s$ of a self-similar tree is identical to its generalised information entropy $H$. By virtue of this definition, $s$ is called the \textit{tree entropy} for the remainder of this paper.

\eq{generalizedshannon} requires a choice for $\alpha>1$. To understand the role of $\alpha$, it helps to consider a self-similar tree of equal-mass mergers, \ie the $n$ merging branches each have identical mass fractions $x_i=n^{-1}$. It is straightforward to show (\aa{equalmass}) that $H(n)$ vanishes for $n=1$ and $n=\infty$ and presents a single maximum ($H=1$) at $n_{\rm c}=e^{1/(\alpha-1)}$.

In this work, we choose $n_{\rm c}=2$, giving $\alpha=1+1/\ln2\approx2.442695$ and $f=e/\ln2\approx3.921652$. This choice follows from our original guideline (\s{introduction}) that maximal merger trees, made of an infinite regression of equal-mass \emph{binary} ($n=2$) mergers, have maximum tree entropy ($s=1$). Physically, this choice can be motivated by the fact that multi-mergers ($n>2$) are likely to arise from a coherent assembly, \ie a correlated inflow of haloes on a time scale smaller than a single merger event. Such coherent assembly is expected to result in more ordered structure (\eg in terms of halo and galaxy spin) than a major binary merger. For instance, simulations of a merger of six equal-mass spiral galaxies in a group \citep{Weil1996} found that such mergers tend to produce remnants with more rotation than typically seen in dry binary mergers \citep[\eg][]{Cox2006}. Hence, with respect to the galaxy kinematics, equal-mass binary mergers are more transformational than multi-mergers.

The optimal choice of $n_{\rm c}$, in the sense of maximising the information pertaining to a particular halo or galaxy property, may depend on the context. 
Choosing $n_{\rm c}=2$ is a pragmatic way to fix the ideas, but alternatives will be discussed in \ss{optimisation}. Incidentally, we note that multi-mergers are subdominant at any redshift in the \lcdm simulation analysed in \s{cdmsim}; and observations of galaxy-galaxy interactions find that the incidence of multi-mergers relative to binaries lies at the percent level \citep{Darg2011}.

\subsubsection{Extension to arbitrary trees}\label{sss:nonselfsimilar}

Next, we need to extend the time-independent global definition $s=H$ for the entropy of self-similar trees (\sss{selfsimilar}) to a \emph{time-dependent} definition of $s(t)$ in arbitrary merger trees. This definition requires a local equation governing the evolution of $s$. Let us first establish a list of physically motivated mathematical conditions for this equation and then develop an Ansatz satisfying these conditions.

Being a measure of the mass assembly hierarchy, $s(t)$ should only change if the mass of a halo changes (condition~1), be it via discrete mergers or smooth accretion. In the case of mergers, the transition from the entropies $\{s_i\}$ of $n$ merging branches to the new entropy $\sn$ of their descendant can only depend on $\{s_i\}$ and on the mass fractions $\{x_i\}$. Following the reasoning of \sss{selfsimilar}, $\sn$ must be a continuous and symmetric function of $\{(x_i,s_i)\}$ (condition~2), bound to $[0,1]$ (condition~3). For consistency with self-similar trees, $\sn$ must asymptote to $H$ for a succession of self-similar mergers (\ie mergers with identical mass fractions $\{x_i\}$, where each merger takes the output entropy of the previous merger as its input). In particular, if all $n$ input entropies $s_i$ are identical ($s_i\equiv s$), the output entropy $\sn$ should be contained in the interval $[s,H]$ (condition~4).

The rule for evolving $s(t)$ under smooth accretion can be determined as the limiting case, where diffuse material of mass $\Delta m$, is added via $p\rightarrow\infty$ minor $n$-mergers ($n\geq2$), each adding $(n-1)$ infinitesimal masses $\delta m=\Delta m/p/(n-1)$ to the main branch. In turn, physical considerations for the evolution of $s(t)$ under smooth accretion will constrain the definition of the merger equation. One such consideration is that the evolution of $s(t)$ under smooth accretion cannot depend on the value of $n$, since it is physically impossible to distinguish binary from multi-mergers in the case of smooth accretion. Similarly, $s(t)$ must \textit{not} depend on the tree entropy $s_{\rm d}$ of the smoothly accreted diffuse material. This is because the macroscopic state of a halo/galaxy cannot depend on whether each smoothly accreted infinitesimal part was itself formed by mergers or not. The only sensible choice is that the smoothly evolving $s(t)$ is independent of $n$ and $s_{\rm d}$ (condition~5) and asymptotically tends to zero as $\Delta m\rightarrow\infty$ (condition~6), consistent with the vanishing entropy of minimal trees. This argument extends to the case, where a halo is formed entirely by a smooth collapse. In particular, if a halo of mass $m$ is formed through the simultaneous coalescence of $n$ progenitors of mass $m/n$, the final tree entropy must be independent of the progenitor entropies and tend to zero as $n\rightarrow\infty$ (condition~7).

All seven conditions above are satisfied by the \textit{Ansatz},
\be\label{eq:weightansatz}
	\sn = H+w\sum_{i=1}^n x_i^2(s_i-H),
\ee
where $H$ depends on $\{x_i\}$ via \eq{generalizedshannon} and $w$ is a function, temporarily taken to be unity. \eq{weightansatz} manifestly satisfies the conditions 1--4. In particular, the difference terms $(s_i-H)$ ensure that $\sn$ is always contained in $[s,H]$ if $s_i\equiv s$ (condition~4). It is necessary to weigh these difference terms by $x_i^q$ with some $q>0$ to ensure that adding haloes with zero mass do not change the result. Condition~7 requires that the sum in \eq{generalizedshannon} vanishes as $n\rightarrow\infty$, which is satisfied iff $q>1$, making $q=2$ the simplest choice. As shown in \aa{smooth}, this choice (but not $q=1$) also satisfies conditions~5 and~6.

The weight $w$ in \eq{weightansatz} specifies how well $\sn$ `remembers' the input entropies $s_i$. A careful choice of $w$ as a function of $\{x_i\}$ allows us to preserve the conditions~1--7, while further specifying two crucial quantities: a rate parameter $\beta\in[0,1]$ determining how fast the most destructive mergers (at $H\approx1$) build up tree entropy and a rate parameter $\gamma\in[0,1]$ determining how fast smooth accretion (at $H\approx0$) removes tree entropy. Explicitly, we demand that a single $H=1$ merger (that is an equal-mass binary merger for our choice of $\alpha$) with zero progenitor entropy results in a tree entropy $\beta$ (condition~8); and that accreting an infinitesimal mass $\d m$ to a halo of mass $m$ and tree entropy $s$ results in an entropy change $(\d s/s)=-\gamma\,(\d m/m)$ (condition~9). In the `no-merger limit', where all but one progenitor haloes have vanishing mass ($x_i=1$, $x_{j\neq i}=0$; thus $H=0$), we require that $\sn=s_i$. Hence, $w$ must be unity in this case. A straightfoward \textit{Ansatz} that satisfies this requirement, while offering two free parameters to accommodate conditions~8 and~9, is the second order polynomial
\be\label{eq:wansatz}
	w = 1+aH+bH^2
\ee
with real constants $a$ and $b$. A short derivation (\aa{smooth}) shows that conditions~8 and~9 translate to $a=(2-\gamma)/f$ and $b=n_{\rm c}(1-\beta)-1-a$. The value of $\sn$ remains bound to $[0,1]$ as long as $\gamma\geq0$.

\eq{weightansatz} guarantees that an infinite regression of self-similar mergers asymptotes to $\sn\rightarrow H$, which is maximal ($H=1$) if the mergers are equal-mass mergers of $n=n_{\rm c}$ haloes. However, \eq{weightansatz} does not guarantee that a \emph{single} merger of $n$ identical haloes ($x_i\equiv n^{-1}$, $s_i\equiv s$) maximises $\sn$ if $n=n_{\rm c}$. For this to be true, $\beta$ must lie above a threshold that depends on $\alpha$ and $\gamma$. A lower bound, which works for all $\gamma\in[0,1]$, reads $\beta\geq\tanh(1.165/(\alpha-1))$. In our case ($n_{\rm c}=2$), the exact limit is about $\beta>0.65$.

Given our choice $n_{\rm c}=2$, a value of $\beta=1$ would mean that equal-mass binary mergers always produce the maximum entropy ($\sn=H=1$), irrespective of their input entropies. Thus, such mergers would completely erase the memory of all past mergers. This choice could be justified on the basis that equal-mass binary mergers dramatically redistribute the particle orbits in a `violent relaxation' process \citep{Lynden-Bell1967} and largely reshape disk galaxies at the halo centres, typically resulting in dispersion-rich featureless spheroids \citep{Toomre1977}. However, detailed simulation-based studies of merging identical disk galaxies (triggered by equal-mass halo-halo mergers) show that the morphological and kinematic memory is not entirely lost. For instance, dissipation-free mergers of two pure stellar disks result in spheroidal remnants with slightly less central density than analogous binary mergers of stellar disk with a bulge \citep{Hernquist1993} and multiple mergers might be needed to fine-tune the fundamental plane \citep{Taranu2015}. Likewise, single dissipational mergers of gas-rich disks might only partially remove the rotation structure \citep{Cox2006} and multiple such mergers seem necessary to explain some of the slowly rotating massive elliptical galaxies \citep{Dubinski1998}. To allow for some memory in equal-mass binary mergers, we adopt $\beta=3/4$ in the definition of the tree entropy.

To choose a value for $\gamma$, we recall that early-type galaxies, likely associated with high-entropy trees (\ss{morphology}), are rarely seen to transform back into late-type ones \citep{Emsellem2007}, despite the predicted continued growth of their haloes (by a factor $\sim3$ since redshift $z=1$; \eg \citealp{Elahi2018}). This can be attributed to feedback from active galactic nuclei, as well as to the fact that the specific angular momentum of haloes increases with cosmic time \citep{Wang2018} and thus late accretion settles less efficiently onto a galaxy in a `biased collapse' scenario \citep{Dutton2012}. Thus, it makes sense to demand that the relative decrease in tree entropy is significantly smaller than the relative increase in mass during smooth accretion, \ie $\gamma\ll1$. Here we choose $\gamma=1/3$, such that \textit{halving} the tree entropy requires a smooth mass increase by a factor $2^3$, which corresponds to \textit{doubling} the halo radius.

Our choices for $\alpha$, $\beta$ and $\gamma$ are, of course, subjective and alternatives will be discussed in \ss{optimisation}.

\subsection{Concise definition of the tree entropy}\label{ss:definition}

The developments of \ss{requirements} have led to an astrophysically motivated definition of a dimensionless parameter $s\in[0,1]$ -- the \emph{tree entropy} -- which quantifies the complexity of hierarchical mass assembly. The minimum entropy ($s=0$) stands for minimal trees without mergers, whereas the maximum entropy ($s=1$) represents maximal trees with a fractal history of equal-mass binary mergers.

We choose the initial entropy of a halo to be $\sinit=0$. This is a natural choice, as it is equivalent to saying that the earliest haloes were assembled from diffuse matter, which is true for many potential CDM and warm DM particles (\eg neutralions; \citealp{Diemand2005}). In practise, the smallest haloes at the free-streaming limit of about an Earth mass \citep{Angulo2010} are rarely resolved in cosmological simulations (but see \citealp{Wang2019b}). Hence, the first haloes to appear in such simulations should arguably have non-zero tree entropy. However, as we will show in \ss{binary} and \fig{pythagoras}, the choice of $\sinit$ is not critical for the results, since the final entropy of a tree does not depend significantly on the initial values for well-resolved merger histories ($\gtrsim$ 10 leaves).


Given an initially vanishing tree entropy, the tree entropy at all later times is calculated successively from previous time steps, using two mutually consistent rules, one governing merger events and one governing the time between successive mergers. In either rule, the tree entropy only changes if the halo mass changes.

\subsubsection{Tree entropy change during mergers}

If $n$ haloes of mass $m_i\geq0$ and tree entropy $s_i\in[0,1]$ ($i=1,...,n$) merge simultaneously into a single halo, the tree entropy $\sn\in[0,1]$ of this new halo is calculated by successively evaluating three equations:
\begin{subequations}
\label{eq:definition}
\begin{align}
        &x_i = m_i/\sum_{i=1}^n m_i~~\text{(normalised masses)},\label{eq:defxi}\\
	&H = -f\sum_{i=1}^n x_i^\alpha\ln x_i~~\text{(generalised information)},\label{eq:defH}\\
	&\sn = H+(1+aH+bH^2)\sum_{i=1}^n x_i^2(s_i-H)\label{eq:defsn},
\end{align}
\end{subequations}
with real constants $f=e\cdot(\alpha-1)$, $a=(2-\gamma)/f$, $b=n_{\rm c}(1-\beta)-1-a$. The default values in this paper are $f=3.921652$, $a=0.424991$, $b=-0.924991$. These constants depend on the three parameters $\alpha$, $\beta$ and $\gamma$, which regulate the dependence of the tree entropy on the type of mergers as detailed in \ss{requirements}. Briefly,
\begin{itemize}
\item $\alpha>1$ (here 2.442695) specifies the relative importance of different orders of mergers (e.g. binary versus triple mergers). The lower the value of $\alpha$, the more entropy is produced by higher order merges relative to lower order ones. Our default choice $\alpha=1+1/\ln2\approx2.442695$ is such that binary mergers produce the most entropy and that maximal merger trees have maximum entropy ($s=1$).
\item $\beta\in[0,1]$ (here 3/4) regulates the rate of entropy change in the most destructive single merger (here an equal-mass binary merger). The higher $\beta$, the closer the output entropy of a single major merger lies to the final entropy $H$ of an infinite cascade of self-similar mergers.\item $\gamma\in[0,1]$ (here 1/3) regulates the rate of tree entropy loss in the smooth accretion limit: the higher $\gamma$, the faster the entropy reduction under smooth accretion.
\end{itemize}

\subsubsection{Tree entropy evolution between mergers}\label{sss:smoothevolution}

The rule for evolving the tree entropy under smooth accretion of a mass $\Delta m$ cannot be chosen freely, but must be derived from these equations. Let us subdivide the smooth accretion into a very large number $p$ of successive mergers, each adding an equal amount of mass of entropy $\sa$. If each merger is thought of as a binary merger ($n=2$), then each joining halo has a small mass $\delta m=\Delta m/p$. For mergers of arbitrary order ($n\geq2$), this mass is $\delta m=\Delta m/p/(n-1)$. Through a short calculation it can be shown that in the limit of $p\rightarrow\infty$ and irrespective of $n$, a $p$-fold repetition of \eqsdef (while increasing the main mass by $\delta m$ at each iteration) leads to an entropy
\be\label{eq:smooth}
	\sn = s\left(\frac{m}{m+\Delta m}\right)^\gamma,
\ee
where $s$ is the tree entropy of the halo of mass $m$ before the smooth accretion. The independence of \eq{smooth} on $\sa$ and $n$ is a crucial requirement for the entropy to be a meaningful measure, since it is conceptually pointless to assign a tree entropy to infinitesimal mass elements and/or to distinguish between binary or higher-order mergers for such elements.


\section{Illustration of tree entropy}\label{s:properties}

The purpose of this section is to illustrate the evolution of the tree entropy $s$, using idealised examples.

\subsection{Tree entropy of single mergers}\label{ss:basic}

\eqsdef governing the tree entropy change during mergers satisfy a list of meta-properties that are necessary in order for $s$ to be a physically meaningful parameter:

\begin{itemize}
	\item \textit{Markovianity:} The tree entropy of a halo only depends on the preceeding time step. Specifically, the new $s$ of a halo immediately after a merger only depends on the masses $m_i$ and tree entropies $s_i$ ($i=1,...,n$) of its $n$ progenitors just before the merger.
	\item \textit{Scale invariance:} Only the ratios, not the absolute values, of the progenitors masses $m_i$ are relevant for the entropy of the descendant. Thus only normalised progenitor masses $x_i\equiv(m_i/\sum m_i)$ appear in \eqsdef.
	\item \textit{Permutation symmetry:} The tree entropy of a descendant halo does not depend on the ordering of its simultaneously merging progenitors.
	\item \textit{No-merger limit:} A merger of $n+k$ haloes with $n$ non-vanishing masses ($m_i>0$, $\forall\,i\leq n$) and $k$ vanishing masses ($m_i=0$, $\forall\,i>n$) results in the same entropy as a merger of the $n$ non-vanishing masses. In particular, if a halo of mass $m_1>0$ and entropy $s_1$ `merges' with some vanishing masses $m_i=0$ ($\forall\,i>1$), the entropy of the resulting halo is $s_1$.
	\item \textit{Continuity:} Vanishing changes in the progenitor masses $\{m_i\}$ and entropies $\{s_i\}$ imply asymptotically vanishing changes in the entropies of the descendant.
\end{itemize}

\begin{figure}
  \includegraphics[width=\colwidth]{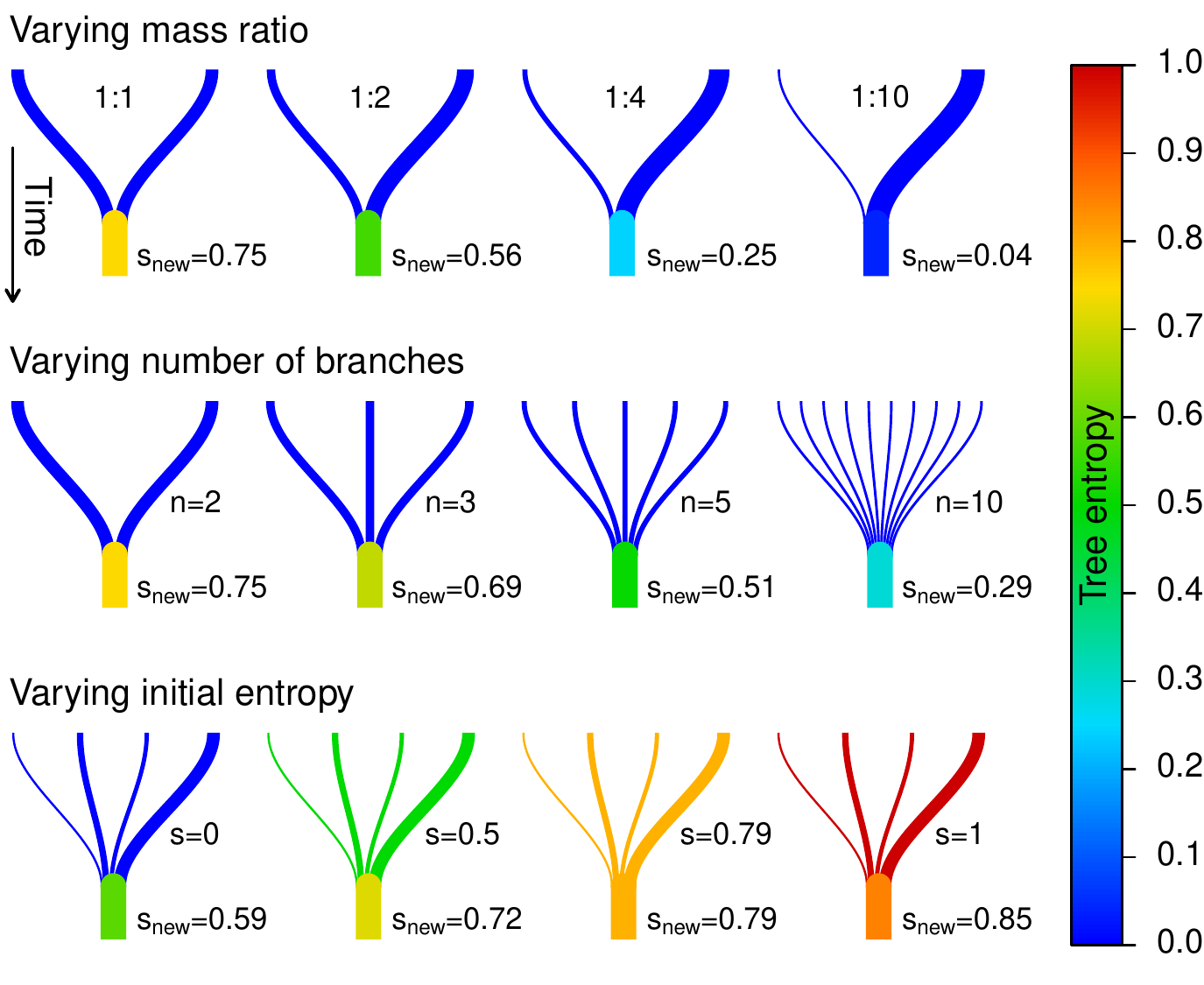}
  \caption{Evolution of the tree entropy $s$ in idealised mergers. In each row, one property of the merger is varied: the binary mass ratio $r$ (top); the number $n$ of equal-mass haloes (middle); the input tree entropy $s$ (bottom) -- see explanations in \ss{basic}. The bottom row uses a 4-merger with a fixed mass ratio of 1:3:2:6.}
  \label{fig:overview}
\end{figure}

To illustrate the qualitative evolution of $s$ during a merger event \fig{overview} shows some contrived cases. In each row, only one aspect of the merger changes, thus illustrating three important properties:

\begin{itemize}
	\item \textit{Mass ratio scaling:} A merger of two haloes of zero tree entropy generates an entropy that is a monotonically increasing function of the mass ratio $r=m/M$. Equal-mass binary mergers generate the largest entropy ($\sn=\beta=3/4$) of any single merger event.
	\item \textit{Multi-merger scaling:} If $n\geq2$ haloes of identical mass and tree entropy merge, the outcome entropy $\sn$ is a decreasing function of $n$ and asymptotes to zero as $n\rightarrow\infty$ (see \ss{smooth} for a visualisation of this limit).
	\item \textit{Entropy linearity:} If $n$ haloes of identical entropies $s$ merge, the outcome entropy $\sn$ is a monotonically, in fact linearly, increasing function of $s$.
\end{itemize}

A key addition to the last point is that s and $\sn$ are not proportional to one another and will generally cross-over at some point. Whether $\sn$ is smaller or larger than $s$ depends on $s$ and the generalised information entropy $H(\{x_i\})$ associated with the mass ratios. If $n$ haloes of identical tree entropy $s$ merge, the entropy
\begin{itemize}
	\item increases towards $H$, if $s<H$;
	\item remains unchanged, if $s=H$;
	\item decreases towards $H$, if $s>H$.
\end{itemize}
Thus, every merger moves the tree entropy towards $H$ (=0.79 in the bottom row of \fig{overview}). By consequence, a cascade of self-similar mergers asymptotically approaches $s=H$. Therefore, the tree entropy gradually `forgets' its past and hence the choice of the initial entropy $\sinit$ becomes irrelevant for sufficiently resolved trees (see illustration in \ss{binary}).

\subsection{Smoothly grown haloes}\label{ss:smooth}
\begin{figure}
  \includegraphics[width=\colwidth]{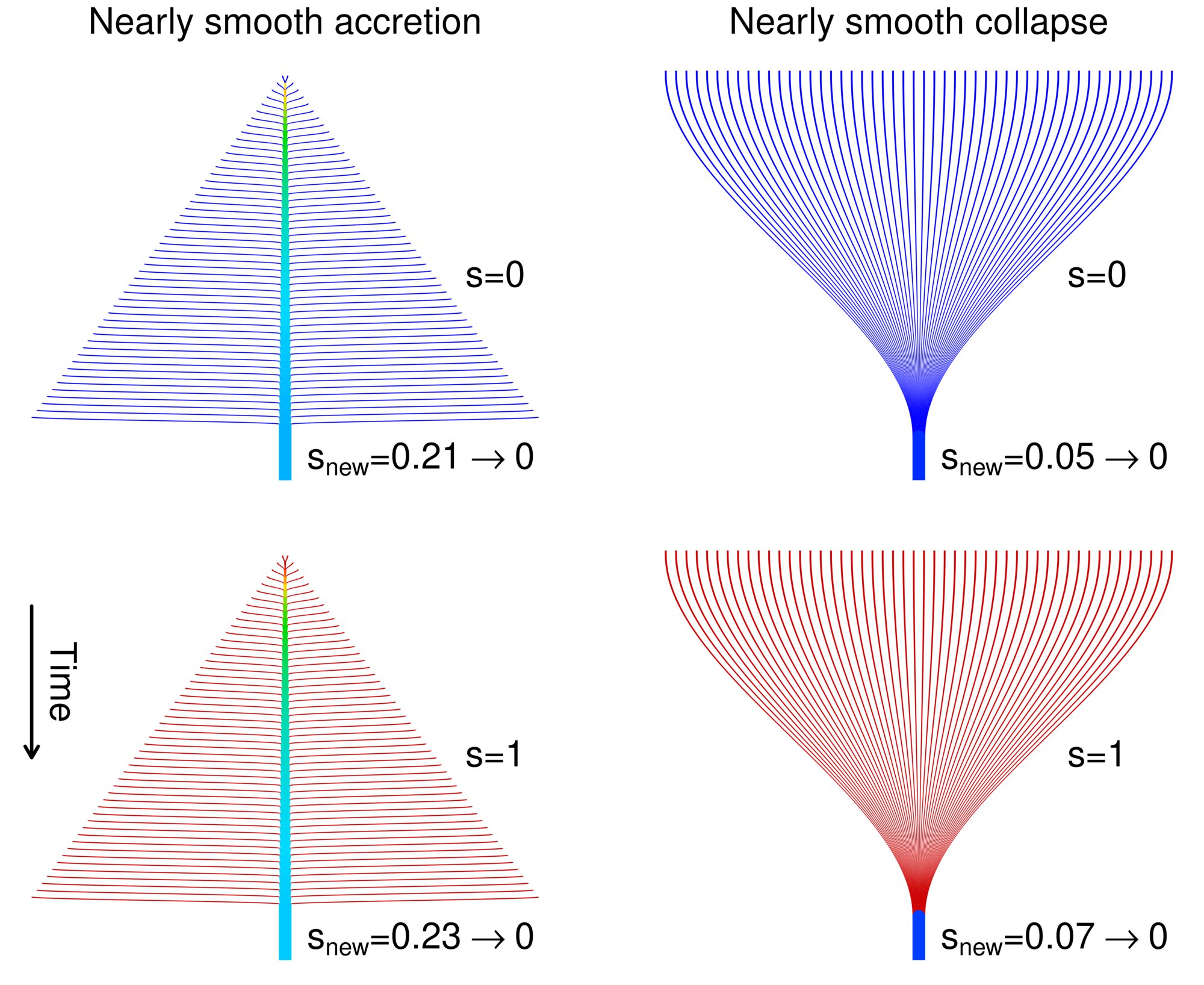}
  \caption{Nearly smooth growth scenarios: the trees on the left grew from $n=50$ very minor accretion events, the ones on the right from a single $n$-merger. In both cases the resulting tree entropy $\sn$ becomes independent of the input entropy $s$ and tends to zero as $n\rightarrow\infty$, as required for consistency with smooth accretion and smooth collapse. (Colour scale as in \fig{overview}.)}
  \label{fig:smooth}
\end{figure}

By definition, haloes grown smoothly without mergers must have vanishing tree entropy, irrespective of whether the halo was grown trough smooth accretion over some time (`minimal trees', see \s{introduction}) or through an instantaneous collapse of diffuse material. Functional continuity then implies that haloes grown almost smoothly, \ie via a large number $n$ of minor accretion events, must have asymptotically vanishing entropy as $n\rightarrow\infty$. The case of $n=50$ is illustrated in \fig{smooth}. The nearly smooth accretion (left) and nearly smooth collapse (right) result in asymptotically vanishing tree entropies (as $n\rightarrow\infty$), which also become asymptotically independent of the progenitor entropy, consistent with the impossibility to assign a tree entropy to diffuse material.

\subsection{Detailed analysis of binary mergers}\label{ss:binary}

Most mergers in halo merger trees are binary mergers. In fact in some EPS-based tree-generating algorithms all mergers are binary \citep[\eg][]{Lacey1993,Moreno2008}. In this case, the mass fractions ${x_i}$ of the merging haloes can be reduced to a single mass fraction $x$, such that $x_1=x$ and $x_2=1-x$, related to the mass ratio, $r=m/M$, via $x=1/(1+r)$. If the merging haloes have equal tree entropy $s$, the outcome $\sn$ only depends on $x$ and $s$.

\begin{figure}
  \includegraphics[width=\colwidth]{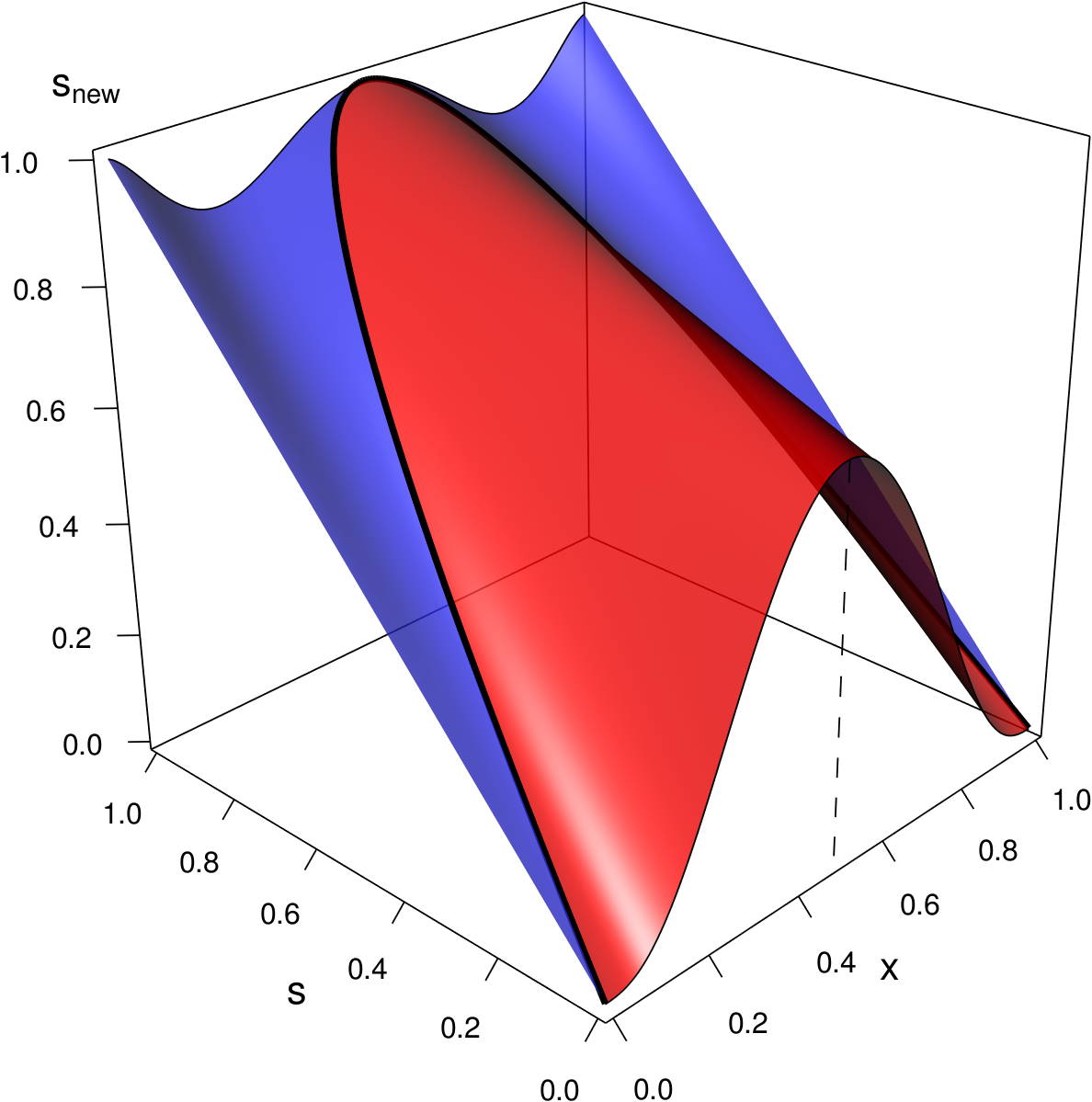}
  \caption{Outcome entropy $\sn$ for a binary merger of two haloes with normalised masses $x$ and $(1-x)$ and identical entropy $s$. Red shows the region where the merger increases the entropy ($\sn>s$); blue shows the region where the entropy decreases ($\sn<s$). The thick solid line is the critical line ($\sn=s=H$). This line gives the values of $s$ towards which self-similar binary merger trees (\fig{pythagoras}) asymptote. The dashed vertical line reaches up to $\sn=\beta=3/4$, the output tree entropy of an equal-mass binary merger with zero initial entropy.}
  \label{fig:3d}
\end{figure}

\newcommand{\myindent}{~~}
\begin{table}
\centering
\begin{tabularx}{\columnwidth}{@{\extracolsep{\fill}}ccccc}
\hline \\ [-2ex]
\myindent$r$ & $x$ & $1-x$ & $\sn$ & $H$ \\ [0.5ex]
\hline \\ [-2ex]
\myindent 0.000 & 1.000 & 0.000 & $0.0000+1.0000s$ & 0.000 \\
\myindent 0.050 & 0.952 & 0.048 & $0.0086+0.9513s$ & 0.177 \\
\myindent 0.100 & 0.909 & 0.091 & $0.0424+0.8687s$ & 0.323 \\
\myindent 0.150 & 0.870 & 0.130 & $0.0988+0.7778s$ & 0.445 \\
\myindent 0.176 & 0.850 & 0.150 & $0.1343+0.7314s$ & 0.500 \\
\myindent 0.200 & 0.833 & 0.167 & $0.1691+0.6905s$ & 0.546 \\
\myindent 0.250 & 0.800 & 0.200 & $0.2450+0.6118s$ & 0.631 \\
\myindent 0.300 & 0.769 & 0.231 & $0.3206+0.5434s$ & 0.702 \\
\myindent 0.333 & 0.750 & 0.250 & $0.3688+0.5034s$ & 0.743 \\
\myindent 0.400 & 0.714 & 0.286 & $0.4569+0.4361s$ & 0.810 \\
\myindent 0.500 & 0.667 & 0.333 & $0.5645+0.3621s$ & 0.885 \\
\myindent 0.600 & 0.625 & 0.375 & $0.6428+0.3127s$ & 0.935 \\
\myindent 0.800 & 0.556 & 0.444 & $0.7283+0.2623s$ & 0.987 \\
\myindent 1.000 & 0.500 & 0.500 & $0.7500+0.2500s$ & 1.000 \\
[0.5ex] \hline
\end{tabularx}
\caption{Tabulated outcome entropy $\sn$ of a single binary merger event of two haloes of identical entropies $s$, as a function of their mass ratio $r=(1-x)/x$. The generalised information entropy $H$ is the stationary value of $s$, where $\sn=s$. This is also the asymptotic entropy attained in an infinite regression of self-similar binary mergers with fixed mass ratios $r$.}
\label{tab:numerical}
\end{table}

The function $\sn(x,s)$ is visualised in \fig{3d}. Its symmetry about $x=0.5$ reflects the permutation symmetry $x_1\leftrightarrow x_2$. At any fixed $x$, $\sn$ is a linear function of $s$. Thus, the sheet in \fig{3d} is made of straight lines, which connect the two extremes $\sn(x,s=0)$ and $\sn(x,s=1)$ (thin solid curves) at fixed $x$. Some values of $\sn(x,s)$ and $H(x)$ have been listed in \tab{numerical}. This table shows, for instance, that $H=1/2$ if $r\approx0.176$. Hence any halo with entropy $H>1/2$ must have had at least one merger with $r>0.176$ in its assembly history. Likewise, haloes with entropy $H>0.743$ must have experienced at least one major merger with $r>1/3$. (The converse is not true.)

\begin{figure}
  \includegraphics[width=0.475\columnwidth]{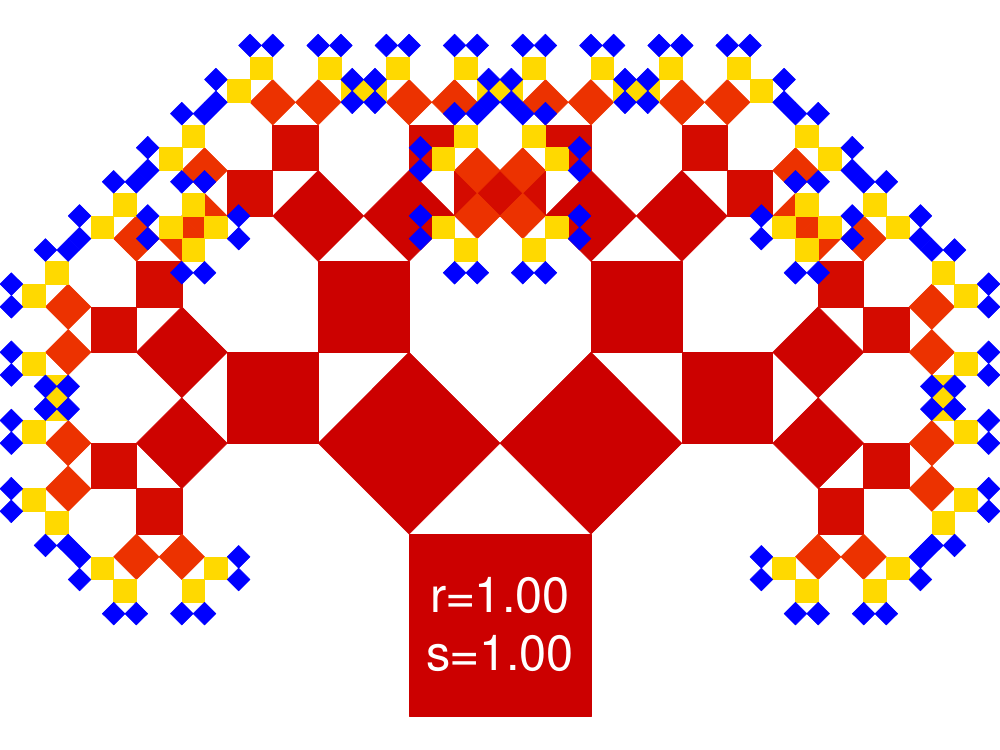}\hspace{2.5mm}
  \includegraphics[width=0.475\columnwidth]{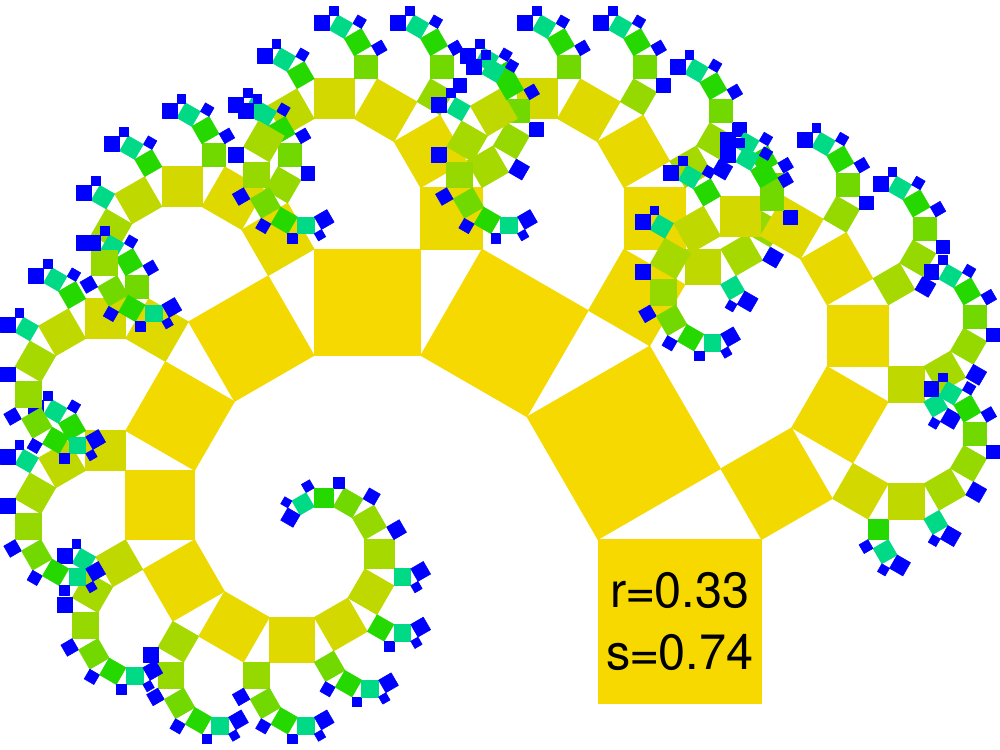}\\
  \includegraphics[width=0.475\columnwidth]{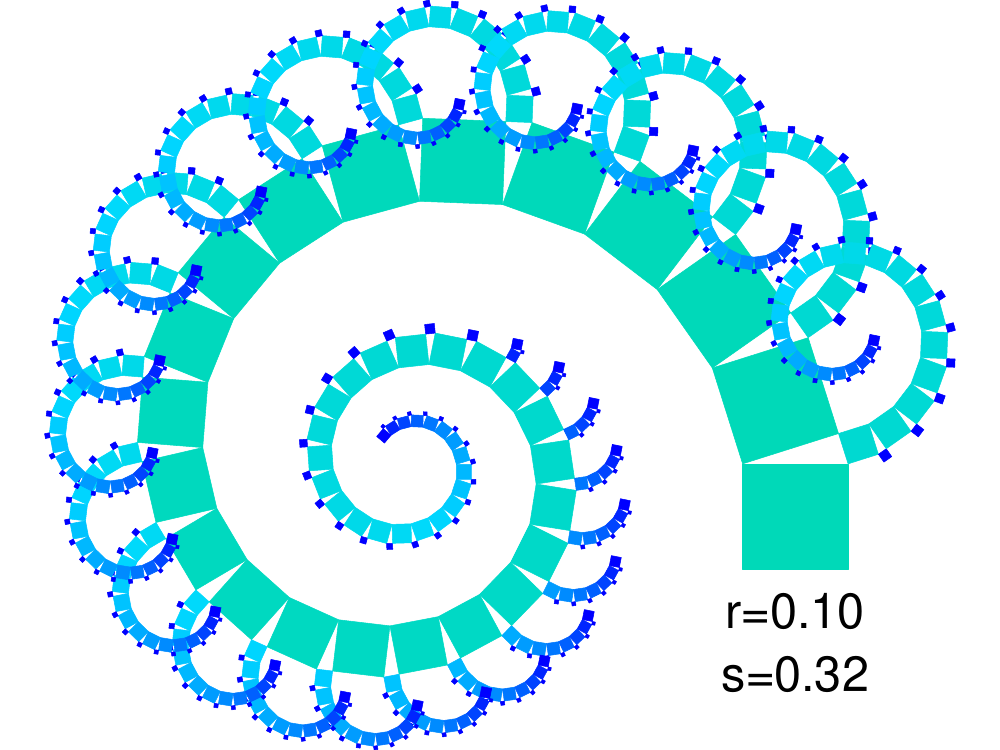}\hspace{2.5mm}
  \includegraphics[width=0.475\columnwidth]{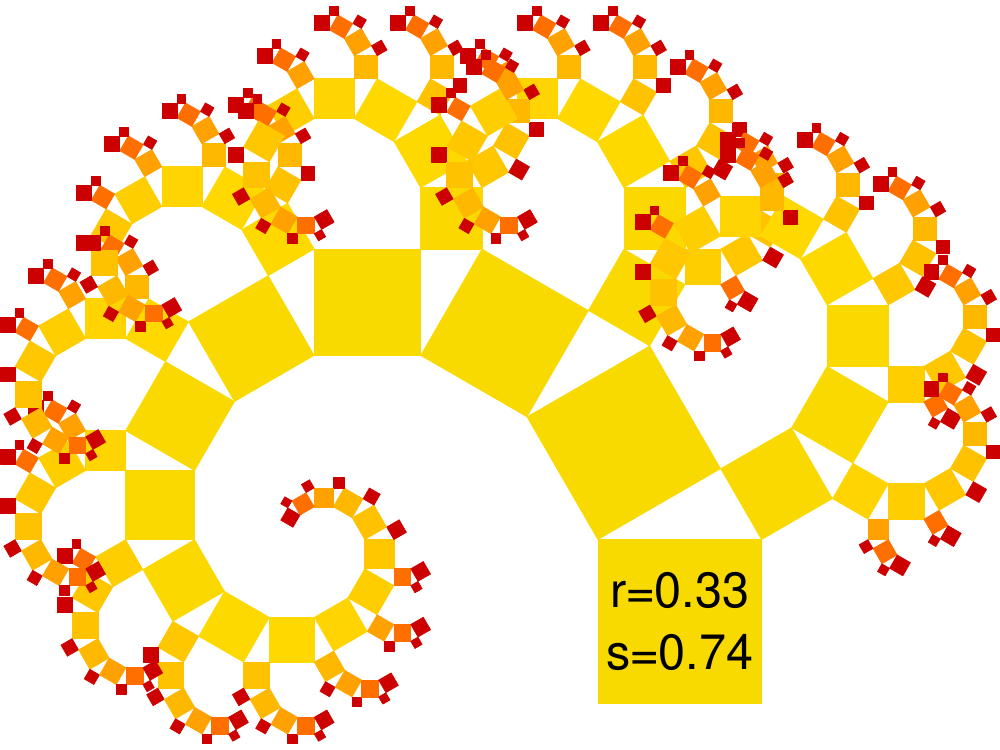}
  \caption{Tree entropy evolution in quasi-fractal binary merger trees, represented as Pythagoras trees where masses are proportional to square areas \citep{Bosman1957}. The trees have different mass ratios $r=m/M$, ranging from 1/1 (top left, a `maximal tree'), to 1/3 (right) and 1/10 (bottom left). The outcome entropy $s$ tends towards the generalised information entropy $H$, which only depends on $r$ (\eq{defH}, values in \tab{numerical}). The final entropy is independent of the initial entropies at the leaves of the tree, as seen in the two trees on the right where $\sinit=0$ (top) and $\sinit=1$ (bottom). (Colour scale as in \fig{overview}.)}
  \label{fig:pythagoras}
\end{figure}

\fig{3d} distinguishes between the regions $\sn>s$ (red) and $\sn<s$ (blue). The dividing line between these regions (thick solid curve) is the geometric location where $\sn(x,s)=s$. According to \eq{defsn}, this line corresponds to $s=H(x)=-f[x^\alpha\ln(x)+(1-x)^\alpha\ln(1-x)]$. The only other case where $\sn(x,s)=s$ is the no-merger limit, where $x=0$ or $x=1$. Hence, the function $\sn(x,s)$ is equal to $s$ if and only if $x=0$, $x=1$ and $s=H(x)$. This explains the shape of $\sn(x,s)$ with one oscillation as a function of $x$ if $s=0$ and two oscillations if $s=1$.

As with any $n$-merger, a sequence of binary mergers with identical mass ratios gradually evolves the tree entropy towards $H(x)$. \fig{pythagoras} shows this asymptotic convergence in the case of quasi-fractal binary trees, \ie trees that are self-similar down to progenitors of a minimal mass (here $10^{-2}$ times the final mass). The two trees on the right show explicitly that the final tree entropy does \textit{not} depend on the initial entropy at the leaves. 


\section{Tree entropies in \lcdm}\label{s:cdmsim}

Let us now investigate the tree entropy of realistic halo merger trees in the concordance $\Lambda$ Cold Dark Matter (\lcdm) cosmology. We will first introduce the $N$-body simulation and post-processing techniques used to build the merger trees and their tree entropies (Sections \ref{ss:surfs} and \ref{ss:computeentropy}) and then discuss the statistics and evolution of the entropy (following subsections). A pure CDM universe without baryonic physics is assumed in this section. Baryons and galaxies will be discussed in \s{galaxies}.

\subsection{Simulated merger trees}\label{ss:surfs}

Our analysis focuses on a run from the \surfs $N$-body simulation suite presented by \cite{Elahi2018}, which uses an up-to-date Planck cosmology \citep[4th column in Table 4 of][]{PlanckCollaboration2016} with characteristic densities $\Omega_{\Lambda}=0.6879$ (vacuum energy), $\Omega_{\rm m}=0.3121$ (all matter), $\Omega_{\rm b}=0.0491$ (baryonic matter, derived), scalar spectral index $n_s=0.9653$, power spectrum normalisation $\sigma_8=0.8150$ and Hubble parameter $H_0=h\cdot100~\rm{km\,s^{-1}Mpc^{-1}}$ with dimensionless Hubble parameter $h=0.6751$.

The \surfs run considered here is ``L210N1536'' (see Table~1 in \citealp{Elahi2018}), a purely gravitational $N$-body simulation of 1536$^3$ particles in a cubic box of side-length $210\, h^{-1}\mpc$ (comoving) with periodic boundary conditions. The implied particle mass is $2.21\cdot10^8\msunh$. The initial conditions at redshift $z=99$ were generated using the second order Lagrangian perturbation theory scheme (2LTP; \citealp{Crocce2012}) with a transfer function generated by CAMB \citep{Lewis2000}. Subsequently the particles were evolved to $z=0$ using a memory lean version of GADGET-2 \citep{Springel2001,Springel2005c}.

Particle positions and velocities were stored at 200 discrete time steps (snapshots), between $z=24$ and $z=0$ in evenly spaced intervals of logarithmic growth factor. This high cadence ensures that adjacent snapshots are separated by less than the free-fall time of virialised overdensities, which is necessary to generate merger trees that accurately capture the evolution of dark matter haloes. Central and satellite subhaloes were identified using VELOCIraptor \citep{Elahi2019}, which first identifies haloes using a 3D FOF algorithm in configuration space \citep{Davis1985} and subsequently identifies substructures using a 6D phase-space FOF algorithm with an unbinding algorithm. Subhaloes have a minimum of 20 particles, which implies a minimum mass of $\mres=4.42\cdot10^9\msunh$.

The subhaloes are then linked across snapshots using the particle correlator code TreeFrog \citep{Elahi2019b}, which can interpolate across several snapshots if necessary. In this work, we used a customised version of the TreeFrog outputs, in which each subhalo has exactly one descendant and satellite subhaloes stay associated with their first central. Since a characterisation of a halo's mass assembly history requires `halo merger trees' rather than `subhalo merger trees' (see \fig{tree_approximation}), we convert the subhalo trees into halo trees by assuming that a halo merges into another halo as soon as it becomes a satellite in the subhalo tree representation. As soon as this happens the two masses are added up to form a single halo and potential `back splashes', where a satellite exits and reenters its central subhalo, are ignored. The mass of haloes is defined as the total mass of their gravitationally bound particles, including bound and associated substructure. To compute the entropies of satellite subhaloes, we construct their effective `halo trees' at the level of satellites, as explained at the end of \ss{mergertrees}.

\subsection{Tree entropy calculation in simulated trees}\label{ss:computeentropy}

All new haloes (\ie without progenitors) in the simulated merger trees are initialised with tree entropy $\sinit=0$. Later tree entropies are then computed snapshot-by-snapshot by applying two computations to each halo. First, if the halo has more than one progenitor, we apply \eqsdef, to compute the entropy change caused by the merger. Second, we compute the entropy change of the halo due to smooth accretion. The latter is accomplished using \eq{smooth} with a small modification that we discuss now.

In general, the mass $m_0$ of a halo differs from the total mass of its $n\geq1$ progenitors by a non-zero amount $\Delta m=m_0-\sum_{i=1}^n m_i$. There can be several reasons for $\Delta m$ to be non-zero. First, haloes can accrete diffuse material that is not resolved into progenitor haloes. Some of this material might be truely diffuse \citep{Genel2010}, whereas some might be bound in small haloes that lie below the resolution limit of the simulation (less than 20 particles). In fact, the smallest (and first) haloes to form in \lcdm have masses of around an Earth mass, corresponding to the small-scale thermal cut-off in the CDM power spectrum \citep{Green2004,Angulo2010}. Since \eq{smooth} for handling smooth accretion also corresponds to the limit of many small mergers, this equation is suitable in either case.

Second, $\Delta m$ can be non-zero due to \textit{numerical} reasons \citep{Contreras2017}: any (sub)halo-finder, including VELOCIraptor, is subject to pseudo-random oscillations in the number of particles that are assigned to a halo at each time step. The continuously changing positions, velocities and energies make it virtually impossible to avoid such oscillations. Such oscillations can be smoothed out by applying \eq{smooth} irrespective of the sign of $\Delta m$. In this way, a spurious numerical mass gain between snapshot $1$ and $2$ of $\Delta m_{1\rightarrow2}>0$, followed by an equal mass loss $\Delta m_{2\rightarrow3}=-\Delta m_{1\rightarrow2}$ results in no net entropy change.

Third, haloes can be stripped of their material, implying a physical mass loss $\Delta m<0$. If blindly applying \eq{smooth} in this case, the tree entropy can increase indefinitely to unphysical values $s>1$. This is particularly problematic for satellite subhaloes, which can be stripped entirely, meaning that all these subhaloes would eventually reach $s>1$, even if the galaxies at their centres remain regular.

A practical way for correctly handling $\Delta m\neq0$ in all the above scenarios is to apply \eq{smooth} for all haloes, irrespective of the sign of $\Delta m$, except for satellites whose mass has dropped below the mass they had when they first became a satellite of another halo. To avoid $s>1$ in some very rare ($\ll 0.1$ percent) cases of massively stripped field haloes, we artificially impose $\sn\leq1$ in \eq{smooth}.

\ap{code} shows how the tree entropy computation is implemented programmatically using pseudo-code.

A key question that needs addressing pertains to numerical convergence: how far from the mass resolution limit of a simulation must a halo be for its tree entropy to be numerically converged? Or, inversely, in which haloes can we trust the entropy value given the resolution of the simulation? This question is addressed in detail in \aa{massconvergence}. Irrespectively of the way smooth accretion is handled (even if ignored), it turns out that the entropy values are sufficiently converged (within $\lesssim0.02$, that is $\lesssim10$\pc) in haloes of mass $\mvir\approx10^{11}\msunh$. At this mass, the smallest resolved (20 particles) progenitors are about 20-times less massive. For the whole discussion that follows we therefore only consider haloes with $\mvir\geq10^{11}\msunh$, corresponding to $>452$ particles. This cut is similar to that of \cite{Correa2015b}, who found that a minimum of $\sim300$ particles is needed for merger trees to be sufficiently resolved for analysing the main branch MAH.

The time-steps between the snapshots in \surfs was deliberately chosen below the free-fall time of the haloes to allow for a robust identification of merger trees. The convergence study of \cite{Benson2012b} found that this cadence is indeed sufficient for merger trees to capture the mass growth of haloes and galaxies with a mass error below 5\pc; in fact, they found this criterion satisfied at 128 snapshots between $z=20$ and $z=0$, whereas \surfs has an even higher cadence of 200 snapshots between $z=24$ and $z=0$. Throughout this work, we can therefore expect $s$ to be well-converged relative to the time-resolution. In \aa{timeconvergence}, we explicitly confirm this assumption by showing that using only every second snapshot only changes the tree entropy values by a marginal amount.

\begin{figure}
  \includegraphics[width=\colwidth]{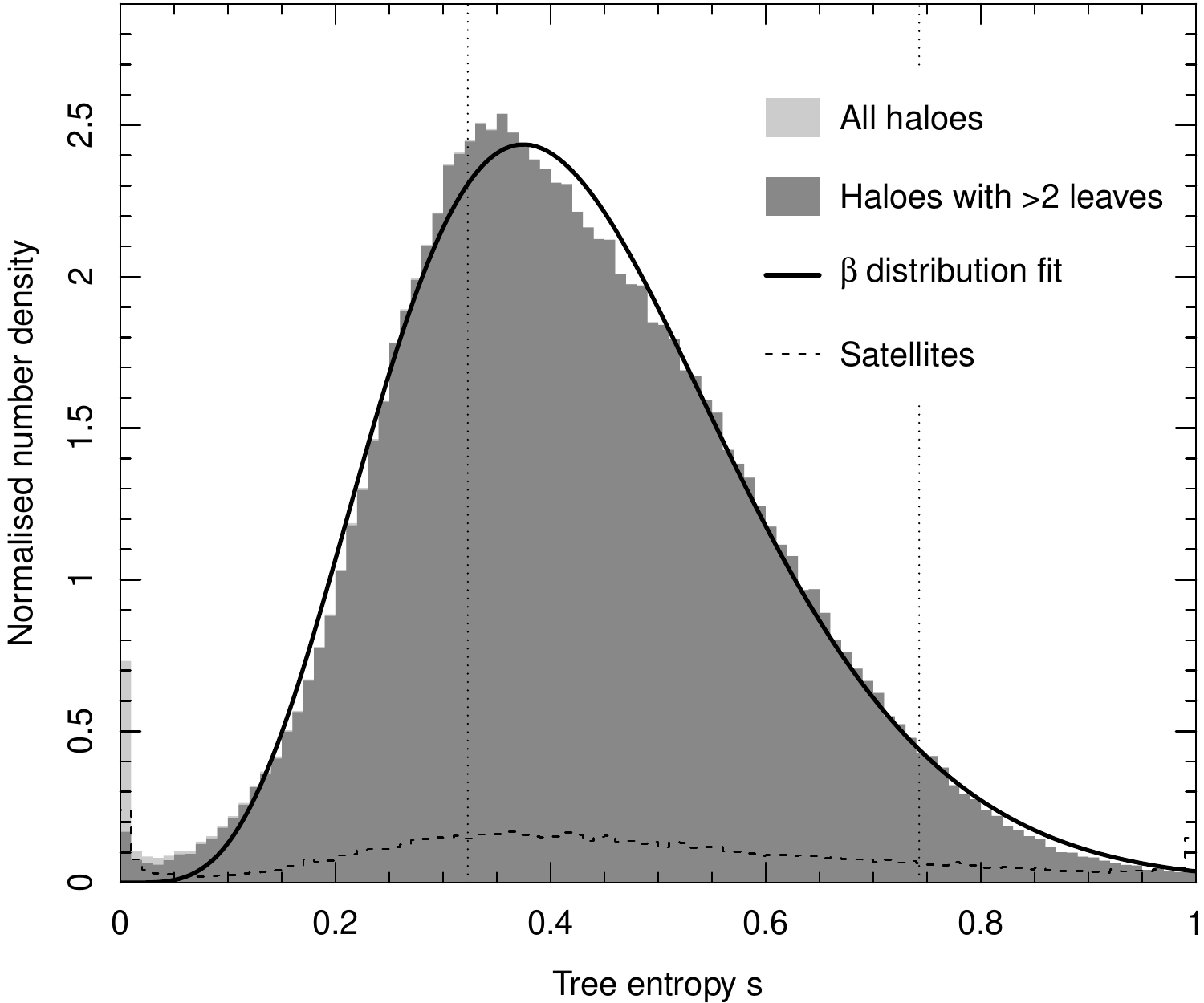}
  \caption{Distribution of the tree entropies of the simulated haloes in \surfs at $z=0$ for haloes with $\mvir\geq10^{11}\msunh$. Dotted vertical lines denote the entropies of self-similar binary trees of mass ratio 1/10 ($s\approx0.323$) and 1/3 ($s\approx0.743$).}
  \label{fig:entropy_stats}
\end{figure}

\begin{figure*}
  \includegraphics[width=1.005\columnwidth]{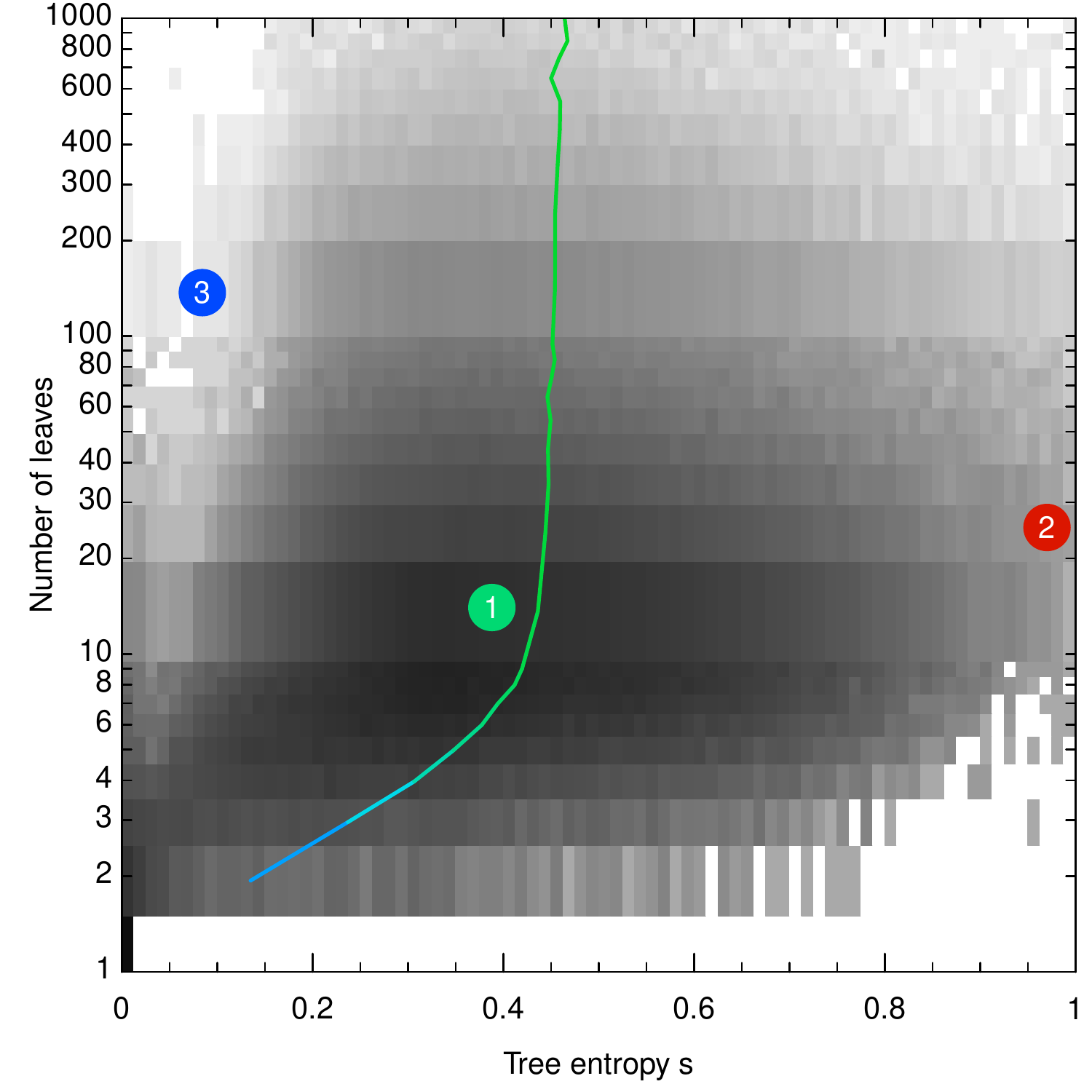}\hspace{5mm}
  \includegraphics[width=1.005\columnwidth]{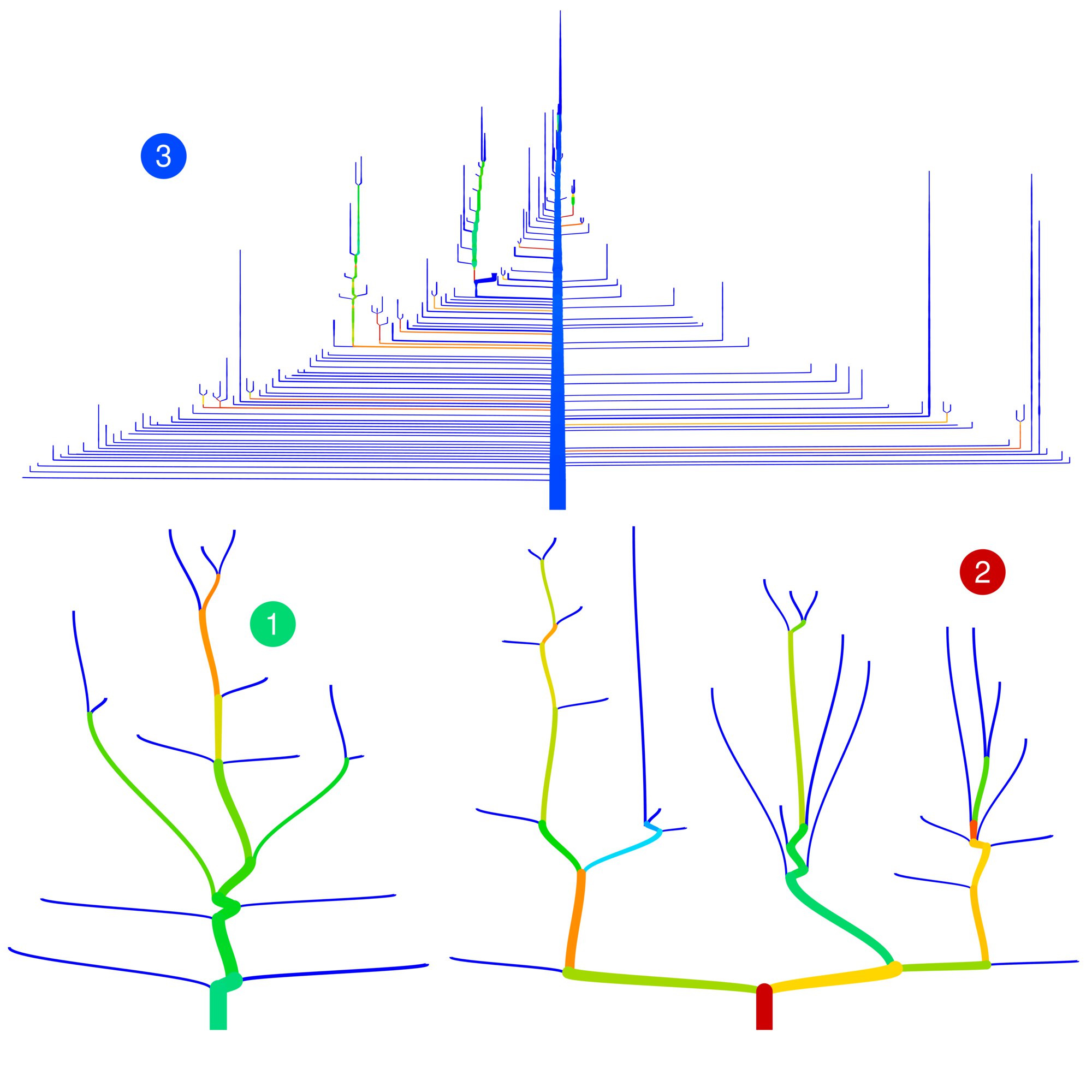}
  \caption{Left: Number density of haloes (at $z=0$) as a function of tree entropy $s$ and number of leaves $n_l$. Densities (per units of $s$ and $\log n_l$) are shown by shades of grey on a logarithmic scale from white (lowest) to black (highest). The curve indicates the mean $\bar s$ as a function of $n_l$. Right: Illustration of three halo merger trees extracted from \surfs, colour-coded as in \fig{overview}. At each merger node, the progenitors are horizontally distributed according to their mass to highlight that major mergers (\eg the last merger in the bottom right tree) have no obvious main progenitor. Curved branches have been chosen in the bottom trees for purely graphical reasons. Coloured circles with numbers indicate the location of these trees in the left panel.}
  \label{fig:entropy_nbranches}
\end{figure*}

\subsection{Global entropy statistics in \lcdm}\label{ss:basicstats}

The tree entropies of haloes ($\mvir\geq10^{11}\msunh$) at $z=0$ span the full range from $s=0$ (minimal merger trees) to $s=1$ (maximal trees). The full distribution of entropies shown in \fig{entropy_stats} is approximated by a $\beta$-distribution, a typical distribution of bound variables resulting from a cascade of random processes (halo mergers in this case).

The distribution of tree entropies peaks near $s=0.4$ (mode $0.35$, median $0.41$, mean $0.43$), approximately matching self-similar trees made of binary minor mergers of mass ratio 1/7 (\tab{numerical}). In this rough sense, we thus expect that median galaxy properties can be modelled using self-similar merger trees with 1/7 ratios. By contrast, only 3.9\pc of the trees fall into the regime ($s\geq0.743$) corresponding to self-similar binary trees made of major mergers ($r>1/3$).

The small excess of haloes with $s\approx0$ (\fig{entropy_stats}) is a numerical artefact, caused by about $0.7$\pc of haloes whose main branch was improperly assigned to a neighbouring halo close to $z=0$. This can happen, for instance, when satellite subhaloes escape from their parent entirely to restart a new life as independent halo \citep[see, \eg][]{Ludlow2009}. In the halo tree representation (\fig{tree_approximation}), the assembly history of these haloes is lost and they seem to just `pop up' shortly before or at $z=0$. Typically, these haloes only have one or two leaves, a low entropy ($s<0.05$) and a young age. Given the low incidence of these odd cases, they do not impinge on the statistical results of this paper.

For completeness, \fig{entropy_stats} also shows the tree entropy distribution of satellite subhaloes at $z=0$, accounting for 8.4\pc of the subhaloes (the remainder being centrals). The distribution is similar to that of the centrals, as expected from self-similarity arguments. On average, the tree entropy of satellites is about 1.1-times higher, likely due to the fact that their late-time growth has been halted/reduced around the time they became a satellite (typically $z\sim1$). This is in quantitative agreement with the cosmic evolution of entropies of (1st generation) haloes (see \ss{evolution}). It is also possible that numerical resolution leads to increased entropies for satellites: more massive satellites are more easily tracked inside their parent than less massive ones, hence skewing the mass ratio of satellite-satellite mergers towards major mergers. In this work, we will not further discuss the tree entropies of satellite subhaloes.

\fig{entropy_nbranches} (left) expands on the relation between tree entropy and leaf number. Trees with only one leaf (minimal trees) have, by construction, zero entropy. Trees with two leaves have had exactly one merger, which means that their entropy can be as high as $\beta=3/4$, bar some rare cases where smooth mass losses lead to slightly higher values. The distribution in \fig{entropy_nbranches} (left) reveals that the entropy distribution remains roughly constant with a mean around 0.43--0.45 (green line), for all trees with more than $\sim10$ leaves. The slight increase in $\bar s$ with number of leaves is a consequence of the weak $s$--mass relation explained in \ss{haloproperties}.

\fig{entropy_nbranches} (right) shows three merger trees extracted from the \surfs simulation. The typical scenario (example 1 in \fig{entropy_nbranches}; green circle) are trees in which the main branch experienced a few (1--3) major mergers, but the total mass contribution of these mergers lies below that of minor mergers and smooth accretion. Such trees typically settle at entropies near $s=0.4$. Trees with exceptionally high entropies (example 2; red circle) normally consist of two or more similarly massive sub-trees, which coalesce in a series of major mergers near $z=0$. In turn, trees with very low entropy (example 3; blue circle), resemble a pine tree, characterised by a strong main branch, without significant major mergers in its recent history.

\subsection{Cosmic evolution of tree entropies}\label{ss:evolution}

\begin{figure}
  \includegraphics[width=\colwidth]{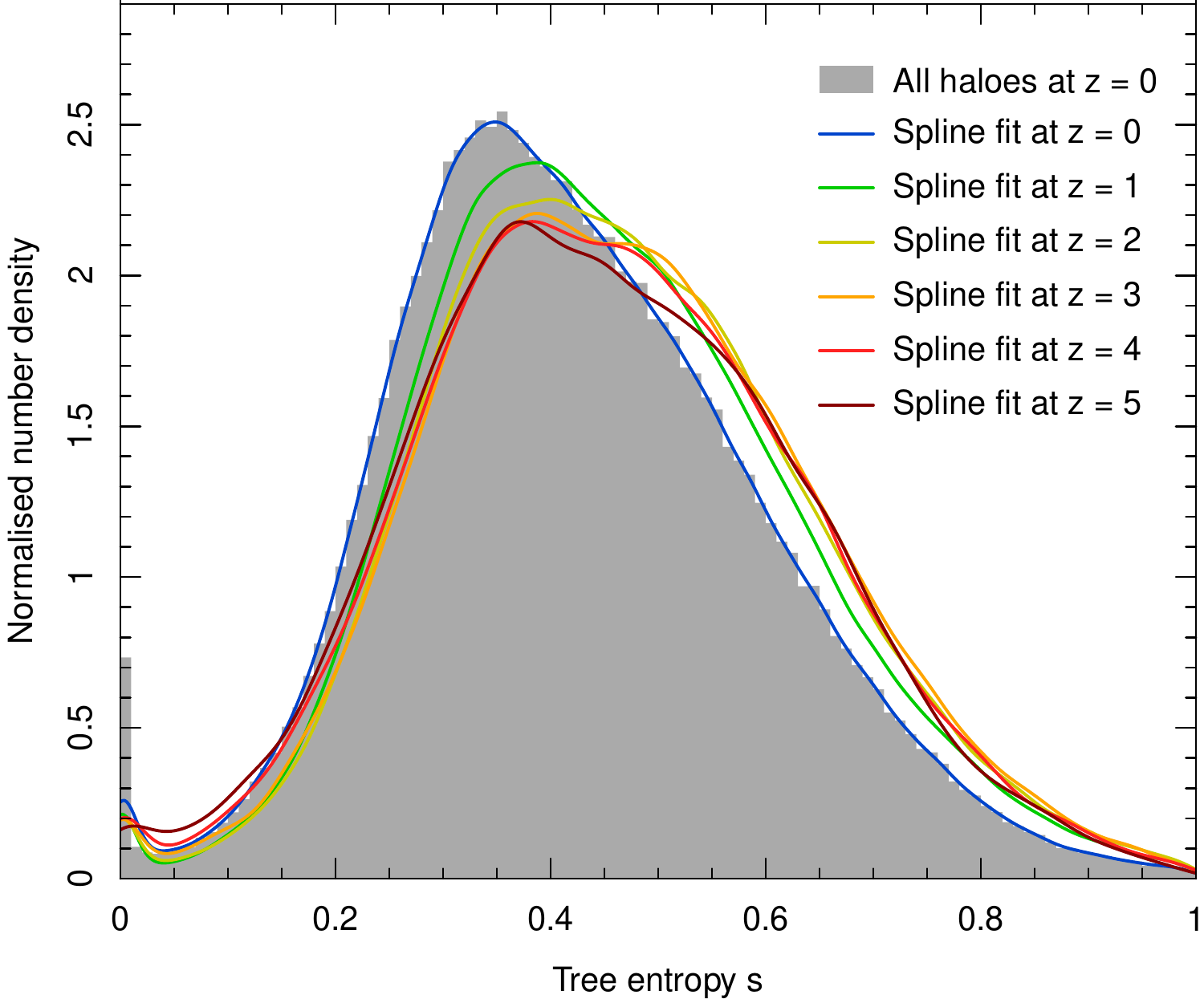}
  \caption{Cosmic evolution of the tree entropy distribution of all simulated haloes ($\mvir\geq10^{11}\msunh$). The shift of the mode is qualitatively consistent with the established weak cosmic decline in the major-to-minor merger ratio in \lcdm.}
  \label{fig:entropy_evolution}
\end{figure}

\fig{entropy_evolution} shows the cosmic evolution of the entropy distribution of \fig{entropy_stats}. The observed increase of the mode of $s$ with redshift $z$ is numerically robust: it persists upon selecting subsamples of massive haloes (\eg $\geq10^{12}\msunh$) and/or with many leaves ($>10$), for which the entropies are expected to be very well converged (\ap{convergence}).

At first sight, it might seem natural that the entropy distribution evolves with redshift in view of the well-established fact that merger rates (at fixed mass, per unit of time) increase strongly with redshift, approximately as $(1+z)^2$ \citep{Stewart2009}. However, the entropy parameter is independent of the overall merger rate, as it only depends on the geometric structure, \textit{not} on the time scale of a merger tree. As long as the relative masses and the chronological ordering of the merging branches remain unchanged, the entropy of a tree remains unchanged, too.

The cosmic evolution towards higher $s$ at higher $z$ can only be explained by an increase in major mergers relative to minor ones. More precisely, the evolution of $s$ shows that merger rates must increase more significantly with redshift for mergers with higher values of $r=m/M$. In fact, in our simulation, the fraction $f_{r>1/3}$ of major binary mergers ($r>1/3$), relative to all binary mergers, varies from 2\pc at $z=0$ to 7\pc at $z=5$, approximately as $f_{r>1/3}\approx 0.02(1+z)^{0.7}$. If major mergers are defined as $r>1/10$, these numbers change to 6\pc at $z=0$ and 25\pc at $z=5$, following $f_{r>1/10}\approx 0.06(1+z)^{0.8}$. This evolution is in qualitative agreement with earlier analyses of $N$-body simulations finding that the strong overall evolution of merger rates shows a weak secondary dependence on $r$, such that major mergers become relatively more prevalent with increasing $z$ \citep{Genel2009,Fakhouri2010}.

\begin{figure}
  \includegraphics[width=\colwidth]{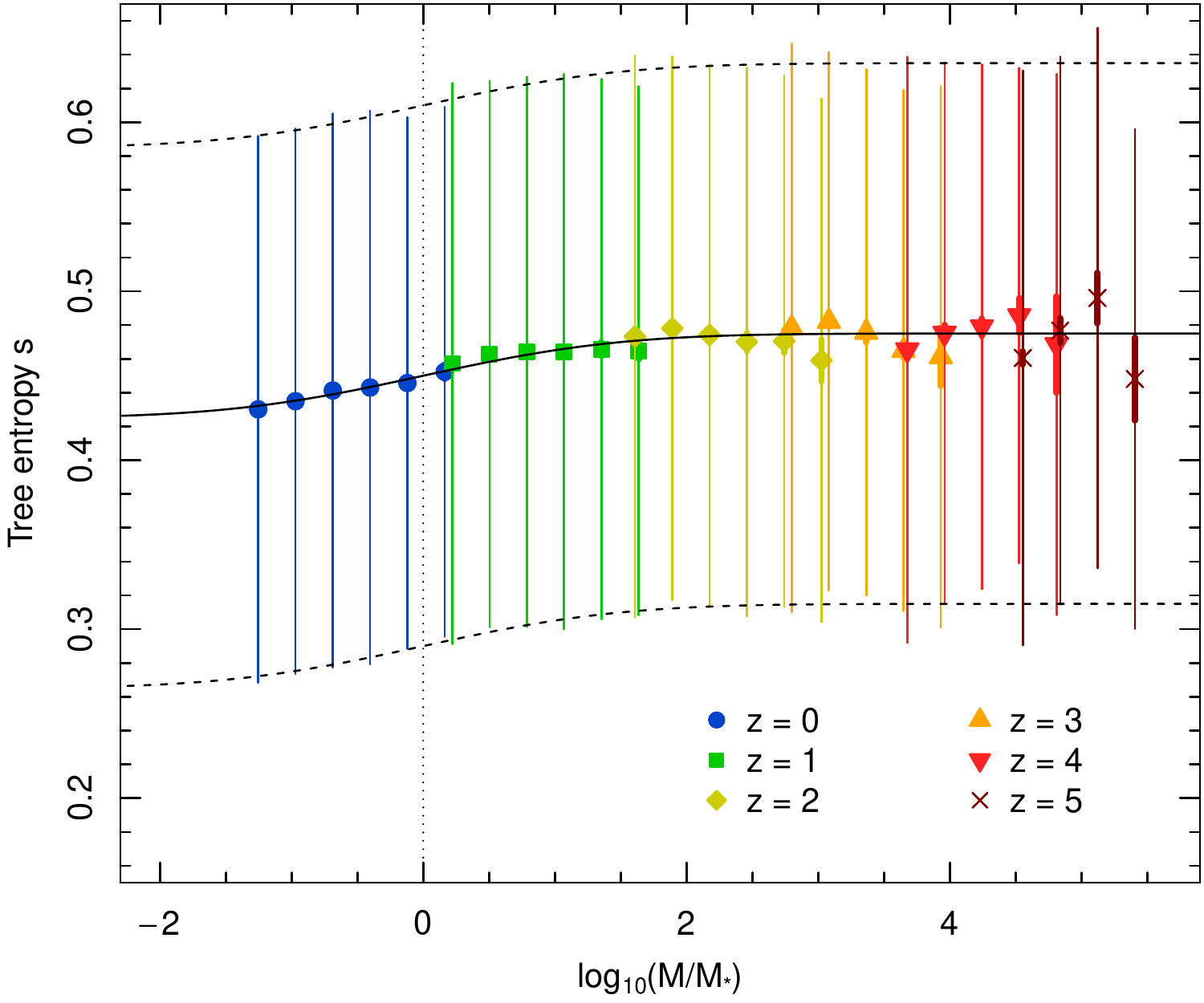}
  \caption{Mean tree entropy plotted as a function of normalised halo mass at different redshifts. Vertical lines show the standard deviation of the $s$-distribution each bin. The statistical uncertainty on the mean is shown by the thicker vertical lines. The black solid curve is the analytic fit of \eq{nufit}.}
  \label{fig:nu}
\end{figure}

The cosmic evolution of $s$ (or $r$) can be understood in terms of the evolving characteristic mass $\mcrit$. This mass, which is related to the break in the halo mass function, is defined as the mass at which the rms $\sigma(\mvir,z)$ of the smoothed density perturbation field $\delta(\mathbf{x},t)$ matches the critical density $\delta_{\rm c}\approx1.69$ for spherical collapse, \ie $\sigma(\mcrit,z)\equiv\delta_{\rm c}$. The characteristic mass defined in this way is directly computable from the matter power spectrum. For masses below $\mcrit$, haloes form efficiently following a power-law statistics, whereas for masses above $\mcrit$, the number density falls of exponentially \citep{Schechter1976}. In a scale-free Einstein-de Sitter universe, $\mcrit$ is the only driver of scale-dependence \citep{Smith2003,Angulo2017}, apart from the \textit{much} smaller free-streaming scale \citep{Angulo2010}, and hence the evolution of merger tree structures should be explainable by the evolution of $\mcrit(z)$. Explicitly, we expect the evolution of $s$ to depend only upon the so-called peak amplitude $\nu\equiv\delta_{\rm c}/\sigma(M,z)$. Equivalently, $s$ expressed as a function of $\mvir/\mcrit(z)$ should be independent of $z$. \fig{nu} demonstrates that this is indeed the case.

Interestingly, the mean of $s$ changes rapidly as a function of halo mass near $\mvir=\mcrit$, but tends to become roughly mass independent as $\mvir\ll\mcrit$ and $\mvir\gg\mcrit$. This behaviour can be approximated by the analytic fit
\be\label{eq:nufit}
	\langle s\rangle=0.45+0.025\,{\rm erf}(x0.6)
\ee
where $x=\log_{10}(\mvir/\mcrit)$. This function is shown as black curve in \fig{nu}. The standard deviation of the $s$-distribution is about $0.16$ with no significant dependence on $x$ (dashed lines in \fig{nu}). On an intuitive level, the increase of $\langle s\rangle$ (and $\langle r\rangle$) with $x$ can be seen as a consequence of the known relation between $x$ and clustering bias, predicted by the theory of peaks in Gaussian random fields \citep{Bardeen1986} and well-confirmed observationally. We reserve a more quantitative exploration of the connection between $s$ and clustering to future work.

The fact that the tree entropy responds to the evolution in the characteristic mass, but not to the strong evolution in the overall merger rate (about an order of magnitude from $z=0$ to $z=2$), is an interesting feature of purely structural estimators such as $s$.

\begin{figure*}
  \includegraphics[width=0.99\textwidth]{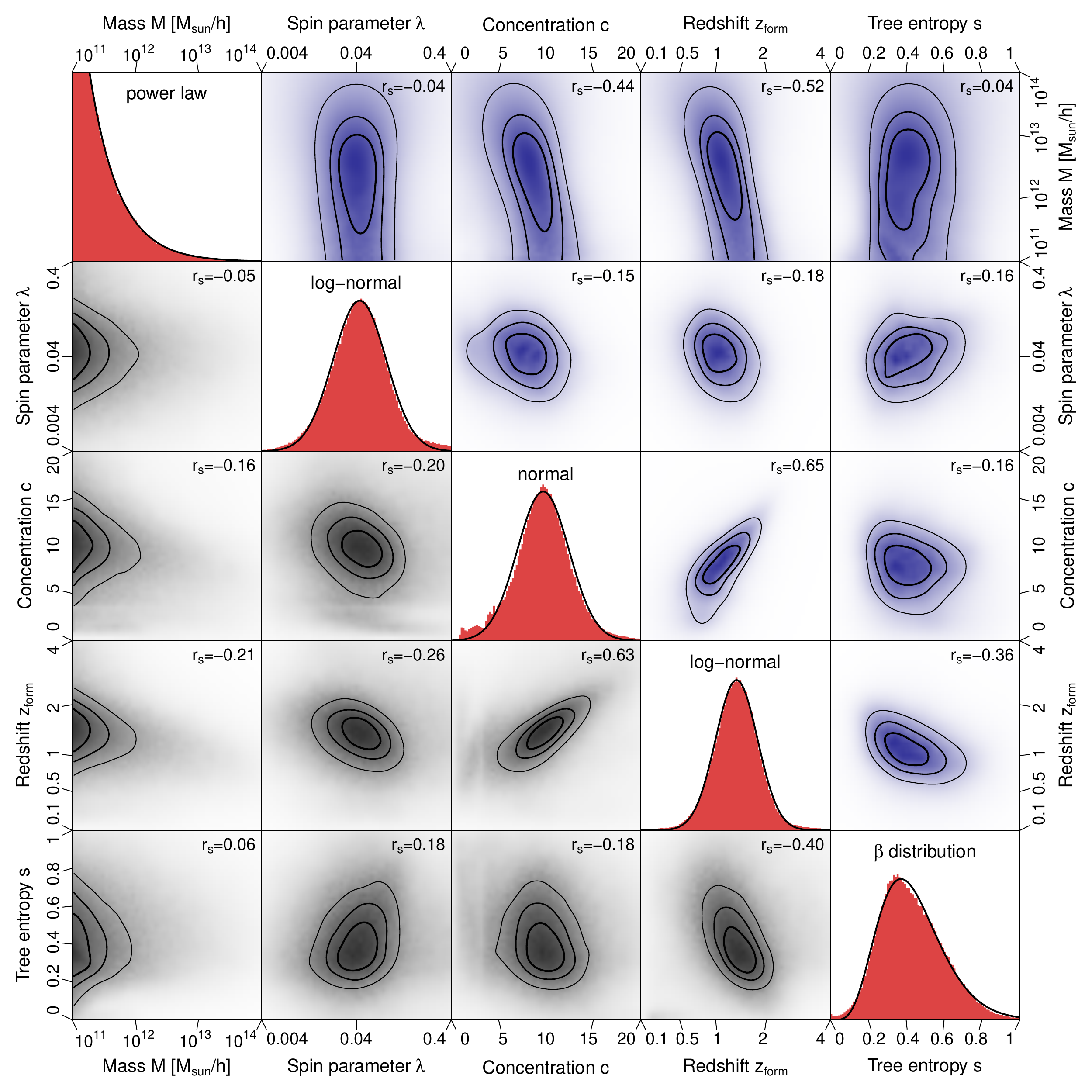}
  \caption{Distributions of five parameters for \surfs haloes with $\mvir\geq10^{11}\msunh$. The diagonal panels show the one-dimensional distributions, fitted with analytic models (black lines, see \ss{haloproperties}). The lower panels (grey) show the two-dimensional \textit{number} density of haloes; whereas the upper panels (blue) show the \textit{mass} density. Contours contain 25/50/75\pc of the counts or mass, respectively. The redshift axis is linear in $\log(1+\zf)$. The $r_s$ values are standard (bottom) and mass-weighted (top) Spearman rank coefficients.}
  \label{fig:parameter_overview}
\end{figure*}

\subsection{Tree entropy relative to other halo properties}\label{ss:haloproperties}

For the new parameter $s$ to be a useful measure, it needs to add some `information' not already captured by simpler halo parameters. It is therefore interesting to question whether/how $s$ relates to established quantities that are commonly used to characterise haloes. We chose to limit this discussion to four key quantities: the virial mass $\mvir$, the spin parameter $\lambda$ (as defined by \citealp{Bullock2001}), the NFW concentration parameter $\cnfw$ (defined by \citealp{Navarro1996}; estimated by VELOCIraptor from the maximum circular velocity) and the formation redshift $\zf$. The latter is defined as the mass-weighted redshift, by which material joins the main branch of the tree, accounting for both mergers and smooth accretion.

The probability distributions and covariances of these quantities and the tree entropy $s$ at $z=0$ are shown in \fig{parameter_overview}. The black solid lines plotted on top of the red histograms show analytical fits to the one-dimensional distributions. The $r_{\rm s}$ values in the covariance plots are the Spearman ranking coefficients. We deliberately chose the Spearman ranking over the standard Pearson coefficient in order to quantify the correlations independently of the scales; \eg the Spearman ranking is invariant to logging an axis, such as $\zf\mapsto\log(1+\zf)$.

The global statistics of $\mvir$, $\lambda$, $\cnfw$ and $\zf$ agree with well-known behaviour in \lcdm:

\begin{itemize}
\item The mass distribution follows a power-law \citep[\eg][]{Murray2013}. The typical exponential truncation around a cut-off mass of $10^{14}\msun$ is present in our data \citep[Fig.~4 of][]{Elahi2018}, but not visible on a linear density scale.
\item The spin parameter distribution is well-approximated by a normal distribution in $\log(\lambda)$ with no significant correlation with $\mvir$, as has been established in various dark matter-only simulations \citep[\eg][]{Knebe2008}. Note that for spherical haloes, $\lambda$ cannot exceed $1/(4\sqrt{2})\approx0.177$; hence the small excess of high-spin objects comes from deformed (and often unrelaxed, interacting) systems.
\item The formation redshift of most haloes falls inside $\zf=1$--$2$ (look-back time $\approx8$--$11\rm~Gyr$). The values are consistent with those quoted by \cite{Elahi2018}, upon accounting for the fact that Elahi's $\zf$ are fixed when the haloes have just 25\% of their final mass, leading to slightly higher values. In line with other pure dark matter simulations \cite[\eg][]{Wechsler2002,Li2008,Zehavi2018}, our formation times closely follow a normal distribution in $\log(1+z)$. Further, $\zf$ shows a negative correlation with $\mvir$, i.e. more massive haloes assemble later, as expected in a hierarchical formation scenario (but see \citeauthor{Li2008} for how this depends on the definition of $\zf$).
\item The concentration parameter $\cnfw$ is approximately normally distributed, with an excess at $\cnfw\lesssim4$ attributed to unrelaxed, predominantly massive haloes, whose three-dimensional density is poorly described by a spherical NFW profile. As first suggested by \cite{Navarro1997}, most of the scatter in $\cnfw$ is explained by differences in collapse times, which are inherited through the assembly history. The earlier a halo is assembled, the higher its concentration, and vice versa; hence the strong positive correlation between $\cnfw$ and $\zf$. The $\mvir$--$\zf$ relation then explains the negative correlation between $\cnfw$ and $\mvir$, whose mode agrees with the detailed studies by \cite{Dutton2014} and \cite{Ludlow2016}. In fact, the full $\zf$--$\cnfw$ relation can be explained in one strike upon realising that the halo density profile reflects the evolving density of the universe weighted by an appropriately defined formation time \citep[\eg][]{Zhao2009,Correa2015a,Ludlow2014,Ludlow2016}.
\end{itemize}

Overall, the tree entropy $s$ only correlates significantly with $\zf$, with a Spearman rank coefficient of $r_{\rm s}=-0.40$ (or $-0.36$ if mass-weighted). This negative correlation naturally arises from the fact that a high tree entropy $s$ normally means that the halo has had a major merger in its recent past. This implies that the main branch grew significantly at a relatively low $z$, hence making the mass-weighted main branch redshift $\zf$ relatively small. Therefore, high values of $s$ often correspond to low values of $\zf$ and vice versa.

We performed a principal component analysis in the space of $(\mvir,\lambda,\cnfw,\zf,s)$ to check if $s$ exhibits any significant correlations with $\mvir$, $\lambda$ and/or $\cnfw$ beyond those already explained via the $s$--$\zf$ relation. However, adding other parameters to $\zf$ only improves the constraints on $s$ by an insignificant amount ($\sim10$\pc).


\section{Connection to galaxies}\label{s:galaxies}

The hierarchical assembly of dark matter haloes naturally drives the mergers of galaxies in the haloes. Theoretical predictions of merger rates, \eg in the ILLUSTRIS and EAGLE simulations \citep{RodriguezGomez2017,Lagos2018a} and semi-analytic modelling \citep{Henriques2015}, are indeed in good agreement with `observed' merger rates deduced from galaxy pair counts \citep{Mundy2017}. This agreement motivates a deeper exploration of the connection between the merger trees and observable galaxy properties.

A full exploration of the galaxy-halo connection lies beyond the scope of this work. However, to illustrate the potential use of $s$ this section discusses a specific example of the information that $s$ carries on galaxies. We limit the discussion to mock galaxies generated by a semi-analytic model, run in post-processing of the CDM trees analysed in \s{cdmsim}; and we restrict the analysis to the stellar bulge-to-total mass ratio at $z=0$ of bulges grown from mergers, reserving a more detailed analysis to future research.

Throughout this section $s$ characterises the merger trees of the CDM haloes, \emph{not} of the galaxies, which merge at different times and with different mass ratios. This is a deliberate choice to investigate the galaxy-halo connection.

\subsection{Galaxy evolution model}\label{ss:sam}

Semi-analytic models (SAMs) of galaxy evolution rely on the assumption that dark matter-dominated structure formation is strictly separable from baryonic processes \citep{Roukema1997,Kauffmann1999}. Exploiting this assumption, halo merger trees are generated first, using purely gravitational modelling or simulations. Mock galaxies are then added in post-processing, using analytic prescriptions for evolving galaxies along the tree branches and nodes. The galaxies are normally approximated as simple axially symmetric systems made of a small number of baryonic components (\eg hot gas, cold gas, stars, black holes). The details of the galaxy models and the physics used to evolve them over time depend on the particular SAM employed.

Here, we use mock galaxies constructed using \shark \citep{Lagos2018b}, a free and flexible software\footnote{Source code available at: https://github.com/ICRAR/shark} framework for running SAMs of nearly any flavour (\ie custom galaxy models and physics). We use the default\footnote{\shark version 1.2.1, timestamp 2019-04-08T10:46:14.} SAM implementation in \shark as detailed by \citeauthor{Lagos2018b}. In this model, galaxies are composed of sub-systems, characterised by their mass, angular momentum and metalicity. These components are split into discs (stellar/atomic/molecular) and spherical systems: hot/cold halo gas, stellar/atomic/molecular bulge, central super massive black hole (SMBH), gas ejected from the halo. The galaxies live in the CDM haloes that form and merge as dictated by the fixed input merger trees. The main baryonic processes that govern the formation/evolution of the galaxies are (1) the accretion of gas onto haloes, modelled via the DM accretion rate; (2) the shock heating and radiative cooling of gas inside haloes onto galactic discs under conservation of specific angular momentum; (3) the formation of molecules and stars in galaxy discs; (4) the suppression of gas cooling by UV background radiation; (5) the chemical enrichment of stars and gas; (6) feedback from stellar winds and supernovae; (7) the growth of SMBHs via accretion of gas and other SMBHs; (8) heating by feedback from SMBHs (`AGN feedback'); (9) galaxy-galaxy mergers driven by dynamical friction inside common DM haloes which can trigger starbursts and the formation and growth of bulges; and (10) the collapse of globally unstable discs that also leads to starbursts and bulges.

\shark was run on the subhalo merger trees of the \surfs L210N1536 simulation (details in \ss{surfs}). Like most SAMs, \shark produces galaxies in three types of CDM environments: \textit{central galaxies} sit at the centre of central subhaloes; \textit{satellite galaxies} sit at the centre of satellite subhaloes, and \textit{orphan galaxies} have lost their halo, for instance through the disruption of a satellite subhalo. The following analysis is performed for the 289,937 central galaxies with subhalo masses $\mvir\geq10^{11}\msunh$. They represent 86\pc of all galaxies in this mass range. Note that including the satellites (6\pc) only changes the numerical results at the percent level.

\begin{figure}
  \includegraphics[width=\colwidth]{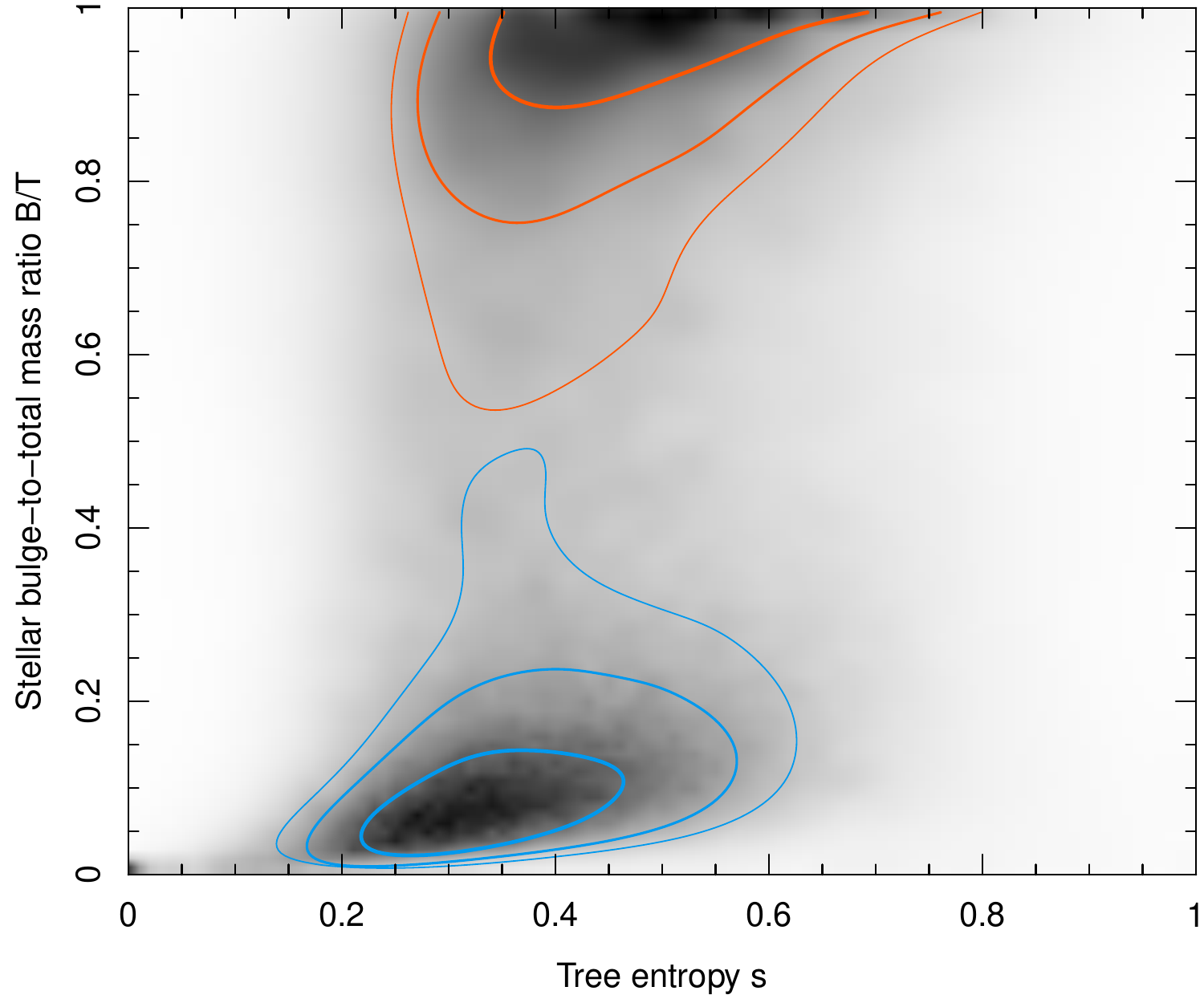}
  \caption{Distribution of galaxies by tree entropy $s$ and stellar $B/T$ mass ratio. The density has been weighted by halo mass in order to make the rarer massive objects with high $B/T$ values visible. Contours contain 25 (thickest), 50 and 75\pc (thinnest) of the total halo mass. To emphasise the association of the bimodal distribution with the blue/red galaxy bimodaility, different colours have been used for upper and lower contours.}
  \label{fig:entropy_bimodalidy}
\end{figure}

\subsection{Tree entropy and morphology}\label{ss:morphology}

Let us now see how the tree entropy $s$ impinges on the mock galaxies. We limit this discussion to evolved galaxies at $z=0$, specifically to their bulge-to-total ($B/T$) mass ratio, already known to depend on the assembly history \citep[\eg][]{Brook2016}. Bulge formation is a complex affair with many facets: dynamic scaling relations and kinematics support a distinction between between dispersion supported `classical bulges' and rotationally supported `pseudo bulges' \citep{Kormendy2004}, which can sometimes coexist \citep{Erwin2014}. Early views that the former are produced by mergers and the latter in-situ evolved into a much more nuanced picture, where classical bulges can also form in-situ \citep{Perez2013}, mergers are an essential driver of pseudo bulges \citep{Guedes2013,Okamoto2013,Gargiulo2019} and mergers transform classical to pseudo bulges \citep{Saha2015}.

\shark distinguishes between bulge stars formed in mergers and bulge stars formed in-situ via global disk instabilities. The instabilities are triggered if the rotational support falls below a classical \citep{Ostriker1973} stability criterion, but currently \shark does not account for tidal instabilities known to precede mergers and which can also form bulges \citep{Gargiulo2019}. Therefore, instability driven bulges probably do not depend as much on the assembly history as they do in reality. To circumvent this discussion, our $B/T$ ratios only include stars formed in mergers and in mergers of progenitor galaxies. These merger-driven bulges are likely a combination of classical and pseudo-bulges, but this distinction is irrelevant here.


\fig{entropy_bimodalidy} shows the halo mass distribution of the mock galaxies in the $(s,B/T)$-plane. The reason for weighting the galaxies by their halo mass is purely graphical: it allows us to visualise low $B/T$ values (found in many low mass galaxies) and high $B/T$ values (found in fewer high mass galaxies) with roughly equal weight. In this scaling the classical bimodality between late-type galaxies (LTG, low $B/T$) and early-type galaxies (ETG, high $B/T$) is very apparent.

There is a clear positive correlation between $s$ and $B/T$. In fact, most (60\pc) of the clear LTGs ($B/T<0.2$) have low entropy ($s<0.4$), while most (74\pc) of the clear ETGs ($B/T>0.8$) have high entropy ($s>1/3$). This statement remains true even if instability driven bulge mass were included (64\pc and 60\pc, respectively). Clearly, the tree entropy encodes important information on the properties of the mock galaxies.

\subsection{Information analysis}\label{ss:information}

We now rigorously quantify the information of $s$ about $B/T$ using the normalised \emph{mutual information} \citep{Strehl2003}. This information measure, contained between 0 and 1, quantifies the information a variable $X$ carries about a variable $Y$ (and vice versa). It is formally defined as
\be\label{eq:mi}
\mathcal{I}_{XY} = \frac{1}{\sqrt{\mathcal{H}_X\mathcal{H}_Y}}\iint\!\!\rho_{XY}(x,y)\ln\!\left[\frac{\rho_{XY}(x,y)}{\rho_X(x)\rho_Y(y)} \right]\!\d x \,\d y,
\ee
where $\rho_{XY}(x,y)$ is the joint 2D probability density, $\rho_X(x)$ and $\rho_Y(y)$ are the individual 1D probability densities, and $\mathcal{H}_X$ and $\mathcal{H}_Y$ are the standard information entropies,
\be
\mathcal{H}_X = \int \rho_X(x)\ln\rho_X(x)\d x.
\ee
Importantly, the mutual information is invariant under non-linear bijective transformations of the variables. For instance, substituting $B/T$ for the so-called numerical Hubble stage $\mathcal{T}$ \citep{deVaucouleurs1977} using the approximation $\mathcal{T}=10-16\sqrt{B/T}$ \citep[Eq. (18) in][]{Obreschkow2009b} or substituting $z$ for $\log_{10}(1+z)$ would have no effect on $\mathcal{I}$.

To put the information $\mathcal{I}_{s,B/T}$ into perspective, we compute $\mathcal{I}_{X,B/T}$ for all five halo properties discussed in \fig{parameter_overview}: the mass $\mvir$, the spin parameter $\lambda$ (as used by \shark), the concentration $\cnfw$ (as output by VELOCIraptor), the formation redshift $\zf$ (defined in \ss{haloproperties}) and the tree entropy $s$. Of these properties, $\mvir$ has the highest mutual information with $B/T$, as one might expect from the well-known mass-morphology relation of galaxies \citep{Calvi2012,Kelvin2014}. For the other four properties, it therefore makes sense to only evaluate the \emph{conditional} mutual information $\tilde{\mathcal{I}}_{X,B/T|\mvir}$ that is not already explained through correlations with $\mvir$. This conditional information is given by $\tilde{\mathcal{I}}_{XY|Z}=\int\mathcal{I}_{XY|Z=z}\,\rho_Z(z)\d z$, where $\mathcal{I}_{XY|Z}$ is the mutual information $\mathcal{I}_{XY}$ at a fixed value of a third property $Z$ (here taken as $Z=\mvir$ or rather $Z=\log\mvir$ for ease of computation). The mutual informations are listed in \tab{information}.

\begin{table}
\centering
\begin{tabularx}{\columnwidth}{@{\extracolsep{\fill}}lc}
\hline \\ [-2ex]
Halo property & Information [percent] \\ [0.5ex]
\hline \\ [-2ex]
Halo mass $\mvir$ & $7.04\pm0.03$ \\
Tree entropy $s$ & $6.08\pm0.11$ \\
Formation redshift $\zf$ & $3.93\pm0.11$ \\
Spin parameter $\lambda$ & $1.31\pm0.11$ \\
Concentration $c$ & $1.01\pm0.11$ \\
[0.5ex]
\hline
\end{tabularx}
\caption{Mutual information between different halo properties and the stellar $B/T$ ratio at $z=0$. For all properties other than $\mvir$, the value is the conditional mutual information that excludes the contribution already explained by correlations with $\mvir$. Properties have been ordered by decreasing information.}
\label{tab:information}
\end{table}

\begin{figure}
  \includegraphics[width=\colwidth]{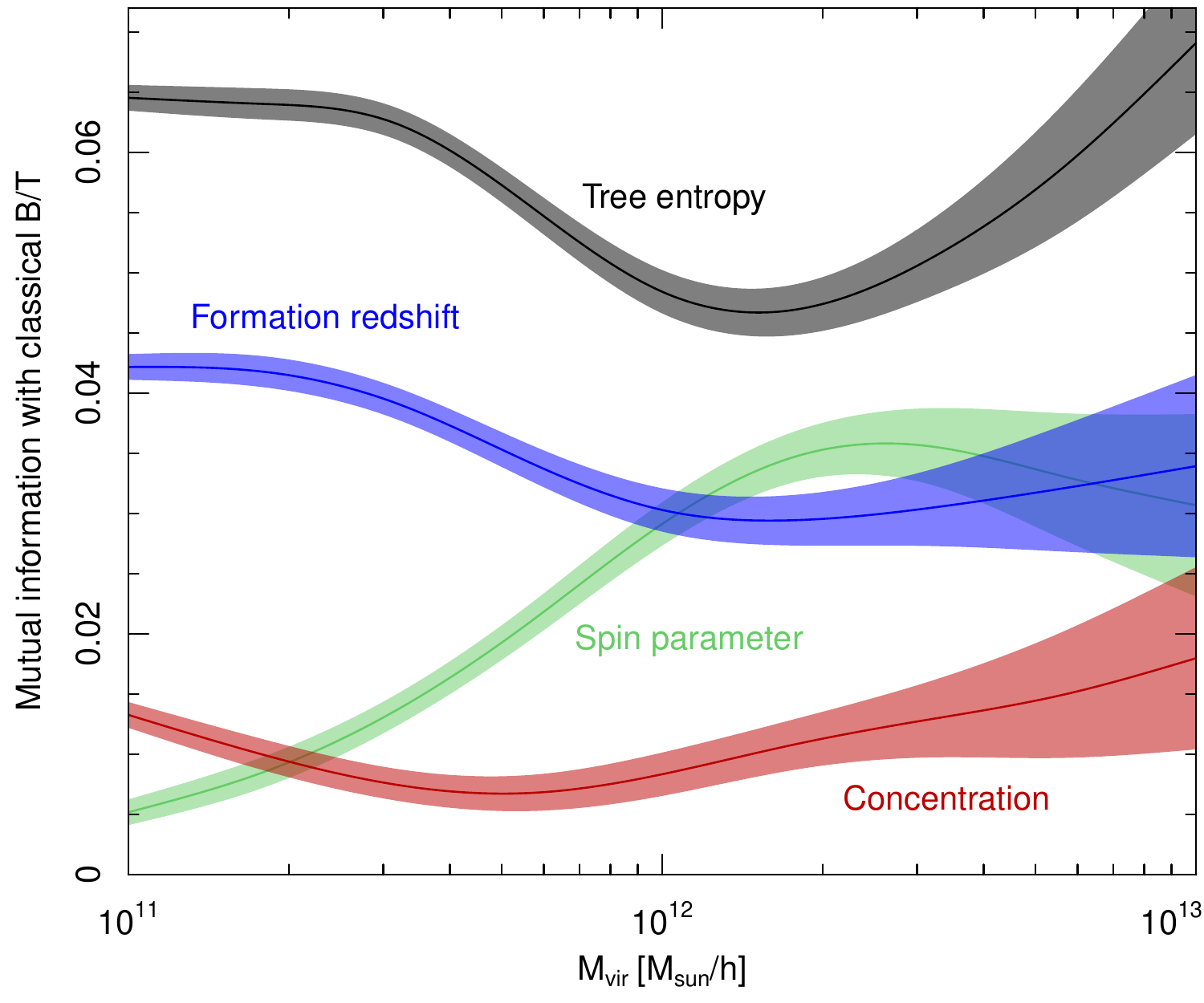}
  \caption{Mutual information, at fixed halo mass $\mvir$, between four halo properties and the stellar $B/T$ ratio at $z=0$. The information measure was evaluated in mass bins of 0.1\,dex and smoothed with a 0.5\,dex top-hat filter. Shaded bands show symmetric 1$\sigma$ confidence intervals determined via bootstrapping.}
  \label{fig:mutual_information}
\end{figure}

Interestingly, the tree entropy $s$ provides the highest amount of independent information on $B/T$, in addition to the information contained in $\mvir$. This statement applies at all values of $\mvir$, as shown in \fig{mutual_information}. 

It is hard to find a halo parameter, assembly related or not, that carries more information on $B/T$ at fixed $\mvir$ than the tree entropy. In fact, in the exhaustive list of halo parameters output by VELOCIraptor \citep[see Table~4 of][]{Elahi2019}, none performs better than $s$. Furthermore, we explicitly searched for other parameters quantifying \emph{single} accretion events: the mass ratio of the last merger, the mass ratio of the last merger above a threshold mass ratio (0.01/0.03/0.1/0.3), the time since the last major merger (defined as $r>1/10$ and $r>1/3$) and the total mass accreted since the last major merger. None of these quantities hold as much information on $B/T$ than $s$.

In summary, the tree entropy $s$ contains a significant amount of information on the $B/T$ ratio. Of course, $B/T$ was chosen deliberately to make this point, since $B/T$ depends on the merger history by construction in \shark. However, the amount of information in $s$ compared to other halo parameters is nonetheless significant. This result demonstrates that galaxies depend on the full structure of the merger trees, not just on isolated merger/accretion events, such as the most recent merger. The exception is the case where the most recent merger was a nearly equal-mass binary merger, since choosing $\beta=1$ (which overwrites all past tree entropies in the case of equal-mass binary mergers) preserves the high information of $s$ on $B/T$ as shown in the following section.

\subsection{Optimisation of free parameters}\label{ss:optimisation}

The tree entropy $s$, as defined in \ss{definition}, is but one way of quantifying merger trees. The question remains whether other definitions satisfying the physical requirements of \ss{requirements} would yield more information, \eg on $B/T$. An exhaustive answer to this question is elusive, but we can at least test other values for the free parameters $\alpha$, $\beta$ and $\gamma$ in the definition of $s$ (Equations~\ref{eq:definition}). As a reminder, $\alpha$ regulates the relative importance of different multi-mergers (\eg binary versus triple mergers), whereas $\beta$ and $\gamma$ regulate the importance of major and very minor mergers (smooth accretion), respectively (see Sections~\ref{ss:requirements}).

\begin{figure}
  \includegraphics[width=\colwidth]{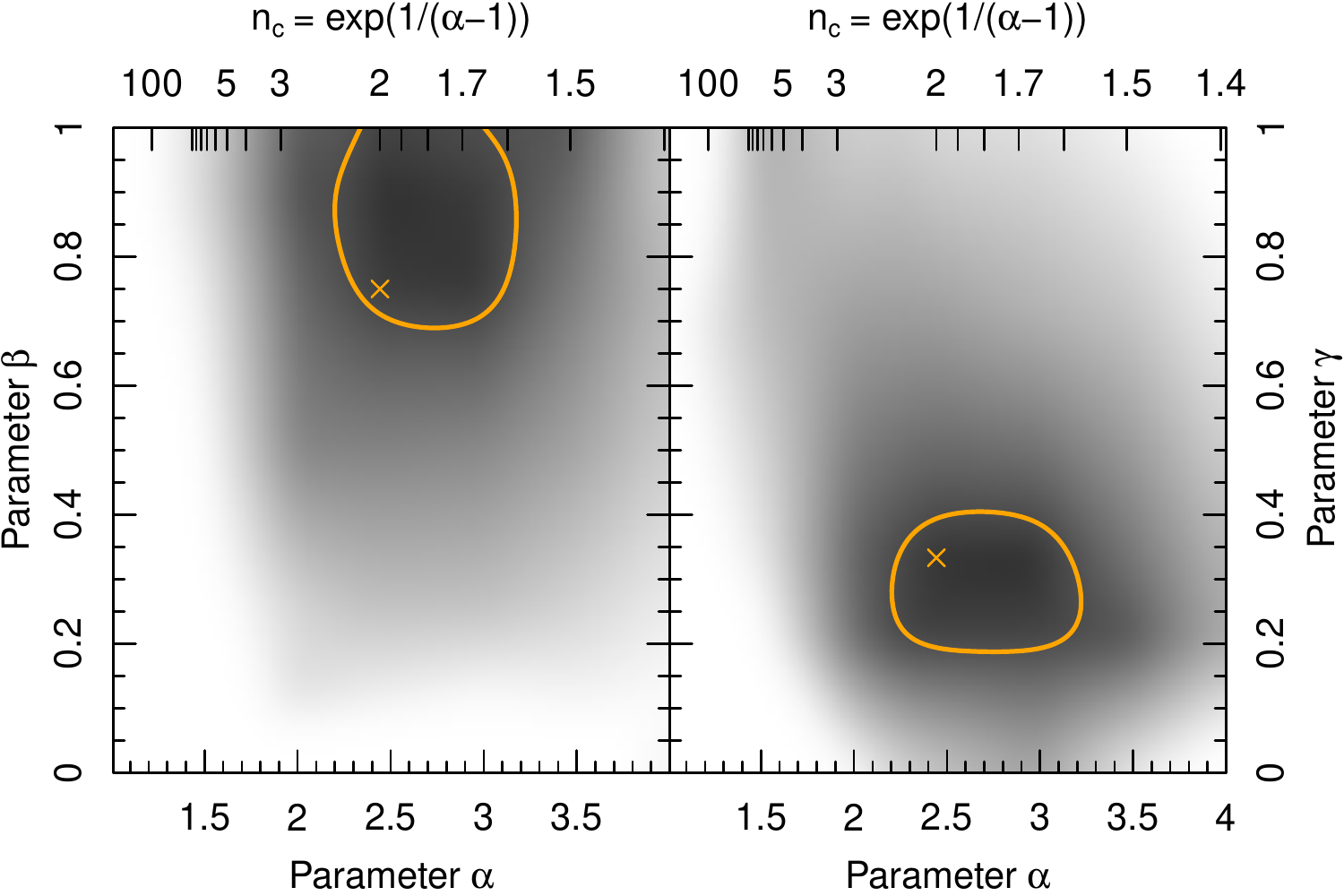}
  \caption{Mutual information $\mathcal{I}_{s,B/T}$ at $z=0$ as a function of $\alpha$, $\beta$ and $\gamma$ in two slices at fixed $\gamma=1/3$ (left) and $\beta=3/4$ (right). White regions denote zero mutual information (\ie $\mathcal{I}$ is at its noise level), while black corresponds to the maximum of $\mathcal{I}$. The contours enclose the region where $\mathcal{I}$ is less than 5\pc from its maximum. Crosses mark the default parameters.}
  \label{fig:optimization}
\end{figure}

\fig{optimization} shows the mutual information $\mathcal{I}_{s,B/T}$ as a function of $\alpha$, $\beta$ and $\gamma$ in two projections. Darker shades denote higher information. The contours enclose the domain, in which $\mathcal{I}_{s,B/T}$ is consistent with its maximum within the statistical noise. Conveniently, our default choice for $\alpha$, $\beta$ and $\gamma$ (orange crosses) approximately maximises the information, justifying this choice with hindsight -- at least for the particular case of $B/T$ at $z=0$ in \shark.

Note that $\alpha$ corresponds to the number of branches $n_{\rm c}=e^{1/(\alpha-1)}$ ($\Leftrightarrow \alpha=1-1/\ln n_{\rm c}$) of the equal-mass merger that generates the highest entropy. Thus, the fact that the default value $\alpha=1+1/\ln 2\approx2.442695$ is a good fit, implies that equal-mass binary mergers ($n=2$) are indeed the `worst' mergers in \shark, in the sense that they cause more massive bulges than other equal-mass multi-mergers.

\section{Conclusions}\label{s:conclusion}

The main purpose of this paper was to define and present a dimensionless parameter to quantify the structure of halo merger trees in an astrophysically meaningful manner. This parameter, the \emph{tree entropy} $s$, extends the mass ratio of binary halo-halo mergers both horizontally, to multi-mergers, and vertically, to trees of sequential mergers.

By construction, minimal trees, grown without mergers, have minimal entropy ($s=0$); and maximal merger trees, assembled exclusively from a hierarchy of equal-mass binary mergers, have maximal entropy ($s=1$). All other trees have intermediate entropies (\fig{overview}).
Consistent with this definition, the leaves (new haloes) and smoothly accreted material are initialised with zero entropy and hence smooth accretion asymptotically decreases the entropy, $s\rightarrow0$. For a single halo, $s$ is an evolving property, whose value only changes if the halo accretes material (smoothly or discretely).

In \lcdm, merger trees exhibit a distribution of tree entropies (\fig{entropy_evolution}), well modelled by a $\beta$-distribution peaked around $s\approx0.4$. The median of $s$ corresponds to that of self-similar merger trees made exclusively of binary mergers of a mass ratio near 1/7. As expected, this demonstrates that typical merger trees fall in between minimal and maximal trees. This result is a manifestation of the fact that major mergers are, in fact, quite rare \citep[\eg][]{Fakhouri2010}.

We showed that $s$ is not reducible to other standard halo parameters (\fig{parameter_overview}) and therefore offers a useful addition to describing the history of haloes. Looking at a single galaxy property, the stellar bulge-to-total ($B/T$) mass ratio, we found that $s$ holds a significant amount of statistical information. In fact, for haloes of fixed mass, $s$ holds more information on this morphology estimator than any other halo parameter we considered (some of which are shown in \fig{mutual_information}). This shows that most galaxies `care' about a large part of their merger history rather than just a single (\eg the last significant) merger/accretion event.

Having introduced a new framework for studying merger trees, this work lends itself to various extensions:
\begin{itemize}
\item A natural idea is to apply $s$ to a deeper exploration of the galaxy-halo connection, firstly by considering a larger range of galaxy properties and secondly by considering other galaxy evolution models. It would be particularly interesting to include modern hydrodynamic simulations, such as EAGLE \citep{Schaye2015} or IllustrisTNG \citep{Nelson2019}, since galactic transformations are not directly hardwired to mergers in such simulations.
\item Likewise, $s$ might help interpret variations in the inner structure of haloes. We have shown that the concentration of the density profile does not significantly correlate with $s$ beyond the established mass--concentration relation, likely because $s$ does not depend on time scales. However, it would be natural to expect that the substructure of haloes depends on the statistics of merger ratios and hence on $s$.
\item The statistics of $s$ offer a new way to compare different combinations of halo finders and tree builders \citep{Onions2012,Srisawat2013}.
\item Similarly, the statistics of $s$ extracted from CDM simulations provides a new benchmark for analytically constructed merger trees \cite[\eg][]{Kauffmann1993,Sheth1999b,Somerville1999,Zhang2008,Parkinson2008}. By applying this benchmark as a tuning factor, Monte-Carlo algorithms for generating merger trees could be tested and improved.
\item There is a lot of room for sharpening our understanding of the statistics of $s$ and its dependence on cosmology. It is likely that this can be achieved analytically using the transformation equations of \cite{Neistein2010}. New insight might also arise from studying the dependence of $s$ on the power spectrum of explicitly scale-free \lcdm simulations.
\item The tree entropy $s$ could also be used to study the direct implications of galaxy-galaxy mergers on galaxy evolution, rather than the indirect implications of halo-halo mergers studied in this work. To do so, $s$ could be computed using the stellar or stellar+cold gas mass assembly trees. 
\end{itemize}

Time will tell if the tree entropy becomes a lasting concept. Maybe modified definitions will turn out to be more fruitful and/or conceptual improvements can be made. It seems, however, likely that the considerations for a physically motivated tree parameter outlined in \ss{requirements} can offer a foundation for forthcoming research.


\section*{Acknowledgements}

We thank the anonymous referee for a very kind and useful report, as well as Paul Schechter and Eric Emsellem for inspiring discussions. We also thank Chris Power, Aaron Robotham and Rodrigo Tobar for their contribution to \surfs, \shark and TreeFrog. Parts of this research were conducted by the Australian Research Council Centre of Excellence for All Sky Astrophysics in 3 Dimensions (ASTRO3D), through project number CE170100013. PE and CL are directly funded by ASTRO 3D. DO and AL are recipients of Australian Research Council Future Fellowships (FT190100083, FT160100250, respectively) funded by the Australian Government. This work was supported by resources provided by the Pawsey Supercomputing Centre with funding from the Australian Government and the Government of Western Australia.



\begin{thebibliography}{127}
\expandafter\ifx\csname natexlab\endcsname\relax\def\natexlab#1{#1}\fi

\bibitem[{Anatole(2007)}]{Katok2007}
Anatole K., 2007, Journal of Modern Dynamics, 1, 545

\bibitem[{Angulo {et~al.}(2017)Angulo, Hahn, Ludlow \& Bonoli}]{Angulo2017}
Angulo R.~E., Hahn O., Ludlow A.~D., Bonoli S., 2017, Monthly Notices of the
  Royal Astronomical Society, 471, 4687

\bibitem[{Angulo \& White(2010)}]{Angulo2010}
Angulo R.~E., White S. D.~M., 2010, Monthly Notices of the Royal Astronomical
  Society, 401, 1796

\bibitem[{{Balakrishnan} \& {Ranganathan}(2012)}]{Balakrishnan2012}
{Balakrishnan} R., {Ranganathan} K., 2012, A Textbook of Graph Theory.
  Springer, New York, NY

\bibitem[{{Bardeen} {et~al.}(1986){Bardeen}, {Bond}, {Kaiser} \&
  {Szalay}}]{Bardeen1986}
{Bardeen} J.~M., {Bond} J.~R., {Kaiser} N., {Szalay} A.~S., 1986, \apj, 304, 15

\bibitem[{{Benson}(2012)}]{Benson2012}
{Benson} A.~J., 2012, New Astronomy, 17, 175

\bibitem[{Benson {et~al.}(2012)Benson, Borgani, De~Lucia, Boylan-Kolchin \&
  Monaco}]{Benson2012b}
Benson A.~J., Borgani S., De~Lucia G., Boylan-Kolchin M., Monaco P., 2012,
  Monthly Notices of the Royal Astronomical Society, 419, 3590

\bibitem[{{Blumenthal} {et~al.}(1984){Blumenthal}, {Faber}, {Primack} \&
  {Rees}}]{Blumenthal1984}
{Blumenthal} G.~R., {Faber} S.~M., {Primack} J.~R., {Rees} M.~J., 1984, \nat,
  311, 517

\bibitem[{{Bond} {et~al.}(1991){Bond}, {Cole}, {Efstathiou} \&
  {Kaiser}}]{Bond1991}
{Bond} J.~R., {Cole} S., {Efstathiou} G., {Kaiser} N., 1991, \apj, 379, 440

\bibitem[{Bosman(1957)}]{Bosman1957}
Bosman A., 1957, Het wondere onderzoekingsveld der vlakke meetkunde. Parcival

\bibitem[{Brooks \& Christensen(2016)}]{Brook2016}
Brooks A., Christensen C., 2016, Galactic Bulges, 317–353

\bibitem[{{Bullock} {et~al.}(2001){Bullock}, {Kolatt}, {Sigad}, {Somerville},
  {Kravtsov}, {Klypin}, {Primack} \& {Dekel}}]{Bullock2001}
{Bullock} J.~S., {Kolatt} T.~S., {Sigad} Y., {Somerville} R.~S., {Kravtsov}
  A.~V., {Klypin} A.~A., {Primack} J.~R., {Dekel} A., 2001, \mnras, 321, 559

\bibitem[{Calvi {et~al.}(2012)Calvi, Poggianti, Fasano \& Vulcani}]{Calvi2012}
Calvi R., Poggianti B.~M., Fasano G., Vulcani B., 2012, Monthly Notices of the
  Royal Astronomical Society: Letters, 419, L14

\bibitem[{{Conselice}(2014)}]{Conselice2014}
{Conselice} C.~J., 2014, \araa, 52, 291

\bibitem[{{Contreras} {et~al.}(2017){Contreras}, {Padilla} \&
  {Lagos}}]{Contreras2017}
{Contreras} S., {Padilla} N., {Lagos} C.~D.~P., 2017, \mnras, 472, 4992

\bibitem[{Correa {et~al.}(2015{\natexlab{a}})Correa, Wyithe, Schaye \&
  Duffy}]{Correa2015a}
Correa C.~A., Wyithe J. S.~B., Schaye J., Duffy A.~R., 2015{\natexlab{a}},
  Monthly Notices of the Royal Astronomical Society, 450, 1514–1520

\bibitem[{Correa {et~al.}(2015{\natexlab{b}})Correa, Wyithe, Schaye \&
  Duffy}]{Correa2015b}
---, 2015{\natexlab{b}}, Monthly Notices of the Royal Astronomical Society,
  450, 1521–1537

\bibitem[{Cowell(1980)}]{Cowell1980}
Cowell F.~A., 1980, European Economic Review, 13, 147

\bibitem[{{Cox} {et~al.}(2006){Cox}, {Dutta}, {Di Matteo}, {Hernquist},
  {Hopkins}, {Robertson} \& {Springel}}]{Cox2006}
{Cox} T.~J., {Dutta} S.~N., {Di Matteo} T., {Hernquist} L., {Hopkins} P.~F.,
  {Robertson} B., {Springel} V., 2006, \apj, 650, 791

\bibitem[{{Crocce} {et~al.}(2012){Crocce}, {Pueblas} \&
  {Scoccimarro}}]{Crocce2012}
{Crocce} M., {Pueblas} S., {Scoccimarro} R., 2012, {2LPTIC: 2nd-order
  Lagrangian Perturbation Theory Initial Conditions}

\bibitem[{{Croton} {et~al.}(2016){Croton}, {Stevens}, {Tonini}, {Garel},
  {Bernyk}, {Bibiano}, {Hodkinson}, {Mutch}, {Poole} \& {Shattow}}]{Croton2016}
{Croton} D.~J., {et~al.}, 2016, \apjs, 222, 22

\bibitem[{{Darg} {et~al.}(2011){Darg}, {Kaviraj}, {Lintott}, {Schawinski},
  {Silk}, {Lynn}, {Bamford} \& {Nichol}}]{Darg2011}
{Darg} D.~W., {Kaviraj} S., {Lintott} C.~J., {Schawinski} K., {Silk} J., {Lynn}
  S., {Bamford} S., {Nichol} R.~C., 2011, \mnras, 416, 1745

\bibitem[{{Davis} {et~al.}(1985){Davis}, {Efstathiou}, {Frenk} \&
  {White}}]{Davis1985}
{Davis} M., {Efstathiou} G., {Frenk} C.~S., {White} S.~D.~M., 1985, \apj, 292,
  371

\bibitem[{{de Vaucouleurs}(1977)}]{deVaucouleurs1977}
{de Vaucouleurs} G., 1977, in Evolution of Galaxies and Stellar Populations,
  {Tinsley} B.~M., {Larson} Richard B.~Gehret D.~C., eds., p.~43

\bibitem[{{Diemand} {et~al.}(2005){Diemand}, {Moore} \& {Stadel}}]{Diemand2005}
{Diemand} J., {Moore} B., {Stadel} J., 2005, \nat, 433, 389

\bibitem[{{Dubinski}(1998)}]{Dubinski1998}
{Dubinski} J., 1998, \apj, 502, 141

\bibitem[{{Dubinski} {et~al.}(1996){Dubinski}, {Mihos} \&
  {Hernquist}}]{Dubinski1996}
{Dubinski} J., {Mihos} J.~C., {Hernquist} L., 1996, \apj, 462, 576

\bibitem[{{Dutton}(2012)}]{Dutton2012}
{Dutton} A.~A., 2012, \mnras, 424, 3123

\bibitem[{Dutton \& Macciò(2014)}]{Dutton2014}
Dutton A.~A., Macciò A.~V., 2014, Monthly Notices of the Royal Astronomical
  Society, 441, 3359–3374

\bibitem[{{Elahi} {et~al.}(2019{\natexlab{a}}){Elahi}, {Ca{\~n}as}, {Poulton},
  {Tobar}, {Willis}, {Lagos}, {Power} \& {Robotham}}]{Elahi2019}
{Elahi} P.~J., {Ca{\~n}as} R., {Poulton} R. J.~J., {Tobar} R.~J., {Willis}
  J.~S., {Lagos} C. d.~P., {Power} C., {Robotham} A. S.~G., 2019{\natexlab{a}},
  arXiv e-prints, arXiv:1902.01010

\bibitem[{{Elahi} {et~al.}(2019{\natexlab{b}}){Elahi}, {Poulton}, {Tobar},
  {Canas}, {Lagos}, {Power} \& {Robotham}}]{Elahi2019b}
{Elahi} P.~J., {Poulton} R. J.~J., {Tobar} R.~J., {Canas} R., {Lagos} C. d.~P.,
  {Power} C., {Robotham} A. S.~G., 2019{\natexlab{b}}, arXiv e-prints,
  arXiv:1902.01527

\bibitem[{{Elahi} {et~al.}(2018){Elahi}, {Welker}, {Power}, {del P Lagos},
  {Robotham}, {Ca{\~n}as} \& {Poulton}}]{Elahi2018}
{Elahi} P.~J., {Welker} C., {Power} C., {del P Lagos} C., {Robotham} A.~S.~G.,
  {Ca{\~n}as} R., {Poulton} R., 2018, \mnras, 475, 5338

\bibitem[{{Emsellem} {et~al.}(2007){Emsellem}, {Cappellari}, {Krajnovi{\'c}},
  {van de Ven}, {Bacon}, {Bureau}, {Davies}, {de Zeeuw}, {Falc{\'o}n-Barroso},
  {Kuntschner}, {McDermid}, {Peletier} \& {Sarzi}}]{Emsellem2007}
{Emsellem} E., {et~al.}, 2007, \mnras, 379, 401

\bibitem[{Erwin {et~al.}(2014)Erwin, Saglia, Fabricius, Thomas, Nowak, Rusli,
  Bender, Vega~Beltrán \& Beckman}]{Erwin2014}
Erwin P., {et~al.}, 2014, Monthly Notices of the Royal Astronomical Society,
  446, 4039

\bibitem[{{Fakhouri} {et~al.}(2010){Fakhouri}, {Ma} \&
  {Boylan-Kolchin}}]{Fakhouri2010}
{Fakhouri} O., {Ma} C.-P., {Boylan-Kolchin} M., 2010, \mnras, 406, 2267

\bibitem[{{Forero-Romero}(2009)}]{ForeroRomero2009}
{Forero-Romero} J.~E., 2009, \mnras, 399, 762

\bibitem[{{Gao} {et~al.}(2008){Gao}, {Navarro}, {Cole}, {Frenk}, {White},
  {Springel}, {Jenkins} \& {Neto}}]{Gao2008b}
{Gao} L., {Navarro} J.~F., {Cole} S., {Frenk} C.~S., {White} S. D.~M.,
  {Springel} V., {Jenkins} A., {Neto} A.~F., 2008, \mnras, 387, 536

\bibitem[{Gargiulo {et~al.}(2019)Gargiulo, Monachesi, Gómez, Grand, Marinacci,
  Pakmor, White, Bell, Fragkoudi \& Tissera}]{Gargiulo2019}
Gargiulo I.~D., {et~al.}, 2019, Monthly Notices of the Royal Astronomical
  Society, 489, 5742–5763

\bibitem[{{Genel} {et~al.}(2010){Genel}, {Bouch{\'e}}, {Naab}, {Sternberg} \&
  {Genzel}}]{Genel2010}
{Genel} S., {Bouch{\'e}} N., {Naab} T., {Sternberg} A., {Genzel} R., 2010,
  \apj, 719, 229

\bibitem[{{Genel} {et~al.}(2009){Genel}, {Genzel}, {Bouch{\'e}}, {Naab} \&
  {Sternberg}}]{Genel2009}
{Genel} S., {Genzel} R., {Bouch{\'e}} N., {Naab} T., {Sternberg} A., 2009,
  \apj, 701, 2002

\bibitem[{{Green} {et~al.}(2004){Green}, {Hofmann} \& {Schwarz}}]{Green2004}
{Green} A.~M., {Hofmann} S., {Schwarz} D.~J., 2004, \mnras, 353, L23

\bibitem[{{Guedes} {et~al.}(2013){Guedes}, {Mayer}, {Carollo} \&
  {Madau}}]{Guedes2013}
{Guedes} J., {Mayer} L., {Carollo} M., {Madau} P., 2013, \apj, 772, 36

\bibitem[{{Guo} \& {White}(2008)}]{Guo2008}
{Guo} Q., {White} S.~D.~M., 2008, \mnras, 384, 2

\bibitem[{Han {et~al.}(2017)Han, Cole, Frenk, Benitez-Llambay \&
  Helly}]{Han2017}
Han J., Cole S., Frenk C.~S., Benitez-Llambay A., Helly J., 2017, Monthly
  Notices of the Royal Astronomical Society, 474, 604

\bibitem[{{Helmi} {et~al.}(2018){Helmi}, {Babusiaux}, {Koppelman}, {Massari},
  {Veljanoski} \& {Brown}}]{Helmi2018}
{Helmi} A., {Babusiaux} C., {Koppelman} H.~H., {Massari} D., {Veljanoski} J.,
  {Brown} A. G.~A., 2018, \nat, 563, 85

\bibitem[{{Henriques} {et~al.}(2015){Henriques}, {White}, {Thomas}, {Angulo},
  {Guo}, {Lemson}, {Springel} \& {Overzier}}]{Henriques2015}
{Henriques} B. M.~B., {White} S. D.~M., {Thomas} P.~A., {Angulo} R., {Guo} Q.,
  {Lemson} G., {Springel} V., {Overzier} R., 2015, \mnras, 451, 2663

\bibitem[{{Hernquist}(1993)}]{Hernquist1993}
{Hernquist} L., 1993, \apj, 409, 548

\bibitem[{Jiang \& van~den Bosch(2014)}]{Jiang2014b}
Jiang F., van~den Bosch F.~C., 2014, Monthly Notices of the Royal Astronomical
  Society, 440, 193–207

\bibitem[{{Jiang} {et~al.}(2014){Jiang}, {Helly}, {Cole} \&
  {Frenk}}]{Jiang2014}
{Jiang} L., {Helly} J.~C., {Cole} S., {Frenk} C.~S., 2014, \mnras, 440, 2115

\bibitem[{{Kauffmann} {et~al.}(1999){Kauffmann}, {Colberg}, {Diaferio} \&
  {White}}]{Kauffmann1999}
{Kauffmann} G., {Colberg} J.~M., {Diaferio} A., {White} S.~D.~M., 1999, \mnras,
  303, 188

\bibitem[{{Kauffmann} \& {White}(1993)}]{Kauffmann1993}
{Kauffmann} G., {White} S.~D.~M., 1993, \mnras, 261

\bibitem[{Kelvin {et~al.}(2014)Kelvin, Driver, Robotham, Taylor, Graham,
  Alpaslan, Baldry, Bamford, Bauer, Bland-Hawthorn \& et~al.}]{Kelvin2014}
Kelvin L.~S., {et~al.}, 2014, Monthly Notices of the Royal Astronomical
  Society, 444, 1647–1659

\bibitem[{{Klypin} {et~al.}(1999){Klypin}, {Gottl{\"o}ber}, {Kravtsov} \&
  {Khokhlov}}]{Klypin1999b}
{Klypin} A., {Gottl{\"o}ber} S., {Kravtsov} A.~V., {Khokhlov} A.~M., 1999,
  \apj, 516, 530

\bibitem[{Klypin {et~al.}(2016)Klypin, Yepes, Gottlöber, Prada \&
  Hess}]{Klypin2016}
Klypin A., Yepes G., Gottlöber S., Prada F., Hess S., 2016, Monthly Notices of
  the Royal Astronomical Society, 457, 4340

\bibitem[{{Knebe} \& {Power}(2008)}]{Knebe2008}
{Knebe} A., {Power} C., 2008, \apj, 678, 621

\bibitem[{{Kormendy} {et~al.}(2009){Kormendy}, {Fisher}, {Cornell} \&
  {Bender}}]{Kormendy2009}
{Kormendy} J., {Fisher} D.~B., {Cornell} M.~E., {Bender} R., 2009, \apjs, 182,
  216

\bibitem[{{Kormendy} \& {Kennicutt}(2004)}]{Kormendy2004}
{Kormendy} J., {Kennicutt} R.~C. J., 2004, \araa, 42, 603

\bibitem[{{Lacey} \& {Cole}(1993)}]{Lacey1993}
{Lacey} C., {Cole} S., 1993, \mnras, 262, 627

\bibitem[{{Lacey} {et~al.}(2016){Lacey}, {Baugh}, {Frenk}, {Benson}, {Bower},
  {Cole}, {Gonzalez-Perez}, {Helly}, {Lagos} \& {Mitchell}}]{Lacey2016}
{Lacey} C.~G., {et~al.}, 2016, \mnras, 462, 3854

\bibitem[{{Lagos} {et~al.}(2018{\natexlab{a}}){Lagos}, {Stevens}, {Bower},
  {Davis}, {Contreras}, {Padilla}, {Obreschkow}, {Croton}, {Trayford} \&
  {Welker}}]{Lagos2018a}
{Lagos} C. d.~P., {et~al.}, 2018{\natexlab{a}}, \mnras, 473, 4956

\bibitem[{{Lagos} {et~al.}(2018{\natexlab{b}}){Lagos}, {Tobar}, {Robotham},
  {Obreschkow}, {Mitchell}, {Power} \& {Elahi}}]{Lagos2018b}
{Lagos} C. d.~P., {Tobar} R.~J., {Robotham} A. S.~G., {Obreschkow} D.,
  {Mitchell} P.~D., {Power} C., {Elahi} P.~J., 2018{\natexlab{b}}, \mnras, 481,
  3573

\bibitem[{{Lewis} {et~al.}(2000){Lewis}, {Challinor} \& {Lasenby}}]{Lewis2000}
{Lewis} A., {Challinor} A., {Lasenby} A., 2000, \apj, 538, 473

\bibitem[{{Li} {et~al.}(2008){Li}, {Mo} \& {Gao}}]{Li2008}
{Li} Y., {Mo} H.~J., {Gao} L., 2008, \mnras, 389, 1419

\bibitem[{Li {et~al.}(2007)Li, Mo, Van Den~Bosch \& Lin}]{Li2007}
Li Y., Mo H.~J., Van Den~Bosch F.~C., Lin W.~P., 2007, Monthly Notices of the
  Royal Astronomical Society, 379, 689–701

\bibitem[{Ludlow {et~al.}(2016)Ludlow, Bose, Angulo, Wang, Hellwing, Navarro,
  Cole \& Frenk}]{Ludlow2016}
Ludlow A.~D., Bose S., Angulo R.~E., Wang L., Hellwing W.~A., Navarro J.~F.,
  Cole S., Frenk C.~S., 2016, Monthly Notices of the Royal Astronomical
  Society, 460, 1214–1232

\bibitem[{Ludlow {et~al.}(2014)Ludlow, Navarro, Angulo, Boylan-Kolchin,
  Springel, Frenk \& White}]{Ludlow2014}
Ludlow A.~D., Navarro J.~F., Angulo R.~E., Boylan-Kolchin M., Springel V.,
  Frenk C., White S. D.~M., 2014, Monthly Notices of the Royal Astronomical
  Society, 441, 378–388

\bibitem[{Ludlow {et~al.}(2013)Ludlow, Navarro, Boylan-Kolchin, Bett, Angulo,
  Li, White, Frenk \& Springel}]{Ludlow2013}
Ludlow A.~D., {et~al.}, 2013, Monthly Notices of the Royal Astronomical
  Society, 432, 1103–1113

\bibitem[{Ludlow {et~al.}(2012)Ludlow, Navarro, Li, Angulo, Boylan-Kolchin \&
  Bett}]{Ludlow2012}
Ludlow A.~D., Navarro J.~F., Li M., Angulo R.~E., Boylan-Kolchin M., Bett
  P.~E., 2012, Monthly Notices of the Royal Astronomical Society, 427, 1322

\bibitem[{{Ludlow} {et~al.}(2009){Ludlow}, {Navarro}, {Springel}, {Jenkins},
  {Frenk} \& {Helmi}}]{Ludlow2009}
{Ludlow} A.~D., {Navarro} J.~F., {Springel} V., {Jenkins} A., {Frenk} C.~S.,
  {Helmi} A., 2009, \apj, 692, 931

\bibitem[{{Lynden-Bell}(1967)}]{Lynden-Bell1967}
{Lynden-Bell} D., 1967, \mnras, 136, 101

\bibitem[{{Mathai} \& {Haubold}(2007)}]{Mathai2007}
{Mathai} A.~M., {Haubold} H.~J., 2007, Physica A Statistical Mechanics and its
  Applications, 385, 493

\bibitem[{{Mesbahi} \& {Egerstedt}(2010)}]{Mesbahi2010}
{Mesbahi} M., {Egerstedt} M., 2010, Graph Theoretic Methods in Multiagent
  Networks. Princeton University Press

\bibitem[{{Moore} {et~al.}(1999){Moore}, {Ghigna}, {Governato}, {Lake},
  {Quinn}, {Stadel} \& {Tozzi}}]{Moore1999}
{Moore} B., {Ghigna} S., {Governato} F., {Lake} G., {Quinn} T., {Stadel} J.,
  {Tozzi} P., 1999, \apjl, 524, L19

\bibitem[{Moreno {et~al.}(2008)Moreno, Giocoli \& Sheth}]{Moreno2008}
Moreno J., Giocoli C., Sheth R.~K., 2008, Monthly Notices of the Royal
  Astronomical Society, 391, 1729–1740

\bibitem[{{Mundy} {et~al.}(2017){Mundy}, {Conselice}, {Duncan}, {Almaini},
  {H{\"a}ussler} \& {Hartley}}]{Mundy2017}
{Mundy} C.~J., {Conselice} C.~J., {Duncan} K.~J., {Almaini} O., {H{\"a}ussler}
  B., {Hartley} W.~G., 2017, \mnras, 470, 3507

\bibitem[{{Murray} {et~al.}(2013){Murray}, {Power} \& {Robotham}}]{Murray2013}
{Murray} S.~G., {Power} C., {Robotham} A.~S.~G., 2013, \mnras, 434, L61

\bibitem[{{Navarro} {et~al.}(1996){Navarro}, {Frenk} \& {White}}]{Navarro1996}
{Navarro} J.~F., {Frenk} C.~S., {White} S.~D.~M., 1996, \apj, 462, 563

\bibitem[{{Navarro} {et~al.}(1997){Navarro}, {Frenk} \& {White}}]{Navarro1997}
---, 1997, \apj, 490, 493

\bibitem[{Neistein {et~al.}(2010)Neistein, Macciò \& Dekel}]{Neistein2010}
Neistein E., Macciò A.~V., Dekel A., 2010, Monthly Notices of the Royal
  Astronomical Society, 403, 984

\bibitem[{{Nelson} {et~al.}(2019){Nelson}, {Springel}, {Pillepich},
  {Rodriguez-Gomez}, {Torrey}, {Genel}, {Vogelsberger}, {Pakmor}, {Marinacci},
  {Weinberger}, {Kelley}, {Lovell}, {Diemer} \& {Hernquist}}]{Nelson2019}
{Nelson} D., {et~al.}, 2019, Computational Astrophysics and Cosmology, 6, 2

\bibitem[{{Obreschkow} {et~al.}(2009){Obreschkow}, {Croton}, {DeLucia},
  {Khochfar} \& {Rawlings}}]{Obreschkow2009b}
{Obreschkow} D., {Croton} D., {DeLucia} G., {Khochfar} S., {Rawlings} S., 2009,
  \apj, 698, 1467

\bibitem[{{Okamoto}(2013)}]{Okamoto2013}
{Okamoto} T., 2013, \mnras, 428, 718

\bibitem[{{Onions} {et~al.}(2012){Onions}, {Knebe}, {Pearce}, {Muldrew}, {Lux},
  {Knollmann}, {Ascasibar}, {Behroozi}, {Elahi} \& {Han}}]{Onions2012}
{Onions} J., {et~al.}, 2012, \mnras, 423, 1200

\bibitem[{{Oser} {et~al.}(2010){Oser}, {Ostriker}, {Naab}, {Johansson} \&
  {Burkert}}]{Oser2010}
{Oser} L., {Ostriker} J.~P., {Naab} T., {Johansson} P.~H., {Burkert} A., 2010,
  \apj, 725, 2312

\bibitem[{{Ostriker} \& {Peebles}(1973)}]{Ostriker1973}
{Ostriker} J.~P., {Peebles} P.~J.~E., 1973, \apj, 186, 467

\bibitem[{{Parkinson} {et~al.}(2008){Parkinson}, {Cole} \&
  {Helly}}]{Parkinson2008}
{Parkinson} H., {Cole} S., {Helly} J., 2008, \mnras, 383, 557

\bibitem[{{Peebles}(1965)}]{Peebles1965}
{Peebles} P.~J.~E., 1965, \apj, 142, 1317

\bibitem[{Perez {et~al.}(2013)Perez, Valenzuela, Tissera \&
  Michel-Dansac}]{Perez2013}
Perez J., Valenzuela O., Tissera P.~B., Michel-Dansac L., 2013, Monthly Notices
  of the Royal Astronomical Society, 436, 259

\bibitem[{{Pillepich} {et~al.}(2015){Pillepich}, {Madau} \&
  {Mayer}}]{Pillepich2015}
{Pillepich} A., {Madau} P., {Mayer} L., 2015, \apj, 799, 184

\bibitem[{{Planck Collaboration} {et~al.}(2016){Planck Collaboration}, {Ade},
  {Aghanim}, {Arnaud}, {Ashdown}, {Aumont}, {Baccigalupi}, {Banday},
  {Barreiro}, {Bartlett} \& et~al.}]{PlanckCollaboration2016}
{Planck Collaboration}, {et~al.}, 2016, \aap, 594, A13

\bibitem[{{Pontzen} \& {Governato}(2013)}]{Pontzen2013}
{Pontzen} A., {Governato} F., 2013, \mnras, 430, 121

\bibitem[{{Richter}(1970)}]{Richter1970}
{Richter} J., 1970, The notebooks of Leonardo da Vinci. Dover, New York

\bibitem[{{Rodriguez-Gomez} {et~al.}(2017){Rodriguez-Gomez}, {Sales}, {Genel},
  {Pillepich}, {Zjupa}, {Nelson}, {Griffen}, {Torrey}, {Snyder},
  {Vogelsberger}, {Springel}, {Ma} \& {Hernquist}}]{RodriguezGomez2017}
{Rodriguez-Gomez} V., {et~al.}, 2017, \mnras, 467, 3083

\bibitem[{{Roukema} {et~al.}(1997){Roukema}, {Quinn}, {Peterson} \&
  {Rocca-Volmerange}}]{Roukema1997}
{Roukema} B.~F., {Quinn} P.~J., {Peterson} B.~A., {Rocca-Volmerange} B., 1997,
  \mnras, 292, 835

\bibitem[{{Roukema} \& {Yoshii}(1993)}]{Roukema1993}
{Roukema} B.~F., {Yoshii} Y., 1993, \apjl, 418, L1

\bibitem[{{Saha}(2015)}]{Saha2015}
{Saha} K., 2015, \apjl, 806, L29

\bibitem[{{Schaye} {et~al.}(2015){Schaye}, {Crain}, {Bower}, {Furlong},
  {Schaller}, {Theuns}, {Dalla Vecchia}, {Frenk}, {McCarthy}, {Helly},
  {Jenkins}, {Rosas-Guevara}, {White}, {Baes}, {Booth}, {Camps}, {Navarro},
  {Qu}, {Rahmati}, {Sawala}, {Thomas} \& {Trayford}}]{Schaye2015}
{Schaye} J., {et~al.}, 2015, \mnras, 446, 521

\bibitem[{{Schechter}(1976)}]{Schechter1976}
{Schechter} P., 1976, \apj, 203, 297

\bibitem[{{Shannon}(1948)}]{Shannon1948}
{Shannon} C.~E., 1948, The Bell System Technical Journal, 27, 379

\bibitem[{{Sheth} \& {Lemson}(1999)}]{Sheth1999b}
{Sheth} R.~K., {Lemson} G., 1999, \mnras, 305, 946

\bibitem[{Shorrocks(1980)}]{Shorrocks1980}
Shorrocks A.~F., 1980, Econometrica, 48, 613

\bibitem[{{Smith} {et~al.}(2003){Smith}, {Peacock}, {Jenkins}, {White},
  {Frenk}, {Pearce}, {Thomas}, {Efstathiou} \& {Couchman}}]{Smith2003}
{Smith} R.~E., {et~al.}, 2003, \mnras, 341, 1311

\bibitem[{{Somerville} \& {Kolatt}(1999)}]{Somerville1999}
{Somerville} R.~S., {Kolatt} T.~S., 1999, \mnras, 305, 1

\bibitem[{{Springel}(2005)}]{Springel2005c}
{Springel} V., 2005, \mnras, 364, 1105

\bibitem[{Springel \& Hernquist(2005)}]{Springel2005b}
Springel V., Hernquist L., 2005, The Astrophysical Journal, 622, L9

\bibitem[{{Springel} {et~al.}(2008{\natexlab{a}}){Springel}, {Wang},
  {Vogelsberger}, {Ludlow}, {Jenkins}, {Helmi}, {Navarro}, {Frenk} \&
  {White}}]{Springel2008b}
{Springel} V., {et~al.}, 2008{\natexlab{a}}, \mnras, 391, 1685

\bibitem[{{Springel} {et~al.}(2008{\natexlab{b}}){Springel}, {White}, {Frenk},
  {Navarro}, {Jenkins}, {Vogelsberger}, {Wang}, {Ludlow} \&
  {Helmi}}]{Springel2008}
---, 2008{\natexlab{b}}, \nat, 456, 73

\bibitem[{{Springel} {et~al.}(2001){Springel}, {Yoshida} \&
  {White}}]{Springel2001}
{Springel} V., {Yoshida} N., {White} S. D.~M., 2001, \na, 6, 79

\bibitem[{{Srisawat} {et~al.}(2013){Srisawat}, {Knebe}, {Pearce}, {Schneider},
  {Thomas}, {Behroozi}, {Dolag}, {Elahi}, {Han}, {Helly}, {Jing}, {Jung},
  {Lee}, {Mao}, {Onions}, {Rodriguez-Gomez}, {Tweed} \& {Yi}}]{Srisawat2013}
{Srisawat} C., {et~al.}, 2013, \mnras, 436, 150

\bibitem[{{Stewart} {et~al.}(2009){Stewart}, {Bullock}, {Barton} \&
  {Wechsler}}]{Stewart2009}
{Stewart} K.~R., {Bullock} J.~S., {Barton} E.~J., {Wechsler} R.~H., 2009, \apj,
  702, 1005

\bibitem[{Strehl \& Ghosh(2003)}]{Strehl2003}
Strehl A., Ghosh J., 2003, J. Mach. Learn. Res., 3, 583

\bibitem[{{Taranu} {et~al.}(2015){Taranu}, {Dubinski} \& {Yee}}]{Taranu2015}
{Taranu} D., {Dubinski} J., {Yee} H.~K.~C., 2015, \apj, 803, 78

\bibitem[{{Toomre}(1977)}]{Toomre1977}
{Toomre} A., 1977, in Evolution of Galaxies and Stellar Populations, {Tinsley}
  B.~M., {Larson} Richard B.~Gehret D.~C., eds., p. 401

\bibitem[{{Toomre} \& {Toomre}(1972)}]{Toomre1972}
{Toomre} A., {Toomre} J., 1972, \apj, 178, 623

\bibitem[{Wang {et~al.}(2019)Wang, Bose, Frenk, Gao, Jenkins, Springel \&
  White}]{Wang2019b}
Wang J., Bose S., Frenk C.~S., Gao L., Jenkins A., Springel V., White S. D.~M.,
  2019

\bibitem[{{Wang} {et~al.}(2011){Wang}, {Navarro}, {Frenk}, {White}, {Springel},
  {Jenkins}, {Helmi}, {Ludlow} \& {Vogelsberger}}]{Wang2011}
{Wang} J., {et~al.}, 2011, \mnras, 413, 1373

\bibitem[{{Wang} {et~al.}(2018){Wang}, {Obreschkow}, {Lagos}, {Sweet},
  {Fisher}, {Glazebrook}, {Macci{\`o}}, {Dutton} \& {Kang}}]{Wang2018}
{Wang} L., {et~al.}, 2018, \apj, 868, 93

\bibitem[{Wang {et~al.}(2017)Wang, Han, Cole, Frenk \& Sawala}]{Wang2017}
Wang W., Han J., Cole S., Frenk C., Sawala T., 2017, Monthly Notices of the
  Royal Astronomical Society, 470, 2351

\bibitem[{Wang {et~al.}(2016)Wang, Pearce, Knebe, Schneider, Srisawat, Tweed,
  Jung, Han, Helly, Onions, Elahi, Thomas, Behroozi, Yi, Rodriguez-Gomez, Mao,
  Jing \& Lin}]{Wang2016b}
Wang Y., {et~al.}, 2016, Monthly Notices of the Royal Astronomical Society,
  459, 1554

\bibitem[{{Wechsler} {et~al.}(2002){Wechsler}, {Bullock}, {Primack}, {Kravtsov}
  \& {Dekel}}]{Wechsler2002}
{Wechsler} R.~H., {Bullock} J.~S., {Primack} J.~R., {Kravtsov} A.~V., {Dekel}
  A., 2002, \apj, 568, 52

\bibitem[{{Weil} \& {Hernquist}(1996)}]{Weil1996}
{Weil} M.~L., {Hernquist} L., 1996, \apj, 460, 101

\bibitem[{{White} \& {Frenk}(1991)}]{White1991}
{White} S.~D.~M., {Frenk} C.~S., 1991, \apj, 379, 52

\bibitem[{{White} \& {Rees}(1978)}]{White1978}
{White} S.~D.~M., {Rees} M.~J., 1978, \mnras, 183, 341

\bibitem[{Zehavi {et~al.}(2018)Zehavi, Contreras, Padilla, Smith, Baugh \&
  Norberg}]{Zehavi2018}
Zehavi I., Contreras S., Padilla N., Smith N.~J., Baugh C.~M., Norberg P.,
  2018, The Astrophysical Journal, 853, 84

\bibitem[{{Zhang} {et~al.}(2008){Zhang}, {Fakhouri} \& {Ma}}]{Zhang2008}
{Zhang} J., {Fakhouri} O., {Ma} C.-P., 2008, \mnras, 389, 1521

\bibitem[{{Zhao} {et~al.}(2009){Zhao}, {Jing}, {Mo} \& {B{\"o}rner}}]{Zhao2009}
{Zhao} D.~H., {Jing} Y.~P., {Mo} H.~J., {B{\"o}rner} G., 2009, \apj, 707, 354

\bibitem[{{Zolotov} {et~al.}(2009){Zolotov}, {Willman}, {Brooks}, {Governato},
  {Brook}, {Hogg}, {Quinn} \& {Stinson}}]{Zolotov2009}
{Zolotov} A., {Willman} B., {Brooks} A.~M., {Governato} F., {Brook} C.~B.,
  {Hogg} D.~W., {Quinn} T., {Stinson} G., 2009, \apj, 702, 1058

\end{thebibliography}


\appendix

\section{Derivations}\label{a:maths}

\subsection{Generalised entropy with equal masses}\label{aa:equalmass}

\begin{figure}
  \includegraphics[width=\colwidth]{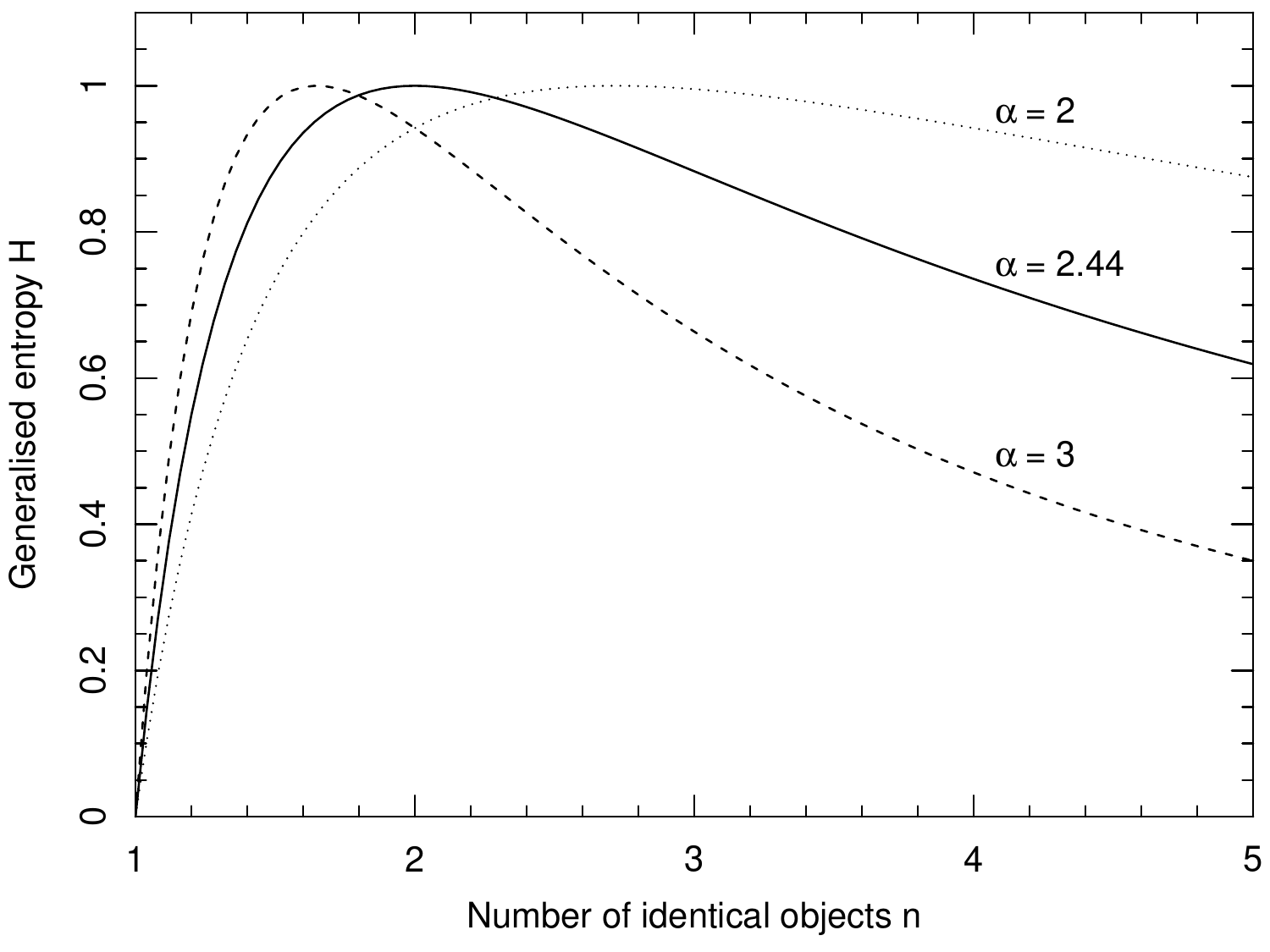}
  \caption{Generalised entropy with $n$ equal masses $x_i=n^{-1}$, here shown in the continuouss extension to real values $n$. The parameter $\alpha=1+1/\ln2\approx2.442695$ produces an entropy that peaks at $n=2$ (solid line).}
  \label{fig:H}
\end{figure}

Let us consider the generalised entropy $H$ of \eq{generalizedshannon}, for equal masses ($x_i=n^{-1}$),
\be
	H(n) = -f\sum_{i=1}^n n^{-\alpha}\ln n^{-1} = f\,n^{1-\alpha} \ln n.
\ee
This function is visualised in \fig{H} for three values of $\alpha$. To find the maximum point let us take the derivative,
\be
	\frac{\d H}{\d n} = f n^{-\alpha}\left[(1-\alpha)\ln n+1\right].
\ee
This equation has one non-trivial root at $n=n_{\rm c}$, where
\be
	n_{\rm c} = e^{\frac{1}{\alpha-1}}~~~\leftrightarrow~~~\alpha = 1+\frac{1}{\ln n_{\rm c}}.
\ee
This root corresponds to the only maximum of $H(n)$. The maximum value is
\be
	H(n_{\rm c}) = \frac{f}{e(\alpha-1)}.
\ee
Thus, the generalised entropy $H$ is normalised to the interval $[0,1]$ by setting,
\be
	f = e(\alpha-1)=\frac{e}{\ln n_{\rm c}}.
\ee

In this work, we chose $n_{\rm c}=2$ (\sss{selfsimilar}), implying $f=e/\ln2\approx3.921652$ and $\alpha=1+1/\ln2\approx2.442695$.

\subsection{Propagation of the tree entropy}\label{aa:smooth}

This section elaborates on selected aspects of the entropy propagation (\eq{defsn}). We first consider the limiting case, where this equation is applied to smooth accretion made of infinitesimal elements. At each infinitesimal time-step, the main branch of mass $m$ and entropy $s$, undergoes an infinitesimal $n$-merger, adding a mass $\dm$ in $k=(n-1)\geq1$ elements, each bringing a mass fraction $\dx=(\dm/k)/(m+\dm)=\dm/(km)+\O(\dm^2)$. The entropy resulting from this event is
\be\label{eq:snsmooth}
	\sn = H+w(1-2k\dx)(s-H)+\O(\dx^2),
\ee
We can expand $H=H'\dx+\O(\dx^2)=kf\dx+\O(\dx^2)$ and $w=1+aH'\dx+\O(\dx^2)=1+akf\dx+\O(\dx^2)$, where $H'=kf$ is the derivative of $H$ with respect to $\dx$ at $\dx=0$. Thus,
\be
	\sn = s+(af-2)sk\dx+\O(\dx^2),
\ee
and invoking $\ds\equiv\sn-s$ and $k\dx=\dm/m+\O(\dm^2)$,
\be\label{eq:ds}
	(\ds/s) = (af-2)(\dm/m)+\O(\dm^2).
\ee
In the infinitesimal limit, $({\d}s/s) = -\gamma({\d}m/m)$ (condition~9 in \sss{nonselfsimilar}) iff $\gamma=2-af\leftrightarrow a=(2-\gamma)/f$. \eq{ds} is independent of the order $n$ (and $k$) of the merger (condition~5). Moreover, the evolution of $\sn$ does not depend on the tree entropy of the diffuse material (2nd part of condition~5), since the latter only appears in the $\dx^2$ term in \eq{snsmooth} due to the $x^2$-terms in \eq{defsn}. Finally, \eq{ds} shows that $s(t)$ tends to zero as the smoothly accreted mass tends to infinity (for $\gamma>0$, condition~6).

To address condition~8 (\sss{nonselfsimilar}), let us consider a merger of $n=n_{\rm c}$ haloes of identical mass ($x_i=n_{\rm c}^{-1}$) and vanishing entropy. By definition $H=1$ in this case and hence \eq{defsn} simplifies to
\be
	\sn = 1+wn_{\rm c}(0-1)n_{\rm c}^{-2}=1-(1+a+b)n_{\rm c}^{-1}.
\ee
The requirement that $\sn=\beta$ (condition~8) is thus satisfied iff $b=n_{\rm c}(1-\beta)-1-a$.


\section{Numerical convergence}\label{a:convergence}

\subsection{Mass resolution}\label{aa:massconvergence}

\begin{figure}
  \includegraphics[width=\colwidth]{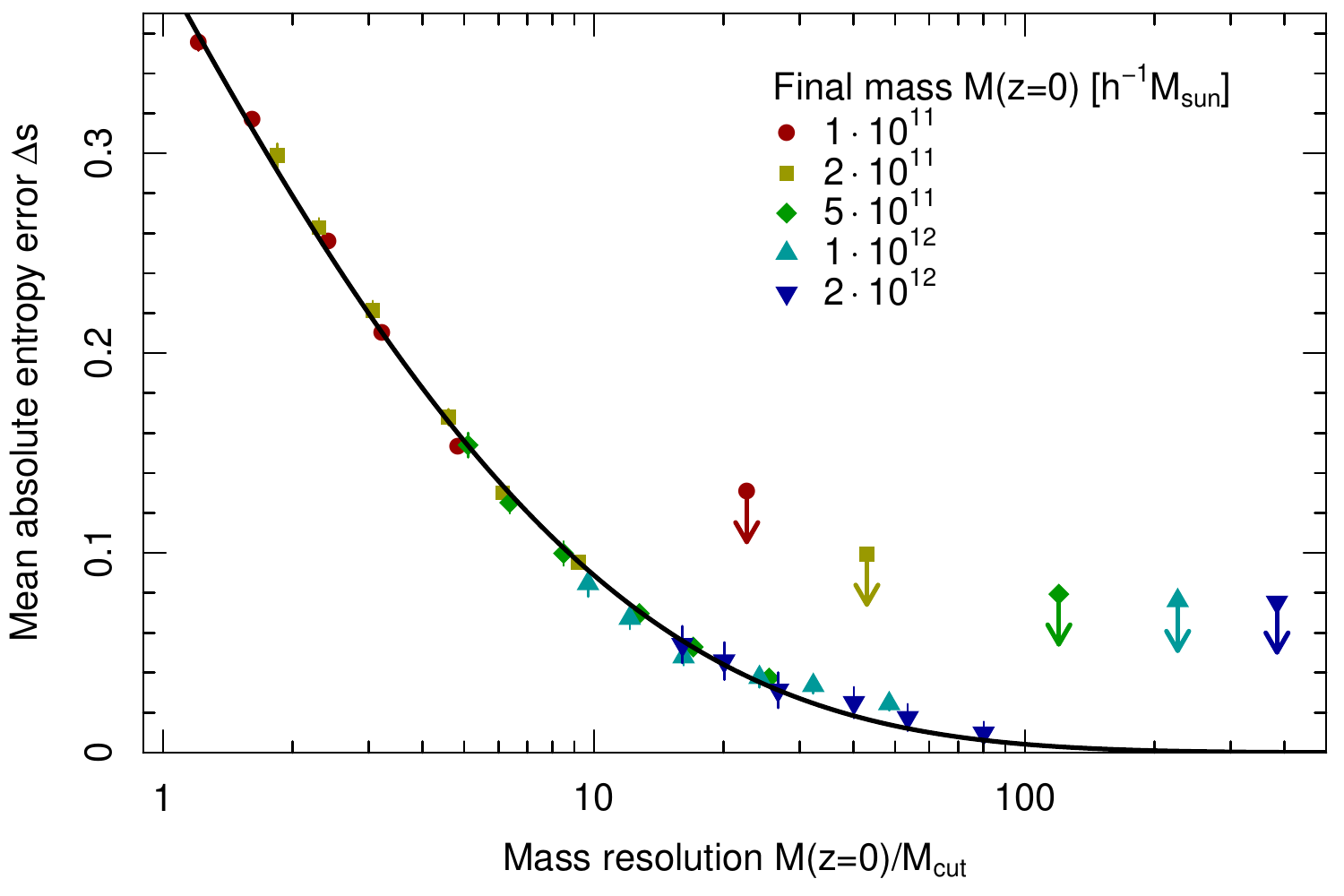}
  \caption{The mean absolute entropy error $\Delta s$ of haloes appears to be a universal function of the relative mass resolution $\mvir/\mcut$. Arrows show the maximum resolution achievable at the five final masses, respectively, given the halo mass resolution of the simulation ($\mres=4.42\cdot10^9\msunh$). In this paper we are interested in keeping relative tree entropy errors below 10\pc (absolute errors $\lesssim0.02$), which is achieved by haloes with final masses $\mvir\geq10^{11}\msunh$ ($>452$ particles).}
  \label{fig:mass_convergence}
\end{figure}

The \surfs simulation used in this work resolves haloes with masses above $\mres=4.42\cdot10^9\msunh$, corresponding to at least 20 particles. Therefore, only progenitors with masses $\geq\mres$ are accounted for in computing the tree entropies. This raises the question as to how well these entropies are numerically converged, \ie how far do they lie from the corresponding values in a simulation with infinite mass resolution ($\mres=0$)? To answer this question, we artificially deteriorated the mass resolution by ignoring all progenitor haloes with masses below some threshold $\mcut>\mres$. For each tree, we then compute the tree entropy error $\Delta s$, defined as the absolute difference in $s$ between the deteriorated tree and the full-resolution tree.

\fig{mass_convergence} shows the mean tree entropy error $\Delta s$ at $z=0$, for different values of $\mvir/\mcut$. Interestingly, this function appears to be universal (solid line), irrespective of the final mass $\mvir$ (different symbols in the figure). We only plotted the symbols with $\mcut>4\mres$, since $\Delta s$ strongly underestimates the real entropy error if $\mcut$ gets close to $\mres$. In fact, $\Delta s$ always vanishes if $\mcut=\mres$.

The universality of the function $\Delta s(\mvir/\mcut)$ justifies the use of this function as an approximation of the true entropy errors for haloes of a given mass. The arrows in \fig{mass_convergence} show the mass resolution at full resolution (\ie if $\mcut=\mres$) for final halo masses $\mvir/[\msunh]=10^{11},\ 2\cdot10^{11},\ 5\cdot10^{11},\ 10^{12},\ 2\cdot10^{12}$. Haloes of $\mvir=10^{11}\msunh$ are predicted to produce entropy errors of about 0.02, which is about the highest entropy error we are willing to accept in this paper as it lies safely (an order of magnitude) below the width of the entropy distribution (\fig{entropy_stats}). 

We therefore chose to limit the analysis of this paper to haloes with $\mvir\geq10^{11}\msunh$ ($>452$ particles). These haloes have a mass resolution of $\mvir/\mres\geq22.6$, which means that even in our least massive haloes ($10^{11}\msunh$), minor mergers down to at least $r\approx0.044$ are resolved.

\subsection{Time resolution}\label{aa:timeconvergence}

\begin{figure}
  \includegraphics[width=\colwidth]{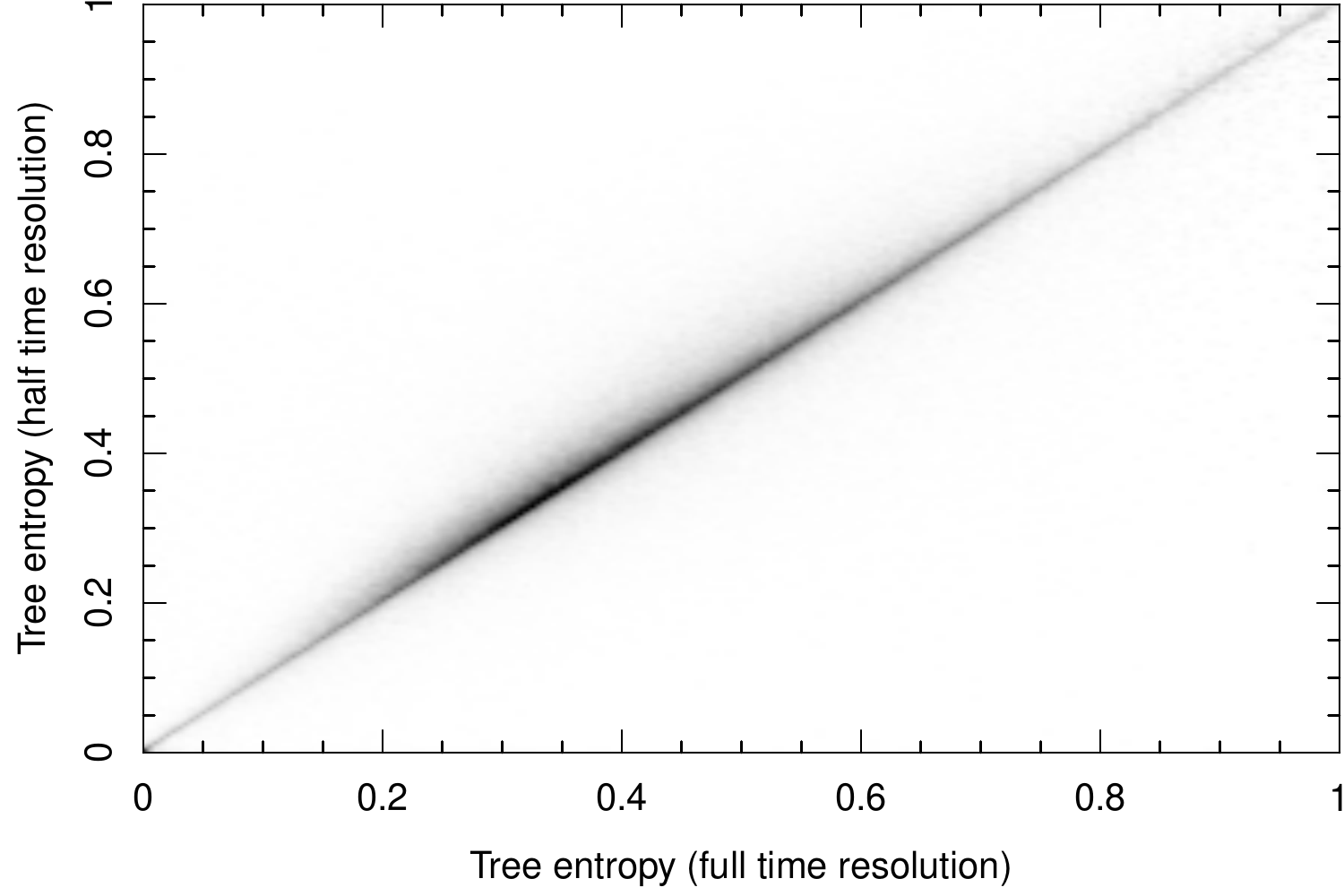}
  \caption{Density plot comparing a degraded tree entropy, computed by skipping every second snapshot, against the `true' tree entropy, computed using all simulation snapshots.}
  \label{fig:time_convergence}
\end{figure}

As argued in \ss{computeentropy}, the time resolution of the \surfs simulation is expected to suffice for a robust extraction of merger trees. To verify this expectation, we artificially degraded the trees, by using only every second snapshot of the simulation (\ie 100 snapshots instead of 200) to build the merger trees and compute the tree entropies. As shown in \fig{time_convergence}, this only marginally changes the individual entropies. The error $\Delta s=s_{\rm half}-s_{\rm full}$ has a mean of $0.014$ and a standard deviation of $0.060$; the corresponding relative error $|\Delta s/s_{\rm full}|$ lies below 10\pc. None of the results in this paper are sensitive to such a small variation.

\section{Tree Entropy Code}\label{a:code}

This pseudo-code can be used to propagate the tree entropy.

\lstset{frame=tb,
  language=C,
  aboveskip=3mm,
  belowskip=3mm,
  columns=flexible,
  basicstyle={\small\ttfamily},
  commentstyle=\color{blue},
  breaklines=false,
  keywordstyle=\bfseries,
  morekeywords={if,return,function}
}

\begin{lstlisting}
// user parameters
nc = 2 // order of most destructive merger
beta = 0.75 // aggressiveness of major mergers
gamma = 1.0/3.0 // aggressiveness of smooth accretion

// derived parameters
alpha = 1/log(nc)+1
f = (alpha-1)*exp(1)
a = (2-gamma)/f
b = nc*(1-beta)-1-a

function snew(m0,mf,m,s) {
   // m0 = mass of new halo (>=0)
   // mf = mass at infall for satellites, 0 for mains
   // m(:) = vector of all progenitor masses (>=0)
   // s(:) = vector of all progenitor entropies [0...1]
   
   // merger (also works for one progenitor w/o merger)
   x = m/(sum(m)+1e-9) // normalised masses
   H = -f*sum(x^alpha*log(x+1e-9)) // generalised entropy
   w = 1+a*H+b*H^2 // weight
   snew = H+w*sum(x^2*(s-H)) // new tree entropy
   
   // smooth accretion (gains and losses)
   if (m0>mf) {snew = min(1.0,snew*(sum(m)/m0)^gamma)}
}

\end{lstlisting}

\end{document}